\documentclass[%
 reprint,
superscriptaddress,
groupedaddress,
 amsmath,amssymb,
aps,
pra,
floatfix
]{revtex4-2}

\usepackage{graphicx}% Include figure files
\usepackage{dcolumn}% Align table columns on decimal point
\usepackage{bm}% bold math
\usepackage{xcolor}
\usepackage[mathlines]{lineno}% Enable numbering of text and display math
\usepackage{multirow} % Para celdas combinadas en tablas
\usepackage{pbox}% self-shrinking parbox widths
\usepackage{tabularx}
\usepackage{makecell}

\usepackage{mathtools} 
\usepackage{amssymb} 
\usepackage{booktabs} 

\newcommand{\tabincell}[2]{\begin{tabular}{@{}#1@{}}#2\end{tabular}}

\newcommand{\ams}{\text{\r{A}}}

\begin{document}

\preprint{APS/123-QED}

\title{Steepest-Entropy-Ascent Framework for Predicting Arsenic Adsorption on Graphene Oxide Surfaces --- A Case Study}% Force line breaks with \\

\author{Adriana Saldana-Robles}
\altaffiliation[Also at ]{Department of Agricultural Engineering, University of Guanajuato.}%Lines break automatically or can be forced with \\
  \email{adrianarobl@vt.edu}
\author{Cesar Damian}%
\altaffiliation[Also at ]{Department of Mechanical Engineering, University of Guanajuato.}%Lines break automatically or can be forced with \\
 \email{cesarasce@vt.edu}
 \author{Michael R. von Spakovsky}%
 \email{vonspako@vt.edu}
  \author{William T. Reynolds Jr.}%
 \email{reynolds@vt.edu}
\affiliation{
Center for Energy System Research, Mechanical Engineering Department, Virginia Tech, Blacksburg, VA 24061
}%


%\affiliation{
%\affiliation{%

\date{\today}% It is always \today, today,

\begin{abstract}
Water contamination by arsenic\,(V) constitutes a major public-health concern, underscoring the need for models that capture both equilibrium and transient adsorption behaviour.  A framework that can do so is the steepest-entropy-ascent quantum thermodynamic (SEAQT) framework, which is used here to describe the uptake of As(V) on graphene oxide (GO) across pollutant concentrations of 25–350\,mg\,L$^{-1}$.  A non-equilibrium equation of motion derived from the steepest-entropy-ascent principle for a five-component system (water, arsenic, two GO functional groups, and $\mathrm{H}^{+}$) is solved with an energy eigenstructure generated by a Replica-Exchange Wang–Landau algorithm and then extrapolated to relevant contaminant concentrations via an artificial neural network.  Without recourse to empirical rate laws, the model predicts the time-dependent adsorption capacity, the stable-equilibrium arsenic concentration, and the pH dependence of removal efficiency.  Equilibrium capacities are reproduced within 5\,\% of experimental isotherms, and the characteristic adsorption time aligns with the reported kinetics.  These results indicate that SEAQT framework provides a thermodynamically consistent, fully predictive tool for designing and optimising adsorbent-based water-treatment technologies.

\end{abstract}

\maketitle

%\tableofcontents

\section{Introduction}
Toxic levels of arsenic contamination in groundwater is a recognized and significant global issue. Arsenic contamination affects over 70 countries with reported groundwater concentrations ranging from 0.10 to 163,000~$\mu$g/L~\cite{kobyareviewdecontaminationarseniccontained2020}. Human consumption of arsenic through drinking water impacts more than 2 million people worldwide~\cite{kobyareviewdecontaminationarseniccontained2020, ullahArsenicContaminationWater2023, adelojuArsenicContaminationGroundwater2021}. Countries such as India, Bangladesh, China, Vietnam, Argentina, Cambodia, Nepal, Bolivia, Mexico, Chile, and Brazil are particularly affected~\cite{rahmanArsenicenrichmentits2021, nguyenReviewTheoreticalKnowledge2023}. Arsenic contamination can originate from both natural and anthropogenic sources with inorganic arsenic, typically in the form of arsenate (a chemical compound of arsenic in its +5 oxidation state such as arsenic acid H$_3$AsO$_4$ and its salts), being the most common form found in groundwater~\cite{rathireviewsourcesidentification2021, kumarScenarioperspectivesmechanism2020}. Human exposure to arsenic degrades health through kidney failure, hair loss, cardiovascular diseases, and cancer~\cite{tabassumHealthriskassessment2019, adebayoGeospatialmappingsuitability2021, malsawmdawngzelaSynthesisnovelclaybased2023}. As a consequence, arsenic has been categorized as a class 1 human toxic element by the International Agency for Research on Cancer (IARC)~\cite{tropeaEnvironmentaladaptationcoli2021}, and the World Health Organization (WHO) has lowered the acceptable level of arsenic in drinking water to 10~$\mu$g/L~\cite{natashaHealthrisksarsenic2021,zhouGroundwaterqualitypotable2021}. 

Several strategies have been developed for arsenic removal from ground water, including ion exchange, coagulation–flocculation, oxidation, and adsorption~\cite{rahmanArsenicenrichmentits2021, rahidulhassanReviewDifferentArsenic2023, litterArsenicArgentinaTechnologies2019}. Among these, adsorption is considered the most scalable method because of its cost-effectiveness, high efficiency, and its simplicity in operation and maintenance. As a result, a wide range of adsorbent materials have been developed for removing arsenic from water.  Graphene oxide (GO) is one such material that is being considered for removing metals like chromium and arsenic from water because it has the potential for being functionalized to bind metals and furthermore has a high porosity and surface area. These properties enable efficient removal, making GO a potentially cost-effective adsorbent material in water-contaminant adsorption applications~\cite{joya-cardenasGrapheneBasedAdsorbentsArsenic2022,ansariSeparationPerchloratesAqueous2019}. Graphene oxide is a two-dimensional (2D) material with functional groups on its surfaces, including hydroxyl (OH$^-$) and carboxyl groups (COOH$^-$)~\cite{reynosa-martinezEffectDegreeOxidation2020,zhangCombinedEffectsFunctional2015}. Several experimental studies have been conducted to assess its adsorption capacity and understand the mechanisms of arsenic adsorption on GO-based materials~\cite{heDesignOptimizationNovel2023,joya-cardenasRemovalPresenceCr2024,hossainCosteffectiveSynthesisMagnetic2024}. Such studies provide insight into the behavior of specific systems, but such comprehensive experimental investigations tend to be expensive~\cite{vazquez-jaimeEffectiveRemovalArsenic2020, lingamdinnePotentialMagneticHollow2021, choiFabricationChitosanGraphene2020, tabatabaieebafrooeeEthylenediamineFunctionalizedMagnetic2021}. On the other hand, computational models of adsorption mechanisms~\cite{maComputationalStudyAdsorption2022, gazzariInteractionTrivalentArsenic2019, srivastavaRemovalArsenicIons2017} often lack experimental validation or the power to predict kinetic paths and the performance of adsorbent materials. 

{  Despite extensive experimental, DFT-level and kinetic-model studies of arsenate removal with graphene-oxide (GO) adsorbents, the current literature still treats equilibrium uptake (e.g., Langmuir/Freundlich isotherms, grand-canonical Monte Carlo) and transient uptake (empirical rate laws, molecular dynamics) in isolation, rarely spanning the experimentally relevant concentration range (25–350 mg L$^{-1}$) without parameter fitting. The absence of a theoretical approach that can encompass transient and equilibrium adsorption without overly restrictive structural assumptions is a significant research lacuna. Here we close this gap by applying the steepest-entropy-ascent quantum thermodynamic (SEAQT) framework -- a unified, first-principles method -- to both the equilibrium adsorption and the non-equilibrium adsorption dynamics of As(V) on GO. The framework predicts time-dependent adsorption capacity, stable-equilibrium isotherms, and pH-dependent removal efficiencies directly from fundamental pair potentials. }

In recent years, the SEAQT framework has been developed to describe a very diverse variety of non-equilibrium processes such as the chemical adsorption of ammonia on GaN (0001)~\cite{kusabaCH4AdsorptionProbability2019,kusabaModelingNonEquilibriumProcess2017}, the folding behavior of polymer chains~\cite{mcdonaldPredictingnonequilibriumfolding2023}, ion sequestration on polymer chains~\cite{mcdonaldPredictingIonSequestration2024}, and many other processes (e.g., ~\cite{mcdonaldPredictingPolymerBrush2023,mcdonaldEntropydrivenmicrostructureevolution2022,yamadaLowtemperatureAtomisticSpin2019,yamadaPredictingcontinuousdiscontinuous2019,yamadaKineticpathwaysordering2020,yamadamethodpredictingnonequilibrium2018,goswamithermodynamicscalinglaw2021,vonspakovskyPredictingChemicalKinetics2020,montanez-barreraDecoherencepredictionssuperconducting2022,montanez-barreraLossofentanglementpredictioncontrolledphase2020,cano-andradeSteepestentropyascentquantumthermodynamic2015,liSteepestentropyascent2018}).  It is a theoretically and mathematically rigorous framework that extends quantum dynamics into the thermodynamic realm of irreversible processes by postulating that isolated systems evolve along a unique thermodynamic path that maximizes the production of entropy at each instant of time~\cite{berettaSteepestEntropyAscent2014,berettaFourthLawThermodynamics2020, berettaMaximumEntropyProduction2010}. The framework employs energy and entropy as fundamental variables and asserts that the principle of SEA or maximum entropy production leads to an equation of motion that uniquely determines the path to stable (thermodynamic) equilibrium. McDonald $et \; al.$~\cite{mcdonaldPredictingIonSequestration2024} applied the SEAQT framework to europium sequestration from an aqueous solution. In the present contribution, a generalized version of that approach~\cite{saldana-roblesModelpredictingadsorption2025} is used to predict the kinetics, adsorption capacity, and thermodynamic properties of As adsorption on graphene oxide. As the first step in applying the SEAQT framework, an energy eigenstructure, which defines a system's density of states (i.e., degeneracies) and discrete energy spectrum, must be constructed for the physical system from which the system energy and entropy are directly calculated. The Replica-Exchange Wang-Landau (REWL) algorithm ~\cite{vogelGenericHierarchicalFramework2013, vogelScalableReplicaexchangeFramework2014,vogelPracticalGuideReplicaexchange2018} is used for this purpose. This algorithm uses 2D or 3D lattices of various sizes to yield the energy eigenlevels (discrete energy spectrum) and their degeneracies for a physical system comprised of a fixed quantity of the component species (e.g., As, H$^+$, and H$_2$O as well as the OH$^-$ and COOH$^-$ functional groups on graphene oxide) while monitoring the average number of adsorbed particles per energy eigenlevel. The results of this process are then used as training data for an artificial neural network (ANN) to predict energy eigenstructures at component concentrations well below what is accessible with the REWL algorithm but in the range of what is seen in experimental studies. {  Since adsorption thermodynamics ultimately emerge from electronic-scale donor–acceptor interactions, natural bond orbital (NBO) analysis can quantify the mesoscopic SEAQT treatment with a NBO analysis on representative GO–arsenate systems.}

{  Recent adsorption research has increasingly leveraged ANN models to capture the complex, multi-variable dependencies that frustrate classical isotherm and rate laws.  Comprehensive surveys now document the success of multilayer feed-forward, adaptive network-based fuzzy inference systems and support vector machine architectures for predicting dye uptake across a wide range of operating conditions \cite{ghaediApplicationsartificialneural2017}.  At the materials level, three-layer back-propagation networks have reproduced and optimised Cu(II) removal by laboratory-synthesized graphene oxide with higher fidelity than thirteen alternative learning algorithms \cite{zhangAppraisalCuii2019}, while fixed-bed studies have shown that Levenberg–Marquardt ANN models can forecast breakthrough curves for Fe(III) on biosourced carbons \cite{dasArtificialneuralnetwork2021} and for Cu(II)/Mn(II) on surfactant-decorated graphene with coefficients of determination ($R^{2}$) > 0.99 \cite{yusufFixedbedcolumn2020}.  The present study embeds a two-hidden-layer ANN within the SEAQT framework not to fit macroscopic uptake directly, but rather to extrapolate REWL eigenstructures into the dilute-concentration regime that is experimentally relevant yet computationally inaccessible to direct Monte-Carlo sampling. This in effect unifies a data-driven flexibility with a first-principles thermodynamic foundation. }{   This strategy is compatible with recent adsorption studies that have begun to pair nanostructured sorbents with optimization tools such as response‐surface methodologies, ANN`s, and adaptive network-based fuzzy inference system models to rapidly tune pH as well as dose and contact time for dyes,\cite{deylamiEfficientphotodegradationdisulfine2023,pournamdariResponsesurfacemethodology2024} humic substances\cite{mombenigoodajdarUltrasonicAssistedNeural2023} and even trace herbicides \cite{geramizadeganMolecularlyImprintedPolymers2023}.}

% In the present work, the SEAQT equation of motion is then applied to these eigenstructures to predict the time-varying occupation probability distribution of the energy eigenlevels. These distributions fully describe the states of the adsorption system at any instant of time, providing a detailed picture of the adsorption kinetics and adsorption capacity changes. The computational predictions are validated against an experimental case of As adsorption on GO. 

{  The paper is organized as follows. In the  section \textit{Computational Methods and Details}, the energetic model and the specific interactions used to quantify the adsorption process are presented as is the REWL algorithm, emphasizing the modifications to the original non-Markovian Monte Carlo algorithm, which include incorporating particle number and the averages of adsorbed molecules. A machine learning (ML) algorithm is then introduced that predicts eigenstructures of systems that are larger than what is practical with the REWL algorithm. An outline of the SEAQT equation of motion is subsequently given as is the procedure for relating the SEAQT simulations to experimental parameters. This section concludes with a validation of the SEAQT model against experimental data for stable equilibrium adsorption and a discussion of the thermodynamic consistency of the SEAQT model.  The next section, \textit{Non-equilibrium Solutions}, explores predicted adsorption kinetics under a range of non-equilibrium conditions and shows how the limiting stable equilibrium case in the SEAQT framework relates to classical adsorption isotherm models. The following section, \textit{pH Variation}, then explores how the variation of the solution pH affects the adsorption capacity. The last section offers some final conclusions.}

\section{Computational Methods and Details}

\subsection{Energetic model}
\label{sec:ener}
Arsenic adsorption to GO is modeled as a non-isolated thermodynamic system interacting with a thermal reservoir and involving multiple pair-interactions among multiple species: As adsorbent molecules, molecular sites on the GO adsorbate, and interacting species in the water. The energetic interactions are described by a collection of Lennard-Jones pair-potentials (short-range interactions) and Coulomb interactions (long-range interactions for charged species) applied to components lying on a 2D lattice. The active sites on GO for As adsorption considered in this study are the hydroxyl groups (OH$^-$) and carboxyl groups (COOH$^-$). {  Throughout the manuscript the adsorbate is the mono-protonated arsenate oxyanion, H$_2$AsO$_4^-$ (As(V)), which exceeds an
$\sim$ 95\% abundance at an experimental pH of 3}. These are selected based upon FTIR data obtained from the GO used for study validation~\cite{zhangCombinedEffectsFunctional2015}. Molecular constituents include the a GO active site for each functional group, namely, one active cite for the OH$^{-}$ and one for -COOH$^{-}$, water (H$_2$O, included as a single interacting molecular component which accounts for its electrostatic interactions due to the dipole effects), and dissociated H$^+$ ions to account for solution pH.  

\begin{figure}[htbp]
   \centering
   \includegraphics[scale=0.35]{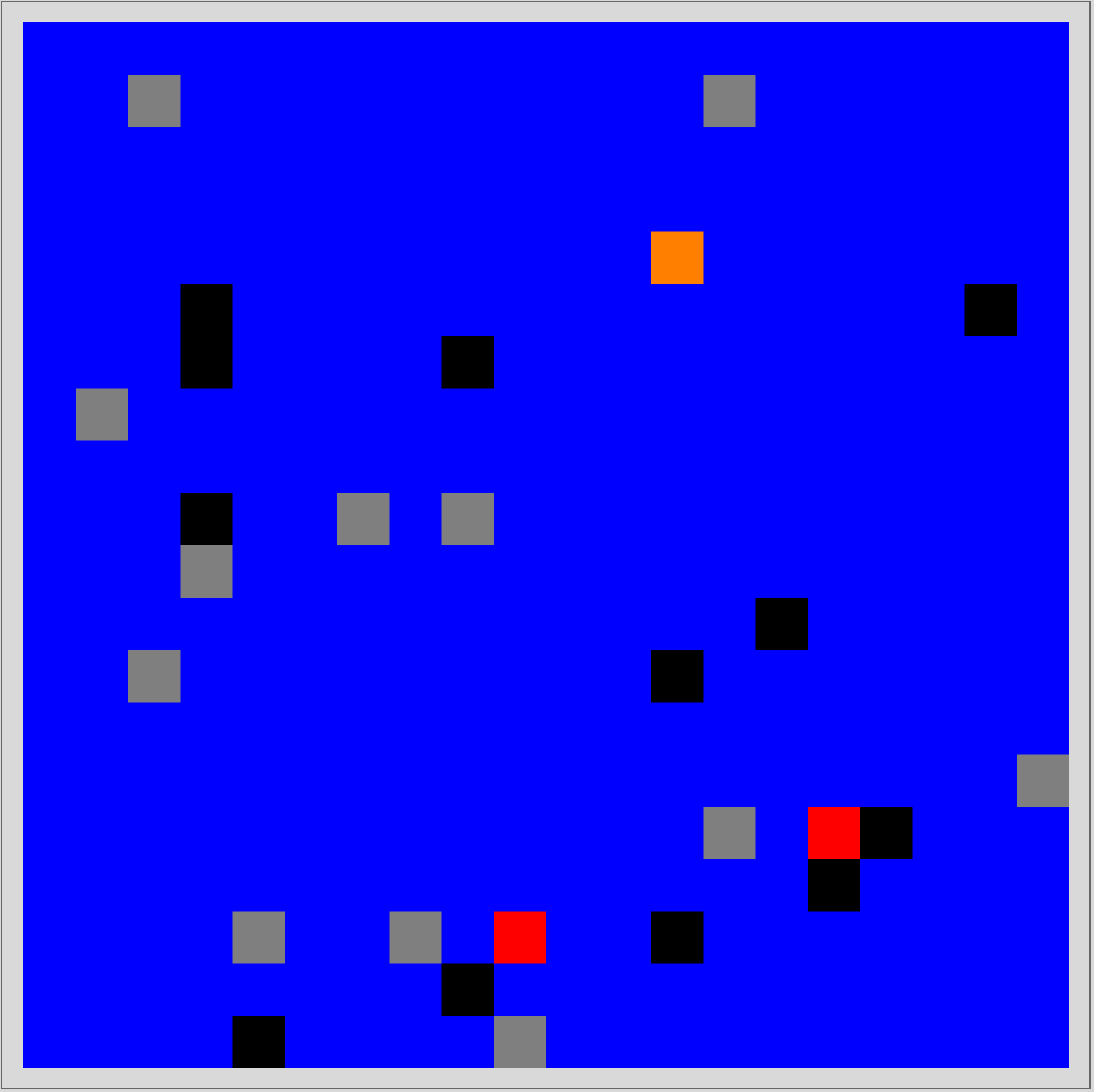} \\
   \caption { Schematic representation of the 2D Wang-Landau simulation domain. The red pixels represent As$^+$ ions, the yellow are H$^+$ ions, the gray OH$^-$ ions, the black COOH$^-$ ions, and the blue H$_2$O molecules. }
   \label{fig:2Dmodel}
\end{figure}

The molecular arrangement consists of a square 2D lattice with periodic boundary conditions on all sides, each side having a length of $X a$, where $a$ is the lattice parameter of the Bond Fluctuation Model (a simple cubic lattice model)  and $X$ is the number of unit cells along each side ($8<X<70$). A schematic arrangement of a molecular lattice is shown in Fig.~\ref{fig:2Dmodel}. For the sake of computational efficiency, a single value of the Lennard-Jones parameter ($\sigma = 2.11 \, \ams$)  was applied to all components. This single value has been calculated as the arithmetic average of all $\sigma$ values considered in Table \ref{tab:pairpot}. This simplification also guarantees that all Lennard-Jones parameters are negative across the entire interaction range. The minimum energy well of the Lennard-Jones potential corresponds to the minimum interaction distance, $2a = 2^{1/6} \sigma = 2.37 \, \ams$, and the equilibrium pair interaction distance. The 2D lattice model constrains the possible interaction distances that can arise in the pair-potential calculations. Three possible distances were included in the energetic calculations, namely, $2a, 2a\sqrt{2}$, and $4a$. The last value is the effective cutoff interaction distance in these simulations. Arsenic is considered to be ``adsorbed'' to GO in the numerical model when an As species is located within the attraction range of one or more of the OH$^-$ or COOH$^-$ functional groups.  The interaction range is any lattice sites within a distance of $4a$. The distance and energy parameters associated with each pair of potentials are provided in Table~\ref{tab:pairpot}. The epsilon parameters of the mixed particle interactions are approximated using the Lorentz and Berthelot equation expressed as
\begin{equation}
\varepsilon_{\alpha \beta}=\sqrt{\varepsilon_{\alpha\alpha} \, \varepsilon_{\beta\beta}}
\end{equation}
where the subscripts $\alpha$ and $\beta$ correspond to each molecule of an interacting pair. The long-range electrostatic interactions are calculated from the charges on the individual ionic species (or a mean value for the water molecule)  for computational efficiency. These are applicable for values of $r\leq 4  a$.

% REVISED VERSION OF TABLE I
\begin{table*}[tp]%[hp]  % La opción [h] intenta colocar la tabla "aquí"
  \caption{Pair potentials used in the As-GO adsorption simulations}
  \label{tab:pairpot}
\begin{tabularx}{\textwidth}{cccccc}
\toprule
 \multicolumn{1}{c}{\boldmath{$V^{\phi}_{n,m}$} }& 
 \multicolumn{1}{c}{\tabincell{c}{\boldmath{$n,m$} \textbf{Molecule} \\ \textbf{Pair Types}}} &
 \multicolumn{1}{c}{\tabincell{c}
 {\textbf{Pair Potential} \\ \textbf{Functions}}} &
% \multicolumn{2}{c}{\textbf{Pair Potential Function}} &
 \multicolumn{2}{c}{\textbf{Parameters}}\\
\midrule
$V_{\text{LJ}}$ & \makecell{van der Waals \\ {\scriptsize (OH:COOH:As:H:O)}} & $\;\;\;\; 4\,\varepsilon _{\alpha\beta}\left(\left(\frac{\sigma }{r_{\text{}}}\right)^{12}-\left(\frac{\sigma }{r_{\text{}}}\right)^6\right)$ & &\makecell{\rule[-0.3\baselineskip]{0pt}{4mm} 
$\varepsilon_{\text{OH-OH}}=0.141 $\\ $\sigma_{\text{OH-OH}}=1.627$ \\[2mm]
$\varepsilon_{\text{COOH-COOH}}=0.152 $\\ $\sigma_{\text{COOH-COOH}}=2.960$ \\[2mm]
$\varepsilon_{\text{O-O}}=0.230 $\\ $\sigma_{\text{O-O}}=1.768$ \\[2mm]
$\varepsilon_{\text{As-As}}=0.230 $\\ $\sigma_{\text{As-As}}=4.00$ \\[2mm]
$\varepsilon_{\text{H-H}}=0.026 $\\ $\sigma_{\text{H-H}}=0.224$ }\vspace{3mm} & \multirow{2}{*}{\textsuperscript{a,b,c}}\\
$V_{\text{Coulomb}}$ & Electrostatic  & $\; \frac{q_1 \, q_2}{4\pi \epsilon_0 r} \;$ \rule[- 0.3\baselineskip]{0mm}{9mm} & &\makecell{\rule[-0.3\baselineskip]{0pt}{7mm} 
$q_{\text{OH$^-$}} = -0.433$\\ $q_{\text{COOH$^-$}} = -0.440$ \\
$q_{\text{O$^-$}} = -0.834$ \\[2mm]
$q_{\text{As$^+$}} = 1.941$ \\ $q_{\text{H$^+$}} = 0.417$}\vspace{2mm} & \multirow{2}{*}{\textsuperscript{d,e}}\\[13mm]
\bottomrule
\end{tabularx}
\flushleft{\footnotesize{\textsuperscript{a} Values are taken from~\cite{ansariSeparationPerchloratesAqueous2019,ansariComputationalStudyRemoval2020} \\
\textsuperscript{b} The $\varepsilon$ quantities are given in units of $\frac{\text{kcal}}{\text{mol}}$, $\sigma$ in units of $\ams$. \\ 
\textsuperscript{c} Mixed molecule interactions are approximated using the Lorentz and Berthelot equation $\varepsilon _{\alpha\beta} =\sqrt{\varepsilon _{\alpha\alpha}\,\varepsilon _{\beta\beta}}$.\\ 
\textsuperscript{d} Values are taken from~\cite{tanakaDifferenceDiffusionCoefficients2013,jorgensenDevelopmentTestingOPLS1996} \\
\textsuperscript{e} The $q$ values are given in elementary charge units.}}
\end{table*}

% ORIGINAL TABLE
%\begin{table*}[!ht]  % La opción [h] intenta colocar la tabla "aquí"
%  \centering
%  \caption{\label{tab:pairpot}Pair potentials used in the adsorption As-Go simulations}
%\begin{tabular}{|p{3cm}|p{5.5cm}|p{4cm}|}
%\hline
%V$_{n,m}^{\phi}$ 		& V$_{LJ}$ 								& V$_{Coulomb}$ \\
%\hline 
%n,m molecule pair 		& van der Waals  				&  Coulomb  \\ \hline 
%					& - OH : -COOH : As : H : O		& -OH : -COOH : As : H : O \\ \hline  
%											
%Potential function		&  4 $\epsilon_{\alpha \beta}((\frac{\sigma}{r})^{12}-(\frac{\sigma}{r})^{6})$	& $\frac{q_1 q_2}{4\phi \epsilon_0r}$  \\
%Parameters			& $\epsilon_{As}=0.230 \frac{kcal}{mol}, \sigma_{As}= 4.00 \ams$
%					   $\epsilon_{OH}=0.141 \frac{kcal}{mol}, \sigma_{OH}= 1.627 \ams$
%					   $\epsilon_{COOH}=0.152 \frac{kcal}{mol}, \sigma_{COOH}= 2.960 \ams$
%					   $\epsilon_{O}=0.23 \frac{kcal}{mol}, \sigma_{O}= 1.768 \ams$			
%					   $\epsilon_{H}=0.026 \frac{kcal}{mol}, \sigma_{H}= 0.224 \ams$													   & q$_As = 1.941$
%					       q$_{OH} = -0.433$
%					       q$_{COOH} = -0.440$
%					       q$_{H} = 0.417$
%					       q$_{O} = -0.834$ \\
%\hline \hline
%\end{tabular}
%\end{table*}

%\begin{figure}[htbp]
%   \centering
%   a) \includegraphics[width=0.8\linewidth]{Figures/VLJ.pdf} \\
%   b) \includegraphics[width=0.8\linewidth]{Figures/Vcoulomb.pdf}
%   \caption {Pair potentials used in the simulations include a) 6-12 Lennard-Jones potentials representing non-%bonding interactions between As and GO and b) Coulomb potentials representing electrostatic interactions between As %and GO.}
%   \label{fig:Vpot}
%\end{figure}

The energy of any specific configuration or arrangement of molecules arranged on a lattice can be calculated from the sum of the interaction potentials among all molecular pairs~\cite{mcdonaldPredictingPolymerBrush2023, mcdonaldPredictingIonSequestration2024}. The potential of each molecular pair depends on the nature of the two interacting molecules and their separation distance. Thus, the discrete energy eigenvalue, $e_j$ of a specific configuration is calculated from the relationship~\cite{mcdonaldPredictingnonequilibriumfolding2023, mcdonaldPredictingIonSequestration2024}
\begin{equation}
% E_j=\frac{1}{2} \sum_{n=1}^{N_{T}} \sum_{m=1, \\ m\neq n}^{N_{T}}V_{n,m}^{\phi}
e_j = \frac{1}{2}\sum_{n=1}^{N_{T}} \; \left. \sum_{\substack{m=1 \\ m\neq n}}^{N_{T}} V_{n,m}^{\phi} \right.
\label{eq:hamiltonian}
\end{equation}
 where $m$ and $n$ refer to two interacting molecules, and the summations run to the total number of molecules in the system, $N_T$. The factor of $\frac{1}{2}$ accounts for double counting of bonding pairs. The parameter $N_T$ includes {  As(V) molecules}, the active sites on the GO (OH$^-$, COOH$^-$), the H$^+$ ion molecule, and the H$_2$O molecule. The $\phi$ of the $V_{n,m}^{\phi}$ factor indicates the type of pair interaction (Lennard-Jones or Coulomb). Both types of potential are included for each interacting pair.

The efficiency of the adsorption process is measured by the standard adsorption capacity at stable equilibrium, $q_{\text eq}$, written as 
\begin{equation}
q_{\text eq}=\frac{(C_o-C_{eq})V}{m}
 \end{equation}
where $C_o$ is the initial concentration of the adsorbate in the liquid phase, $C_{eq}$ is the stable equilibrium concentration of the adsorbate in the liquid phase, $V$ is the volume of the liquid phase, and $m$ is the mass of the adsorbent.

\subsection{Replica-exchange Wang-Landau algorithm}
\label{sec:WL}
\begin{figure}[htbp]
    \centering
    a) \includegraphics[width=0.7\linewidth]{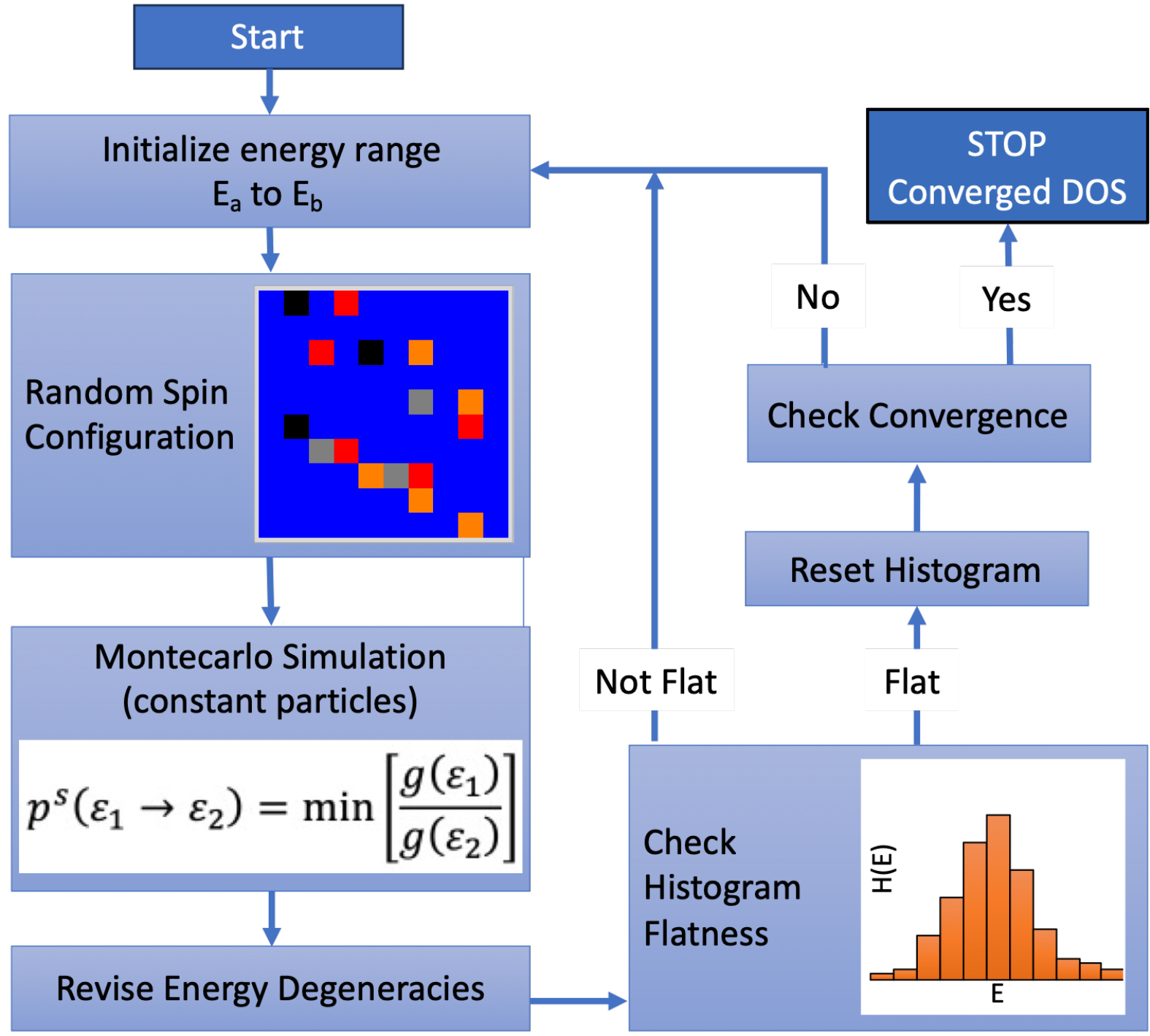}\\[10mm]
    b) \includegraphics[width=0.6\linewidth]{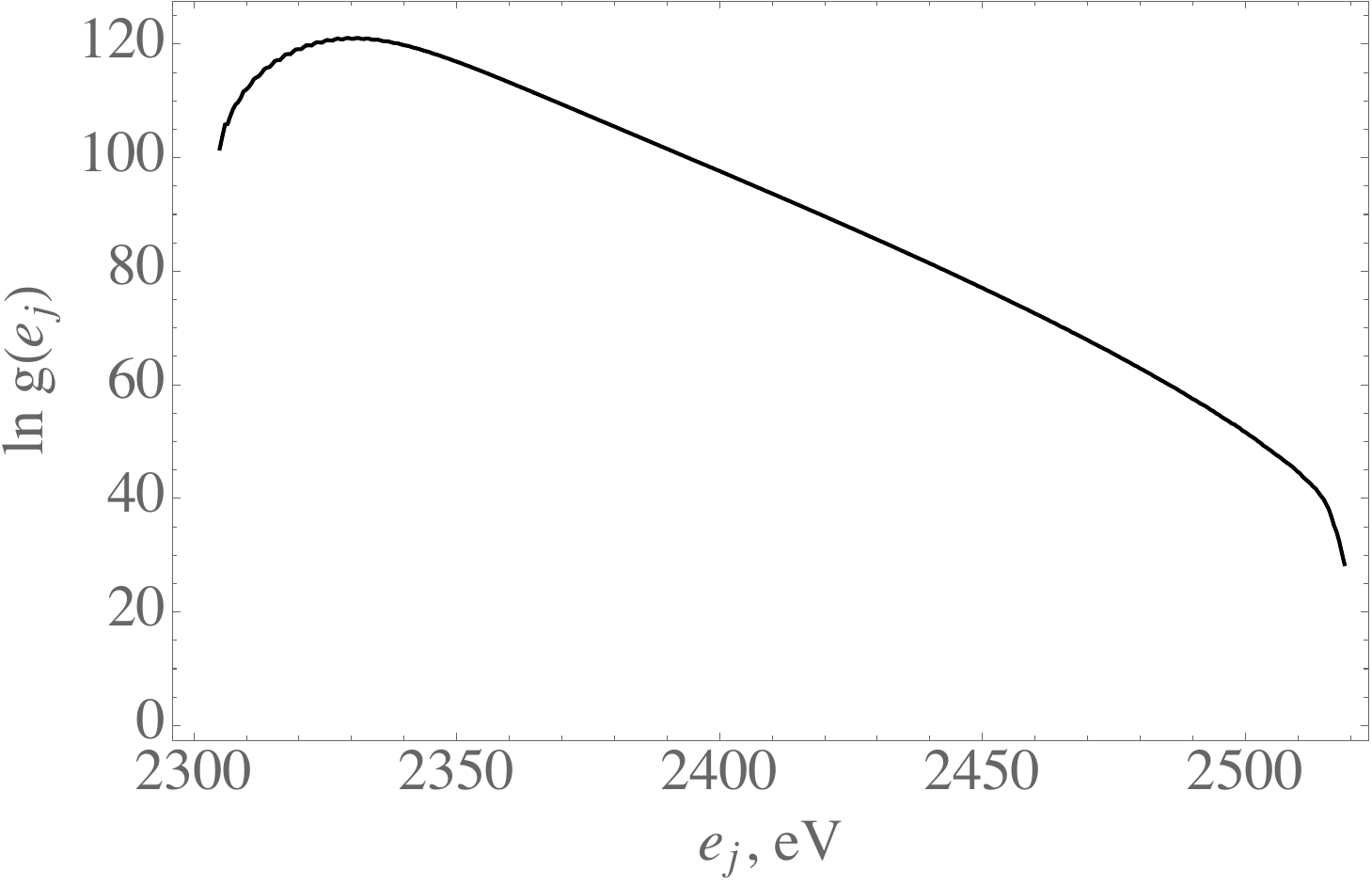}
    \caption{a) Flowchart of the REWL algorithm employed to construct the energy eigenstructure of b) the density of states for a sample lattice with 12 COOH$^-$ and 12 OH$^{-}$. %{  Also, in the dark blue box of a) make the lettering for "Converged DOS" white.}
    }
\label{fig:REWL}
\end{figure}

{  In the SEAQT description, solving the equation of motion first requires knowledge of the energy spectrum and the corresponding degeneracies (i.e., the density of states). This information is encoded in the energy
eigenstructure, which is represented as a finite set of many-body pairs
$\{e_j, g_j\}_{j=1}^{\Omega}$. Here $e_j$ is the energy of eigenlevel $j$ and $g_j$ its degeneracy. For every level we also store the eigenvalue $n^{ad}_j$ of the adsorbed-particle-number operator.} The Wang-Landau algorithm is a non-Markovian Monte Carlo method for estimating the degeneracies, $g_j$, of the discrete energy eigenlevels, $e_j$, of a system~\cite{wangDeterminingDensityStates2001,  wangEfficientMultipleRangeRandom2001}.  The REWL algorithm~\cite{vogelGenericHierarchicalFramework2013, vogelScalableReplicaexchangeFramework2014, liNewParadigmPetascale2014, vogelPracticalGuideReplicaexchange2018} is a parallelized, computationally efficient variant of the Wang-Landau method.  It proceeds by partitioning the whole range of the system's discrete energy spectrum (defined by Eq.~(\ref{eq:hamiltonian})) into smaller overlapping sub-ranges, or energy windows, and performing a parallel series of random walks through the energy levels of each window.  The Monte Carlo transition probabilities in each window are adjusted by updating estimates of the degeneracies until a histogram of walker visitations to each energy eigenlevel becomes flat and the degeneracies converge.  The density of states of adjacent energy windows are periodically replicated and exchanged to construct the energy eigenstructure for the whole energy range of the system. The probability of allowing a replica exchange for configurations $x$ and $y$ across walkers $i$ and $j$, is specified by
\begin{equation}
\mathbb{P} \left( x \leftrightarrow y \right) = \text{min}\, \left[ 1, \frac{g_i (e_x) }{g_i (e_y)} \frac{g_j (e_y)}{g_j (e_y)} \right]
\end{equation}
where $g_i (e_x)$ is the degeneracy level at level $x$, which works as an estimator of the density of states. Each energy window undergoes one or more Wang–Landau sampling operations until its walker visitation histogram reaches a flatness criterion of 0.8. The simulation is terminated when the degeneracies of the slowest converging energy window are modified by a factor of less than $10^{-7}$. A schematic flowchart of the REWL algorithm is shown in Fig.~\ref{fig:REWL} a), while an example of a density of states curve is provided in Fig.~\ref{fig:REWL} b).  The random walkers of the Monte Carlo algorithm vary the locations of the chemical species distributed over the 2D lattice representing As in water with functionalized GO.  The energy of each random configuration is calculated from Eq.~(\ref{eq:hamiltonian}), and the REWL algorithm provides an estimate of the energy eigenstructure for the whole energy range.

The algorithm is applied to a system with a fixed number of particles of each species (As, OH$^-$, COOH$^-$, H$^+$ and H$_2$O). Additionally, a descriptor is introduced to differentiate between As in solution and As that is adsorbed by GO. This descriptor, which is equivalent the the particle number eigenvalue for each eigenlevel, counts the number of configurations where As is located within the electrostatic attraction region of a GO functional group and calculates the arithmetic average of this number for each energy eigenlevel of the eigenstructure.

% \begin{widetext}
% \end{widetext}

The number of energy eigenlevels (the resolution of the discrete energy spectrum) determines the number of simultaneous SEAQT equations of motion that must to be solved (see Section~\ref{sec:SEAQT}). Thus, the number of energy eigenlevels is adjusted by dividing the energies obtained from Eq.~(\ref{eq:hamiltonian}) by 100 and truncating the remaining decimals. This procedure effectively increases the spacing of the energy eigenlevels by reducing the precision of these levels. Alternatively, in cases where the potential configurations or conformations of the system result in numerous closely-spaced eigenlevels, the levels can be grouped into intervals (binned) to create a coarse-grained pseudo-eigenstructure for the system ~\cite{liSteepestentropyascentQuantumThermodynamic2016}.

\subsection{Machine learning}
\label{sec:ML}
In principle, the energy eigenstructure predicted by the REWL algorithm is all that is required to make predictions with the SEAQT equation of motion.  However, to investigate adsorption in very dilute As solutions comparable to experimentally relevant concentrations, the size of the lattice must be made intractably large. However, using the energy eigenstructures of tractable lattices produced by REWL as training data, the eigenstructures of much larger lattices can be predicted using machine learning. An artificial neural network (ANN) with two hidden layers is selected here for that purpose. The number of molecules of As, functional groups, and the energy eigenlevels are selected as inputs and the corresponding degeneracies for the energy eigenlevel are the predicted outputs. The training cases are constructed from REWL simulations for configurations of lattice size 70$\times$ 70 with the number of {  As(V) molecules} and functional groups ranging from 1 to 5. $80\%$ of the data generated, i.e., a total of 25 energy eigenstructures, is used as validation or training sets. To predict the energy eigenstructure, the topology of the ANN is chosen to have 100 neurons in a first layer, 30 neurons in the hidden layer, and an activation function given by
\begin{equation} \label{eq:logs}
  f(x) =  -2 +\frac{4}{1+e^{-x}}
\end{equation}
where $x$ is a variable proportional to the product of the weights and the parameters used in the training set. This expression represents a sigmoid function in the range of $-2$ to $2$. This customized activation function allows for the extrapolation of the data. The training data sets are scaled between $-0.1$ to $1.9$ to make the extrapolation range of the activation function $-1.9$ to $-0.1$. 

\begin{figure}[htbp]
\vspace{3mm}
    \centering
    a) \includegraphics[width=0.5\linewidth] {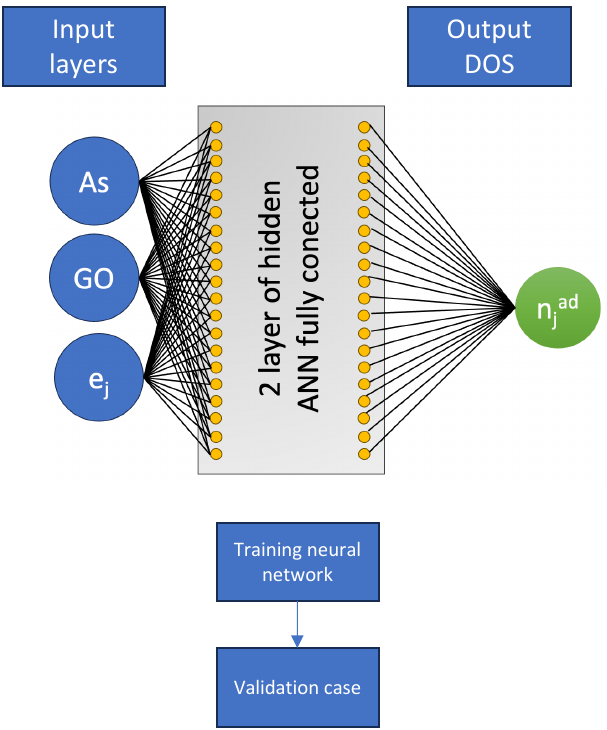}\\[10mm]
    b) \includegraphics[width=0.7\linewidth] {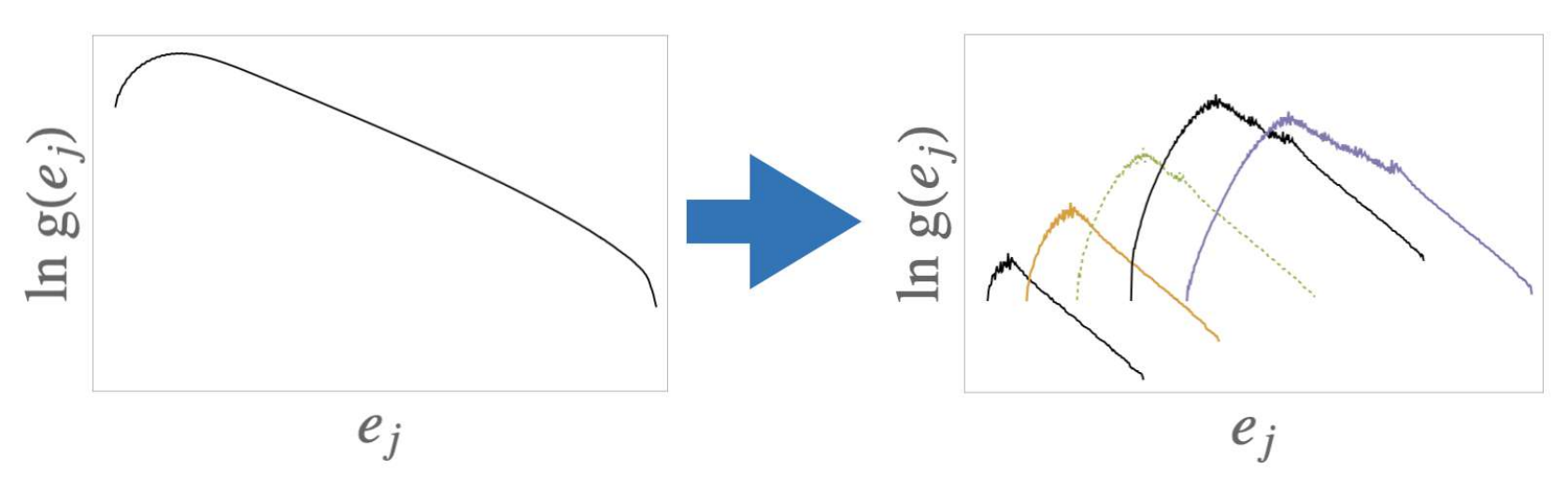}\\[10mm]
     c) \includegraphics[width=0.7\linewidth] {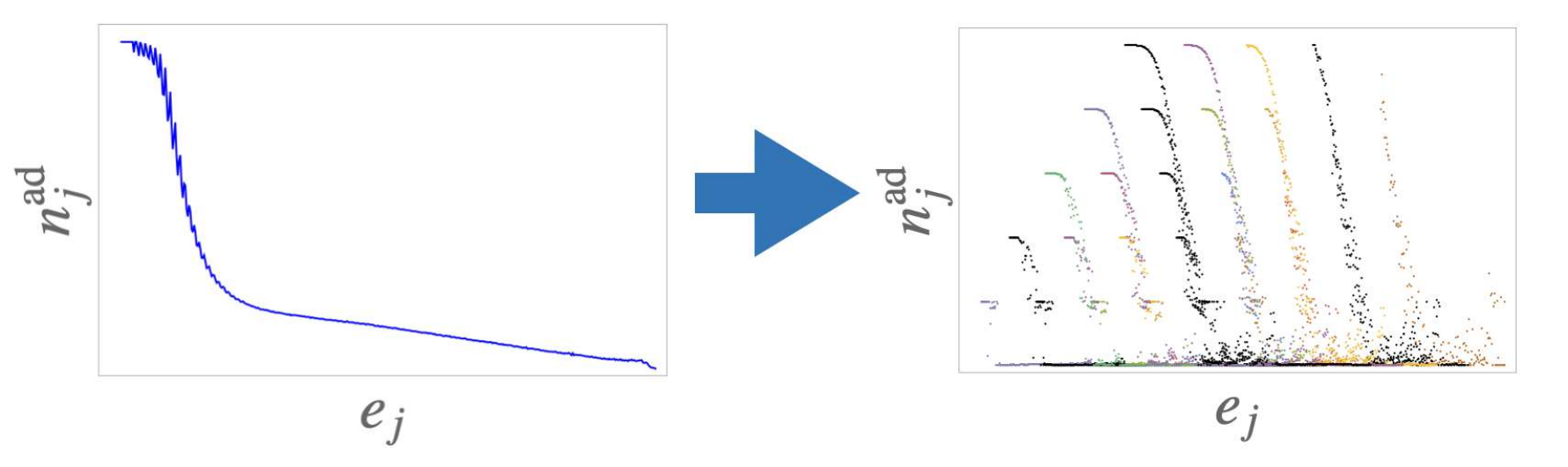}
    \caption{a) Flowchart of the machine learning algorithm; b) the input  energy eigenstructure generated with the Replica Exchange Wang Landau algorithm and the corresponding neural network-predicted output energy eigenstructures; and c) the Replica Exchange Wang Landau-generated adsorbed particle number eigenvalues and the corresponding predicted neural network adsorbed particle number eigenvalues. }
    \label{fig:MachineLearning}
\end{figure}

The average number of adsorbed molecules per energy eigenlevel is treated as a mixture of the activation functions because of the nature of the training cases. Namely, considering two hidden layers each one with 30 neurons and the first layer with the activation function
\begin{equation}
f(x) = x \tanh \log(1+ e^{x})
\end{equation}
leads to non-activation for $x\ll 0$ and to linear increases for $x>0$.  The second layer has an activation function as given by Eq.~(\ref{eq:logs}).  The choice of this activation function leads to a better prediction of the non-adsorption probabilities at the high energies expected of typical adsorption processes. The structure of the ANN is shown in Fig.~\ref{fig:MachineLearning}~a). Example input REWL generated energy eigenstructures and adsorbed particle number eigenvalues are shown on the left sides of Figs.~\ref{fig:MachineLearning} b) and c), respectively.  The colored curves on the right sides of Figs.~\ref{fig:MachineLearning} b) and c) are predictions of the ANN for the energy eigenstructures and adsorbed particle number eigenvalues. {  The hyper-parameters were fixed \emph{a priori}: weights initialised with the He–normal, the Adam optimiser with $\eta = 10^{-3}$, a mini-batch size $of 128$, and an early-stopping patience of $50$ epochs. The data set (25 eigenstructures, $1.2\times10^{5}$ samples) was split $80\,\%/10\,\%/10\,\%$ into training/validation/test subsets. The network achieved a test-set mean-squared error of $1.1\times10^{-4}$. Ten-fold cross-validation confirmed that deeper or wider topologies improved accuracy by $< 2\,\%$.}

%\begin{widetext}
%\end{widetext}

The trained artificial neural network employs a loss function (mean squared error, MSE) of the form
\begin{equation}
    \text{MSE} = \frac{1}{N} \sum_{i}^N \left( Y_i - \hat Y_i \right)^2
\end{equation}
where the sum runs through all the 
$N$ energy eigenlevels, and the $Y_i$ are the elements of the training set, while the $\hat Y_i$ are the predicted values of the network.  The loss function in the training sets reaches a value of $10^{-4}$. 

\subsection{SEAQT equation of motion}
\label{sec:SEAQT}
After establishing the energy eigenstructure of the system, the applicable form of the SEAQT equation of motion, which is based on the steepest-entropy-ascent principle, must be derived for the particular application at hand.  The solution of this matrix differential equation (i.e., system of first-order ODEs in time) provides a unique  kinetic (i.e., non-equilibrium thermodynamic) path from any arbitrary initial state to stable equilibrium. The variational principle upon which the equation of motion is based guarantees that the kinetic path reaches the globally stable equilibrium state corresponding to maximum entropy. A notable feature of this equation is that it does not require an $a \; priori$ specification of the underlying rate-limiting mechanisms of the kinetic processes involved. It depends only upon the physical principle that systems naturally evolve in a way that maximizes the entropy production.  In its most general form, this equation is expressed as
\begin{equation}
 \frac{d \hat{\rho}}{dt} = \frac{1}{i\hbar}[{\hat{\rho}}, \hat{H}]+\frac{1}{\tau  (\hat{\rho})}{\hat{D}(\hat{\rho})}
 \label{eqn:generalEOM}
 \end{equation}
where $\hat \rho$ is the density or ``state'' operator, $t$ the time, $i$ the imaginary unit, $\hbar$ Planck's modified constant, $[{\hat{\rho}},\hat{H}]$ the commutator of $\hat{\rho}$ and $\hat{H}$, $\hat{H}$ the Hamiltonian operator, $\tau$ the relaxation parameter, and $\hat{D}(\hat{\rho})$ the dissipation operator. For a classical (non-quantum) system, the density and Hamiltonian operators are diagonal in the energy eigenvalue basis so that the first term on the right side of Eq.~(\ref{eqn:generalEOM}), the so-called symplectic term, is zero~\cite{ liSteepestentropyascentQuantumThermodynamic2016,liGeneralizedThermodynamicRelations2016,berettaSteepestEntropyAscent2014,liSteepestentropyascentmodelmesoscopic2018}. 

Now, the dissipation operator consists of a ratio of determinants with the size of each determinant depending on the number of generators of the motion that are active for a given application (e.g., for more details see~\cite{saldana-roblesModelpredictingadsorption2025}). For the system considered here, there are four generators of the motion: the identity operators, $\hat{I}_S$ and $\hat{I}_R$, of the adsorbing system and the thermal reservoir with which the system interacts, respectively; the Hamiltonian operator, $\hat{H}$, of the composite of the adsorbing system and reservoir; and the particle number operator, $\hat{n}$, for the species being adsorbed in the system. The observables corresponding to these generators of the motion ($\tilde{C}_i = \tilde{I}_S, \tilde{I}_R, \tilde{H}, \tilde{n}$) are conserved by the equation of motion. 

Note that the SEAQT description is always that of an isolated system so that if an interaction (e.g., an energy or mass interaction) is present between two systems as in this case between a system and a thermal reservoir, a composite system must be formed of the two in order to capture the interaction. As a result, the Hilbert space or state space of the composite system is factored into two subspaces and a 2$^{nd}$-order hypoequilibrium description~\cite{liSteepestentropyascentQuantumThermodynamic2016,liGeneralizedThermodynamicRelations2016} used to described the subspaces. Normally, an equation of motion is written for each subspace. However, since one of the subspaces is that of a thermal reservoir, the equation of motion for the reservoir ($R$) is not required because any changes to the total energy and entropy of the reservoir are by definition negligible and, thus, do not affect its state. Therefore, the reservoir expectation values associated with the energy and entropy that appear in the equation of motion represent the energy and entropy added to or subtracted from the reservoir. The equation of motion for the present application is then written as  
 
\begin{equation} \label{eq:eqomot}
\frac{dp_j}{dt} = 
\frac{1}{\tau} 
\frac{
    \left|
\begin{array}{ccccc}
 p_j s_j & p_j & 0 & e_j p_j & n_j^{ad} p_j \\
 \langle s\rangle_S & P_S & 0 & \langle e \rangle_S  & \langle n \rangle_S \\
 \langle s \rangle_R & 0 & P_R & \langle e\rangle_R & 0 \\
 \langle es\rangle_R+\langle es\rangle_S & \langle e\rangle_S & \langle e\rangle_R & \langle e^2 \rangle_R+\langle e^2\rangle_S & \langle en \rangle_S \\
 \langle ns\rangle_S & \langle n\rangle_S & 0 & \langle en\rangle_S & \langle n^2 \rangle_S \\
\end{array}
    \right|
}{
    \left|
\begin{array}{cccc}
 P_S & 0 & \langle e \rangle_S  & \langle n\rangle_S   \\
 0 & P_R & \langle e\rangle_R  & 0  \\
 \langle e \rangle_S  & \langle e\rangle_R & \langle e^2 \rangle_R +\langle e^2 \rangle_S & \langle en \rangle_S \\
 \langle n \rangle_S  & 0  & \langle en\rangle_S  & \langle n^2 \rangle_S  \\
\end{array}
    \right|
}
 \end{equation}
where the determinant in the denominator is a Gram determinant, which ensures the linear independence of the generators of the motion and $p_j$ and $s_j = -\ln\left(p_j/g_j \right)$ are the occupation probability and non-dimensional entropy of the $j^{th}$ energy eigenlevel, respectively, $g_j$ is the degeneracy of the $j^{th}$ energy eigenlevel, and $\langle \cdot \rangle_S$ and $\langle \cdot \rangle_R$ are expectation values of the absorbent system and reservoir, respectively. The quantity, $P_S$, is the sum of the probabilities for the absorbent system, which in this case must equal 1, and $P_R$ is a real number related to the degrees of freedom of the reservoir. The system (S) expectation values for the energy, entropy, particle number, and products thereof appearing in Eq. (\ref{eq:eqomot}) are expressed as follows:
\begin{eqnarray}
         \langle e \rangle &=&\sum_{j}p_je_j \\
         \langle s \rangle &=&  \sum_{j}p_j s_j  \\
         \langle e^2 \rangle &=& \sum_{j}p_j e_j^2 \\ \langle es \rangle &=&  \sum_{j}p_j e_j s_j  \\
         \langle en \rangle &=& \langle en \rangle_{ad} + \langle en \rangle_{un} \nonumber \\
         &=& \sum_{j} p_j e_j n_j^{ad} + \langle en \rangle_{un}\\ 
        \langle ns \rangle &=& \langle ns \rangle_{ad} +\langle ns \rangle_{un} \nonumber \\
        &=& \sum_{j} p_j n_j^{ad} s_j + \langle ns \rangle_{un} \\
        \langle n^2 \rangle &=& \langle n^2 \rangle_{ad} + \langle n^2 \rangle_{un} \nonumber \\
        &=& \sum_{j}p_j (n_j^{ad})^2 + \langle n^2 \rangle_{un}
\end{eqnarray}

Expanding the determinants in the numerator and denominator of Eq.~(\ref{eq:eqomot}) and grouping terms into various thermodynamic quantities (for details see~\cite{saldana-roblesModelpredictingadsorption2025}), the equation of motion for the system (S) reduces to the following compact form:
\begin{equation} \label{eq:CompactEOM}
\frac{dp_j}{dt} = \frac{\beta_R}{\tau} \left(  p_j \langle \Phi \rangle -p_j \Phi_j \right)
\end{equation}
 where $\Phi_j = e_j - \beta^{-1}_R \, s_j - \gamma^{-1}_R \, n_j^{ad}$ is a non-equilibrium analogue of the grand canonical potential. The inverse temperature of the reservoir is the only reservoir property that appears in this expression for the system (S). The expectation value $\langle \Phi \rangle$ is found from
 \begin{equation}
\langle \Phi \rangle = \sum_j p_j \, \Phi_j \,,
 \end{equation}
while the intensive properties $\beta_R$ (the thermodynamic beta of the reservoir) and $\gamma_R = 1/\mu_{R_n} = \beta_R / \tilde{\mu}_n$ are defined as
\begin{align} \label{eq:intensive}
\beta_R &= \frac{1}{k_B \, T_R} \, \nonumber \\[2mm]
\quad \gamma_R &= \frac{A_{ns} A_{ee} - A_{es}A_{en}}{A_{en}^2 -A_{nn} A_{ee}}
\end{align}
where $T_R$ is the temperature of the thermal reservoir and $\tilde \mu_n$ is the dimensionless chemical potential that converges to the usual chemical potential at stable equilibrium. Note that in general, i.e., at non-equilibrium and stable equilibrium, 
\begin{equation} \label{eq:generalBeta}
\beta = \frac{A_{ns} A_{ee} - A_{es}A_{nn}}{A_{en}^2 -A_{nn} A_{ee}}
\end{equation}
and the quantities, $A_{xy}$, are fluctuation parameters given by $A_{xy} = \langle xy \rangle - \langle x \rangle \langle y \rangle$. The fluctuations provide a connection between the extensive and intensive thermodynamic properties in the SEAQT equation of motion and the underlying physics of the process experienced by the system.\\

As can be seen from Eq.~(\ref{eq:CompactEOM}), at stable equilibrium, the fluctuation in the grand potential represented by the quantity in the parentheses on the right goes to zero. As a consequence, the probability distribution of the state of the system reduces to the grand canonical distribution. Thus, in the non-dissipative limit, the behavior of the system is fully compatible with equilibrium thermodynamics, highlighting the consistency of the SEAQT framework with classical thermodynamic principles in the equilibrium limit. At stable equilibrium, the probability distribution predicted by the SEAQT equation of motion is equivalent to the grand canonical expression given by
\begin{equation}
p_j = \frac{\exp \left( -\beta_R \,\Phi_j \right)}{\sum_i \exp \left( -\beta_R \,\Phi_i \right)} \,,
\end{equation}

In the following sections, numerical solutions of the equation of motion, Eq.~(\ref{eq:CompactEOM}), are found and used to calculate the number of particles adsorbed at each instant of time as well as at stable equilibrium. 

\subsection{  Model validation against experimental data}
\label{sec:validation}
% Comment from Bill: I swapped the order of the presentation in this paragraph so the choice of model numbers follows from the experiments.  The original:
% To accurately validate a specific point on the experimental adsorption isotherm using numerical results, a minimum lattice size of 150 $\times$ 150 is necessary. This setup includes 1 molecule of arsenic (As) with a concentration of 348 mg L$^{-1}$ and 1 molecule of each functional group found in graphene oxide (GO), specifically COOH and OH, giving a concentration of 10 g L$^{-1}$. Additionally, to maintain a constant pH as observed experimentally, it is essential to include 1 particle of H$^+$ in all scenarios. In contrast, experimental data from the technical literature report using an initial As concentration of 350 mg L$^{-1}$, a GO concentration of 0.8 g L$^{-1}$, and a pH level of 3. This indicates that the GO concentration utilized in numerical modeling is approximately 12.5 times greater, and although the pH level is slightly lower, these differences can significantly impact the expected outcome, leading to imprecise validation. This discrepancy underscores a limitation of the Replica Exchange Wang-Landau (REWL) method. To accurately simulate dilution and approach lower (more realistic) concentrations, large lattice sizes are necessary, which complicates the application of these methods for validating and simulating real-world scenarios. Despite these challenges, such approaches remain prevalent for predicting the intrinsic properties of both adsorbents and adsorbates.

To validate the model against experimental adsorption isotherm data {  found in the literature}, it is important to ensure the relevant concentrations used in the REWL  algorithm fall within the range of experimental values~\cite{suHighperformanceIronOxide2017}. Typical values reported in the literature are 350 mg L$^{-1}$ for the initial As concentration, a GO concentration of 0.8 g L$^{-1}$, and a pH of 3. Approximating this As concentration on a discrete lattice requires one {  As(V) molecule} on a lattice size of at least 150 $\times$ 150. Furthermore, to maintain a constant pH in the model, it is essential to include one H$^+$ ion.  If the {  As(V) molecule} in the lattice model is associated with one GO particle with the minimum number of functional groups (one each of COOH$^-$ and OH$^-$), the corresponding GO concentration in the model is 10 g L$^{-1}$, a value roughly 12.5 times greater than the assumed experimental concentration. This concentration difference can significantly affect adsorption and lead to ambiguous validation at best. 

The difference between experimental concentrations and those accessible with a discrete lattice and integer numbers of the component molecules underscores a limitation of the REWL method since the larger lattices required to accurately simulate low (realistic) As concentrations are computationally burdened. This problem can be circumvented, however, by combining the energy eigenstructure information generated with the REWL algorithm on smaller lattices with the machine learning approach described above. Rather than calculating the energy eigenstructures of large lattices with the REWL algorithm, the machine learning algorithm can be used to determine the required low-concentration eigenstructures using training data generated with REWL on  smaller lattices. This strategy also makes it possible to use fractional numbers of the component species and estimate the energy eigenstructure of dilute, experimentally relevant concentration regimes. These low-concentration eiegenstructures are then used by the SEAQT equation of motion to make predictions, which are validated for solution concentrations in the range of the experimental values~\cite{suHighperformanceIronOxide2017}. A machine-learning lattice size of 70$\times$70 was used with 0.2 molecules of As and 0.017 molecules of each functional group of GO.

To relate the number of  H$^+$ ions to the pH, the relationships for the pH, the mass of hydrogen ions and water, and the volume of water are used, i.e.,
\begin{align} \label{eq:pH}
\text{pH} &= -\log_{10} \frac{m_{\text{H}^+}}{V_{\text{H}_2 \text{O}}} \\[2mm]
m_{\text{H}^+} &= \frac{1}{N_A}\left(n_{\text{H}^{+}}\, M_{\text{H}^{+}}\right)  \\[2mm]
m_{\text{H}_2 \text{O}} &= \frac{1}{N_A}\left(n_{\text{H}_2 \text{O}}\, M_{\text{H}_2 \text{O}}\right)  \\[2mm]
V_{\text{H}_2 \text{O}} &= m_{\text{H}_2 \text{O}} \; \rho_{\text{H}_2 \text{O}} \;, 
\end{align}
where $n_{\text{H}^{+}}$ and $M_{\text{H}^{+}}$ are the number and molecular weight of $\text{H}^{+}$ ions, respectively, $n_{\text{H}_2 \text{O}}$ and $M_{\text{H}_2 \text{O}}$ are the number and molecular weight of water molecules, respectively, $N_A$ is Avogadro's number, and $\rho_{\text{H}_2 \text{O}}$ and $V_{\text{H}_2 \text{O}}$ are the density and volume of water, respectively. Eq.~(\ref{eq:pH}) provide a connection between the number of hydrogen ions and the pH. To keep the adsorbent concentration fixed and close to the experimental values, the number of GO particles is maintained constant with one functional group of GO$-$OH$^-$ and one functional group of GO$-$COOH$^-$. The concentrations of these species are determined using
\begin{equation} \label{eq:conv}
\left[ \text{GO}-\text{X} \right] = \frac{n_{\text{GO}-\text{X}}\, M_{\text{GO}}}{V_{\text{H}_2 \text{O}}\, N_A} \,,
\end{equation}
where $n_{\text{GO}-\text{X}}$ represents the number of either GO$-$OH$^-$ or GO$-$COOH$^-$ functional groups. 

In Eq.~(\ref{eq:eqomot}) , the number of adsorbed As ions per energy eigenlevel, $n_j^{ad}$, are constant values determined from the Wang-Landau algorithm (or the artificial neural network), whereas the expected value, $\langle n \rangle_{ad}$, is time-varying via the occupation probabilities, $p_j$. Solution of the SEAQT equation of motion provides the distribution of occupation probabilities as a function of time so that the expected adsorbed As number changes with time via the expression
\begin{equation} \label{eq:nt}
\langle n \rangle_{ad} (t) = \sum_{j} p_{j}(t)\, n_{j}^{ad}
\end{equation}
{  For a complete description of this expression and its use, refer to}~\cite{saldana-roblesModelpredictingadsorption2025}.

\begin{figure}[htbp]
    \centering
    \includegraphics[width=0.6\linewidth]{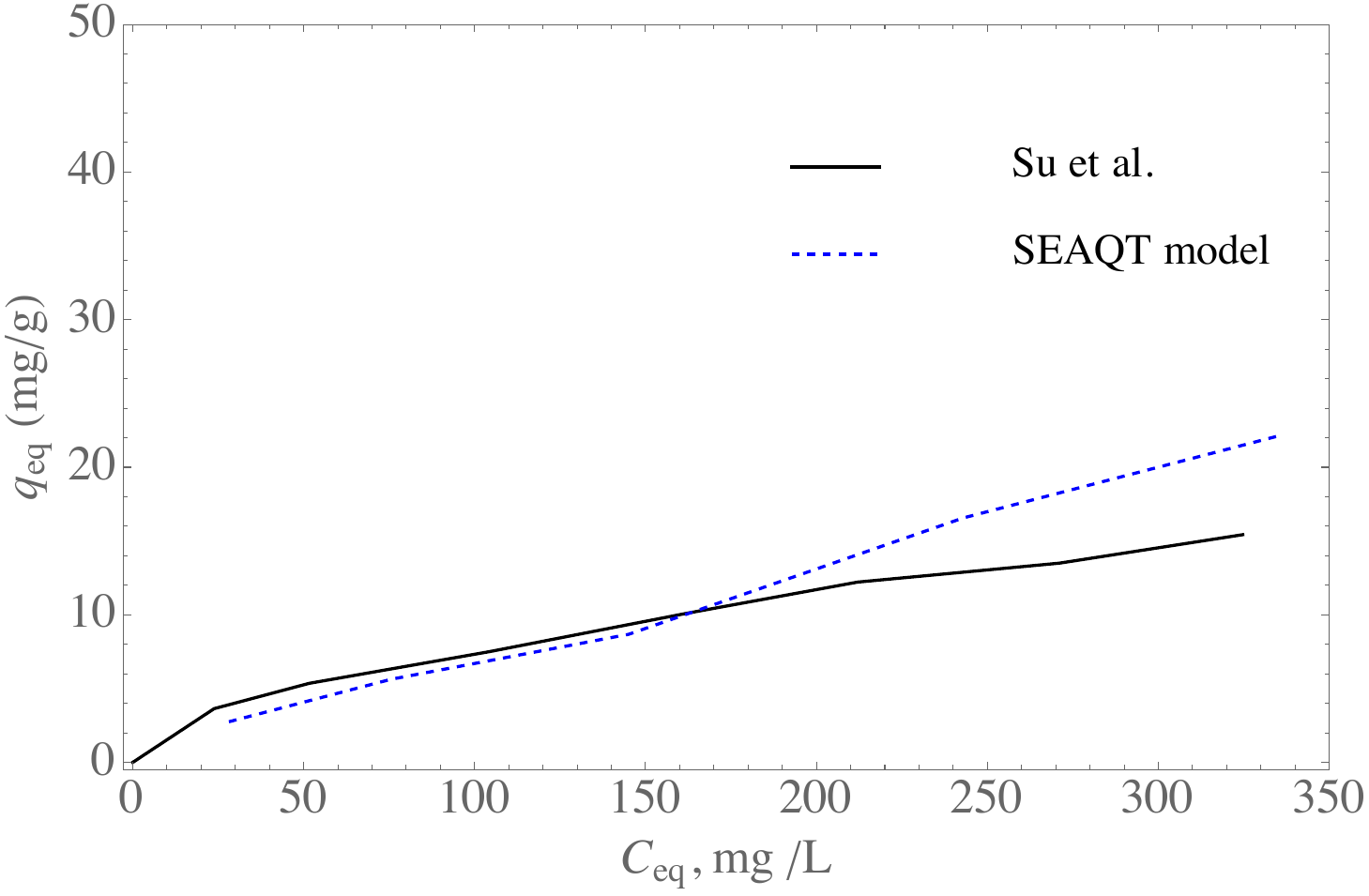}
    \caption{{ The SEAQT predicted adsorption capacity $q_{\text{eq}}$ as a function of the equilibrium As concentration in the aqueous solution yields a mean-squared error (MSE) of 3.5780 mg g$^{-1}$ relative to the experimental data.} The black curve represents a Langmuir model fit to experimental stable equilibrium adsorption for a range of As concentrations with 0.8 g L$^{-1}$ of GO and a pH of 3~\cite{suHighperformanceIronOxide2017}. The individual points are SEAQT predictions.}
    \label{fig:Validation}
\end{figure}

The adsorption capacity predicted at stable equilibrium by the SEAQT equation of motion using an energy eigenstructure obtained with REWL and the artificial neural network model described in Section~\ref{sec:ML} is shown in Fig.~\ref{fig:Validation} where it is validated against an experimentally-fitted Langmuir adsorption isotherm~\cite{suHighperformanceIronOxide2017}.  The Figure plots the stable equilibrium amount of As bound to GO, denoted by the adsorption capacity, $q_{\text{eq}}$ (in mg $\text{g}^{-1}$) (or equilibrium adsorption uptake), versus the equilibrium concentration of unbound As in the aqueous solution, $C_{\text{eq}}$.  The black line in Fig.~\ref{fig:Validation}, from reference~\cite{suHighperformanceIronOxide2017}, is the Langmuir model fitted to the experimental adsorption isotherm for GO loaded to $0.8\, \text{mg}\,\text{mL}^{-1}$ in a solution with pH 3 and initial As concentrations ranging from $25$ to $350\, \text{mg}\,\text{L}^{-1}$.  The blue dashed line in the figure are the stable equilibrium adsorption capacities predicted by the SEAQT equation of motion. The SEAQT adsorption capacity is slightly below the experimental adsorption isotherm curve for stable equilibrium solution concentrations below about $150\, \text{mg}\,\text{L}^{-1}$ and slightly above the experimentally fitted Langmuir curve for concentrations from 150 to $350\, \text{mg}\,\text{L}^{-1}$. At the highest equilibrium concentration $C_{\text{eq}} $ of 336 mg L$^{-1}$, the SEAQT model predicts a value for $ q_{\text{eq}} $ of 21.14 mg g $^{-1}$, whereas the Langmuir model predicts approximately 15.42 mg g$^{-1}$. At the lowest predicted $ q_{\text{eq}} $ by the SEAQT model, corresponding to a $ C_{\text{eq}} $ of 29 mg L$^{-1}$, the SEAQT model yields $ q_{\text{eq}} $ as 1.96 mg g$^{-1}$, while the Langmuir model predicts about 3.6 mg g\(^{-1}\). At intermediate concentrations, the predictions of both models are much closer together. The deviations between the two models are attributed to the inherent differences in their assumptions. The Langmuir model assumes monolayer adsorption, where all adsorption sites have the same energy and only one layer of molecules can adsorb onto the surface. In contrast, the SEAQT model, through the eigenstructure derived from the Wang-Landau algorithm, allows for a more complex scenario where adsorption involves variable adsorption energies due to electrostatic interactions between adsorption sites. This difference in assumptions leads to the varying predictions for $q_{\text{eq}} $ across the concentration range. {  In the chosen concentration window the system resides in its Henry-law regime, so both the Langmuir isotherm and the SEAQT prediction reduce to
$ q_{\mathrm{eq}}\;\propto\;C_{\mathrm{eq}}\;,$
producing an apparently straight trend.  The linearity is thus emergent, not imposed.  }

{  Reported As(V) capacities for char carbon (34.46 mg g$^{-1}$), iron (III) oxide-loaded melted slag (18.8 - 78.5 mg g$^{-1}$), and phosphorylated crosslinked orange waste (68 mg g$^{-1}$) under similar acidic conditions\cite{pattanayakparametricevaluationremoval2000,zhangIronoxideloadedslag2005,ghimireAdsorptiveseparationarsenate2003} place the graphene oxide ($\approx$21 mg g$^{-1}$, this work) in the middle of the current performance spectrum. Furthermore, while the present simulations fix the number of surface functional groups to reflect typical GO site densities, future studies may explore how varying the total number and distribution of adsorption sites modifies the configuration hierarchy and average binding energy, particularly under high-density or cooperative regimes.
}

{ 
\subsection{Thermodynamic consistency with adsorption energy data}
In the SEAQT framework, the evolution of the system is governed by a deterministic equation that drives it toward a thermodynamic equilibrium state. The system is modeled as a discrete ensemble of microstates, each corresponding to a configuration with $n_{j}^{ads}$ adsorbed molecules. The average occupancy of the adsorbed molecules at equilibrium is given by
\begin{equation}
    \theta^{eq} = \frac{\langle n\rangle_{ad}(t)}{N_{sites}}
\end{equation}
where $\langle n\rangle_{ad}(t)$ is the expected number of adsorbed molecules at time $t$, and $N_{sites}$ is the total number of adsorption sites. For monolayer adsorption, the equilibrium constant is defined as 
\begin{equation}
    K = \frac{\theta^{eq}}{(1-\theta^{eq})C_{eq}}\,.
\end{equation}
from which the Gibbs free energy can be obtained via
\begin{equation}
\Delta G_{ads} = - R T \ln K
\end{equation}
Here, $R = 8.314$ J mol$^{-1}$ K$^{-1}$ is the universal gas constant and $T$ the stable equilibrium temperature. For this case, the ML algorithm is adapted to give a concentration of 1.2 $\times$10$^{5}$ sites in the lattice. Thus, at stable equilibrium, this corresponds to an average value of adsorbed As (V) molecules of 3.84$\times 10^{4}$ for which $\theta^{eq} = 0.32$. Since the remaining solute concentration in the bulk is $C_{eq} = 1.25 \times$ 10$^{-6}$ mol L$^{-1}$, the Gibbs free energy of adsorption is $\Delta G_{ads} \approx -31.8$ kJ mol$^{-1}$. This value is in agreement with \cite{suHighperformanceIronOxide2017} with a reported value ranging from -29 to -35 kJ mol$^{-1}$.
} 

{  
Note that the SEAQT formalism is not limited to neutral–aqueous equilibria.
If one extends the density-of-states to include an electron reservoir
(with the electrochemical potential of the electrons), the same SEAQT equation predicts the equilibrium
distribution of oxidation states. Redox potentials then emerge directly from the requirement that the chemical potential of the As(III) is equal to the chemical potential of the As(V) at a given pH, reproducing the lines that constitute a Pourbaix diagram. Although a complete SEAQT‐generated Pourbaix diagram is beyond
the scope of the present adsorption study,
the capability is inherent in the framework and will be explored
in forthcoming work.}

{  Finally, several density-functional studies employing Natural Bond Orbital (NBO) analysis have already demonstrated the critical role of charge transfer in graphene-oxide metal interactions.  For instance, Xia \emph{et al.}\,\cite{xiaUnexpectedlyefficiention2022} quantified electron donation from surface O-lone pairs to Co$^{2+}$ and Mn$^{2+}$ aqua complexes and linked the natural charge transfer to adsorption enthalpies. Javarani \emph{et al.}\,\cite{javaraniRemovalHeavyMetal2022} reported second-order donor–acceptor stabilisation energies $E^{(2)}$ that rationalise the Zn$^{2+}$ < Cd$^{2+}$ < Hg$^{2+}$ affinity trend on amine-functionalised carbon nanosheets, while Dimakis \emph{et al.}\,\cite{dimakisLiNaAdsorption2019} showed that epoxide and defect sites enhance Li/Na adsorption by increasing the net charge transferred into the sheet.  These precedents suggest that the larger adsorption energies and faster uptake predicted here for arsenate at carboxyl sites would correlate with stronger LP(O)$\rightarrow\sigma^*_{\text{O–H}}$ donor–acceptor interactions and higher natural charges on GO, thereby providing an electronic-structure rationale consistent with the mesoscopic SEAQT picture. A dedicated NBO investigation along these lines is planned for future work.
}

\section{Non-equilibrium Solutions \label{sec:nonequilibrium}}

\subsection{Time evolution of adsorption}
\begin{figure}[htbp]
    \centering
    a)\includegraphics[width=0.6\linewidth]{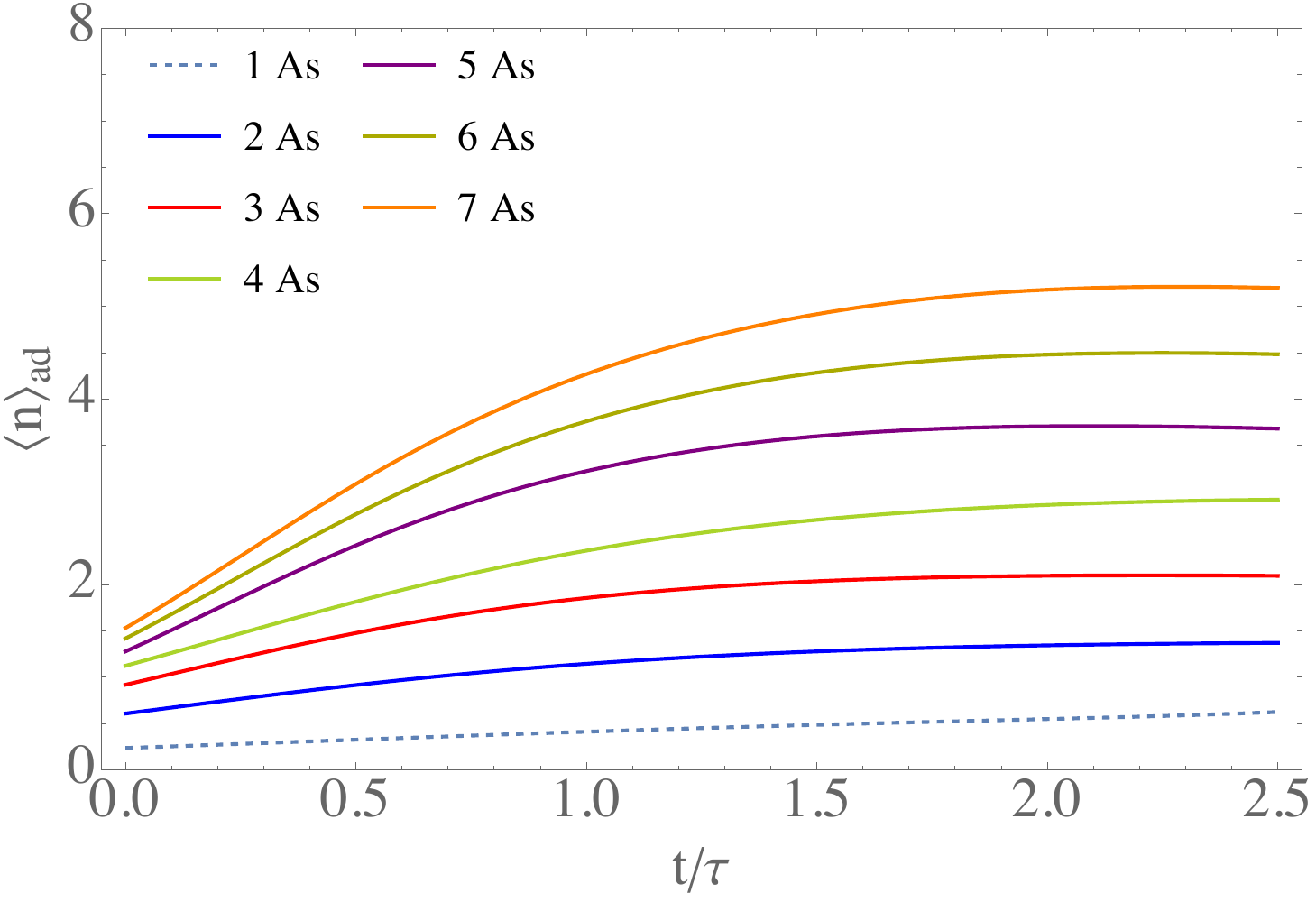}
    b)\includegraphics[width=0.6\linewidth]{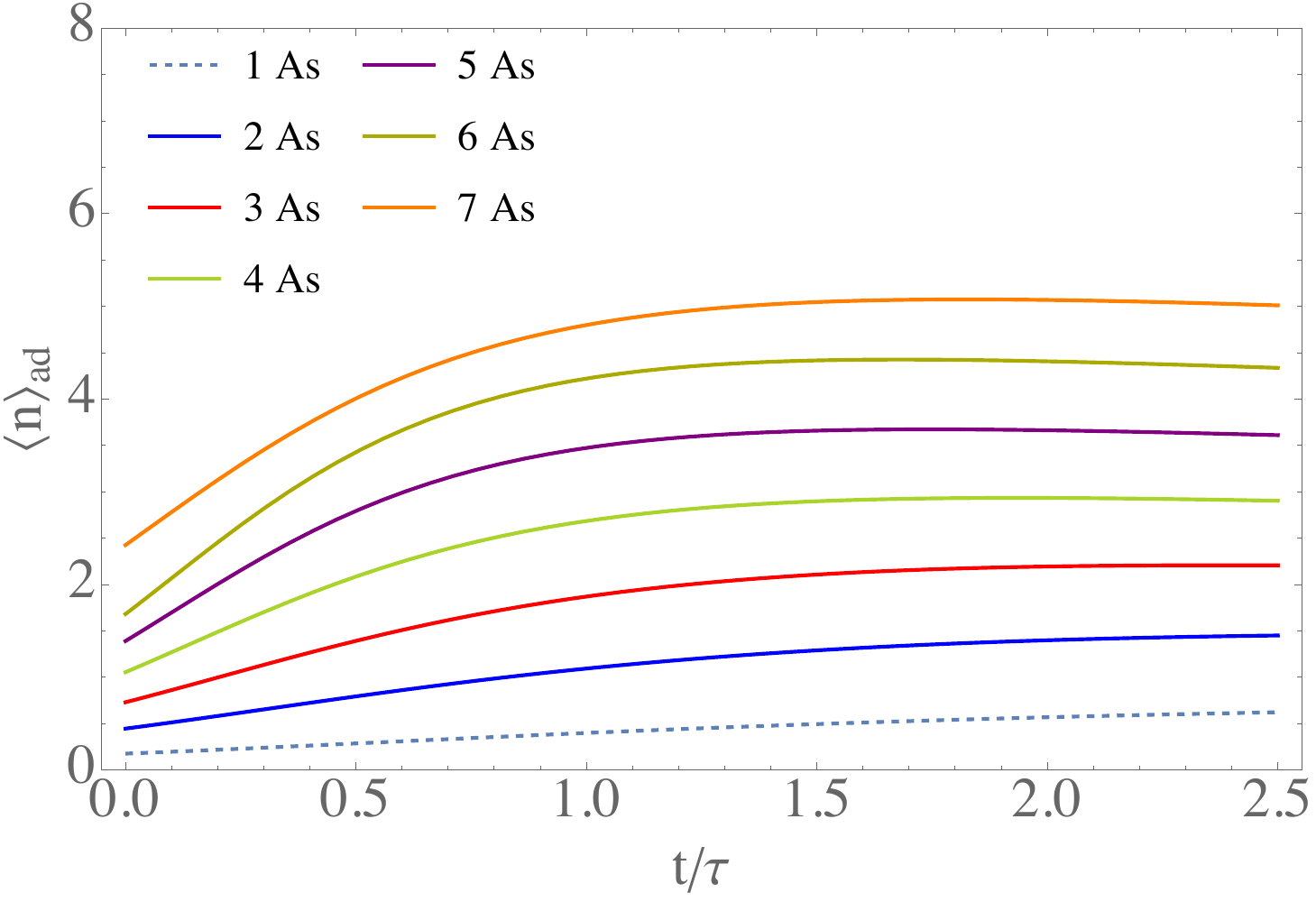}
    c)\includegraphics[width=0.6\linewidth]{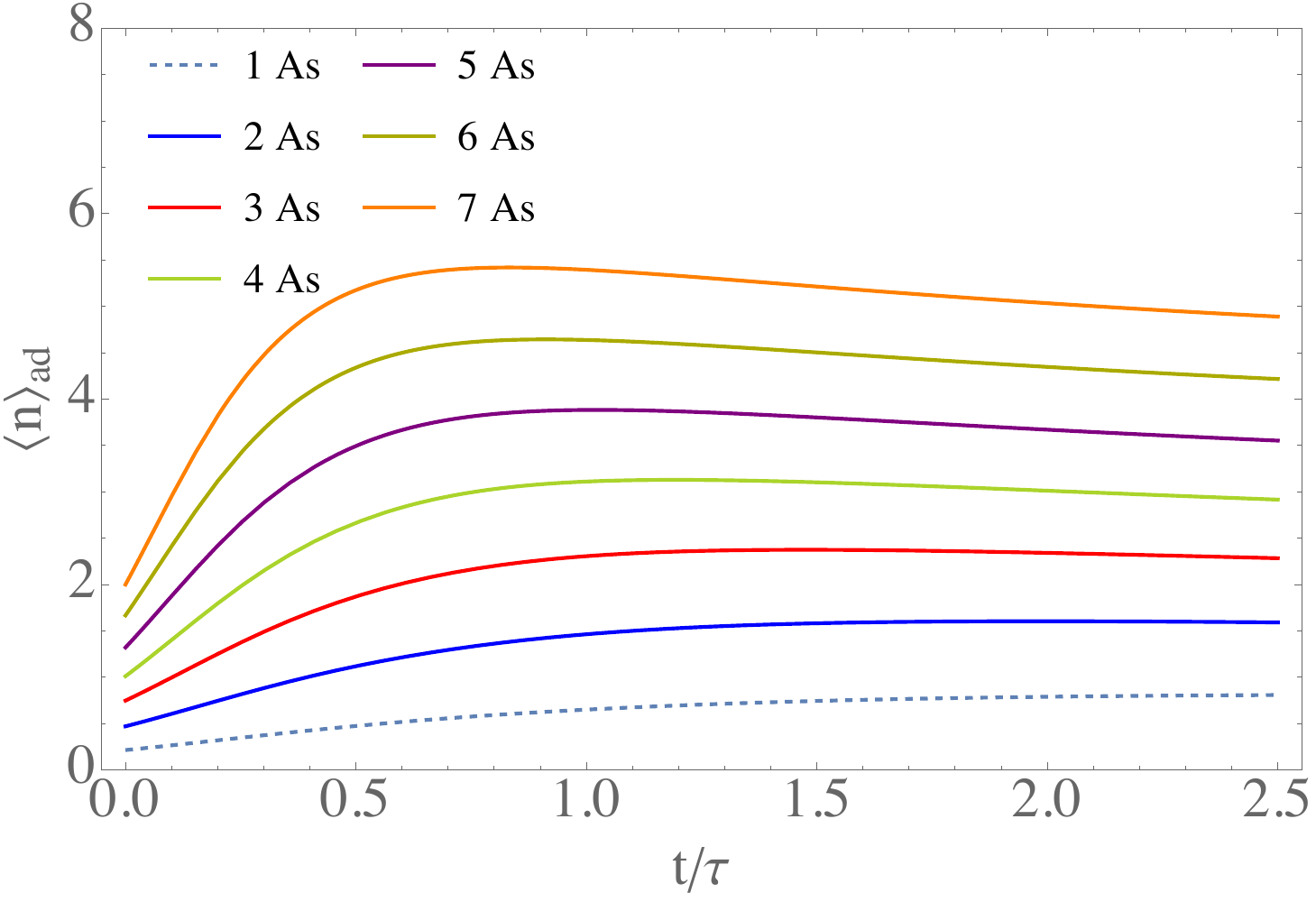}
    \caption{Time evolution of the number of absorbed {  As(V) molecules} by a) 3 GO, b) 6 GO and c) 12 GO.}
    \label{fig:Evol}
\end{figure}

The preceding section validated the equilibrium predictions of SEAQT model by comparing them with experimentally determined stable equilibrium As adsorption values. The present section considers the kinetics or the time-dependent evolution of As adsorption from some initial non-equilibrium state to stable equilibrium. Fig.~\ref{fig:Evol} shows the non-equilibrium time evolution determined by the SEAQT framework for a lattice size of 20$\times$20 for three different GO loadings: 3 GO functional groups in Fig.~\ref{fig:Evol}a), 6 in Fig.~\ref{fig:Evol}b), and 12 in Fig.~\ref{fig:Evol}c). In each case, seven different initial aqueous concentrations of As, labeled As 1 to As 7, are considered. For the sake of providing a physical insight into the concentrations related to the above particle As number content, Table \ref{tab:conc-part} shows its equivalence to concentration. These values were determined by employing Eq. \ref{eq:conv}. For instance, for a particle content of 3 GO (3 GO-OH$^{-}$ and 3 GO-COOH$^-$) and 1 As, the arsenic concentration is equivalent to 19.97 g L$^{-1}$. The number of H$^+$ ions is kept constant in all cases, which corresponds to a constant pH with 1 particle of H$^{-}$.

\begin{table}
    \centering
    \caption{  Conversion between lattice occupancy and bulk concentration.\\}
    \label{tab:conc-part}
    \begin{tabular}{c|cccc}
    \toprule
                &  {\bf 3 GO} & {\bf 6 GO} & {\bf 12 GO} & \\
\midrule
        {\bf 1 As} &  19.97  &  20.28 &  20.93 & g $L^{-1}$\\
        {\bf 2 As}  &  40.05&  40.67 & 41.98 & g $L^{-1}$ \\
        {\bf 3 As}  & 60.23 &  61.17 & 63.14 & g $L^{-1}$\\
        {\bf 4 As}  &  80.51 &  81.77 & 84.42 & g $L^{-1}$\\
       {\bf 5 As}  &  100.90 & 102.4 & 105.81 & g $L^{-1}$\\
       {\bf 6 As}  &  121.39 &  123.3 & 127.31 & g $L^{-1}$\\
       {\bf 7 As}  &  142.00 &  144.2& 148.94 & g $L^{-1}$\\
       \bottomrule
    \end{tabular}
    
\end{table}

In Fig. \ref{fig:Evol} the horizontal axis of these figures is dimensionless time, starting at the initial state, $t/\tau=0$, and arriving at stable equilibrium on the far right of each plot. By comparing the curves for the seven As concentrations in Fig.~\ref{fig:Evol}, it is evident that a greater As concentration in the system leads, of course, to a larger number of adsorbed {  As(V) molecules}. This intuitive observation holds true for each moment of time during the adsorption process and for each level of GO loading (Figs.~\ref{fig:Evol} a), b), and c) ).  Somewhat surprisingly, while the stable equilibrium number of adsorbed {  As(V) molecules} for the three loading cases (right sides of Figs.\ref{fig:Evol} a), b), and c)) does depend upon the amount of As in solution, the adsorbed number is insensitive to the GO loading. Evidently, for a fixed As concentration in the solution, a higher GO loading does not increase the equilibrium number, $\langle n \rangle_{ad}^{eq}$, of As that is ultimately adsorbed. This behavior can be explained by the fact that, in the variation from 3 to 12 GO molecules, the adsorbed As was very similar for each of the concentrations studied (3, 6, and 12 GO molecules). This is because, starting with just 3 GO molecules, there is already an excess of available adsorption sites. Therefore, at equilibrium, the amount of adsorbed As remains nearly the same for 3, 6, and 12 GO molecules, as the adsorption site saturation is achieved with a smaller number of GO molecules.     

For the case of low GO loading, Fig.~\ref{fig:Evol} a), the approach with time to stable equilibrium is monotonically increasing. For higher amounts of GO loading, Figs.~\ref{fig:Evol} b) and c), the initial As adsorption is more rapid, particularly for solutions with a higher As concentration (e.g., As 5, As 6, and As 7).  For these cases, the number of adsorbed {  As(V) molecules} increases quickly, reaching a maximum slightly higher than the stable equilibrium value, after which it gently decays to equilibrium. This ``overshooting'' of the stable equilibrium As adsorption number is unexpected. It may be related to how strongly the adsorption process is driven because excess As adsorption above the stable equilibrium number only appears for the highest As concentrations (As 5 to As 7) and for the two cases of higher GO loading (Figs.~\ref{fig:Evol} b) and c)). From a technology standpoint, this adsorption overshoot may be relevant since it is evident that the maximum adsorption can occur at a transient state along the non-equilibrium path.   

{ 
The SEAQT trajectory provides more than macroscopic kinetics. In particular, the occupation probability of each eigenlevel yields a real-time map of which molecular configurations dominate.  At $t / \tau < 0.2$, the system occupies primarily those levels in which As(V) forms inner-sphere, bidentate complexes with adjacent –COOH groups, consistent with the strong electrostatic attraction ($q_{\mathrm{As}}=1.94$ versus $q_{\mathrm{COOH}}=-0.44$) and the larger donor–acceptor stabilisation energies reported by DFT/NBO studies of GO–metal systems \cite{xiaUnexpectedlyefficiention2022,javaraniRemovalHeavyMetal2022}.  For $t / \tau >0.2$ the population gradually shifts to eigenlevels where As(V) is shared between a –COOH and a neighbouring –OH site, forming mixed inner/outer-sphere complexes. This redistribution produces the small overshoot in Fig.\,\ref{fig:Evol} c) since it maximises the level degeneracy and hence the entropy production.   
}

\begin{figure}[htbp]
    \centering
    a) \includegraphics[width=0.6\linewidth]{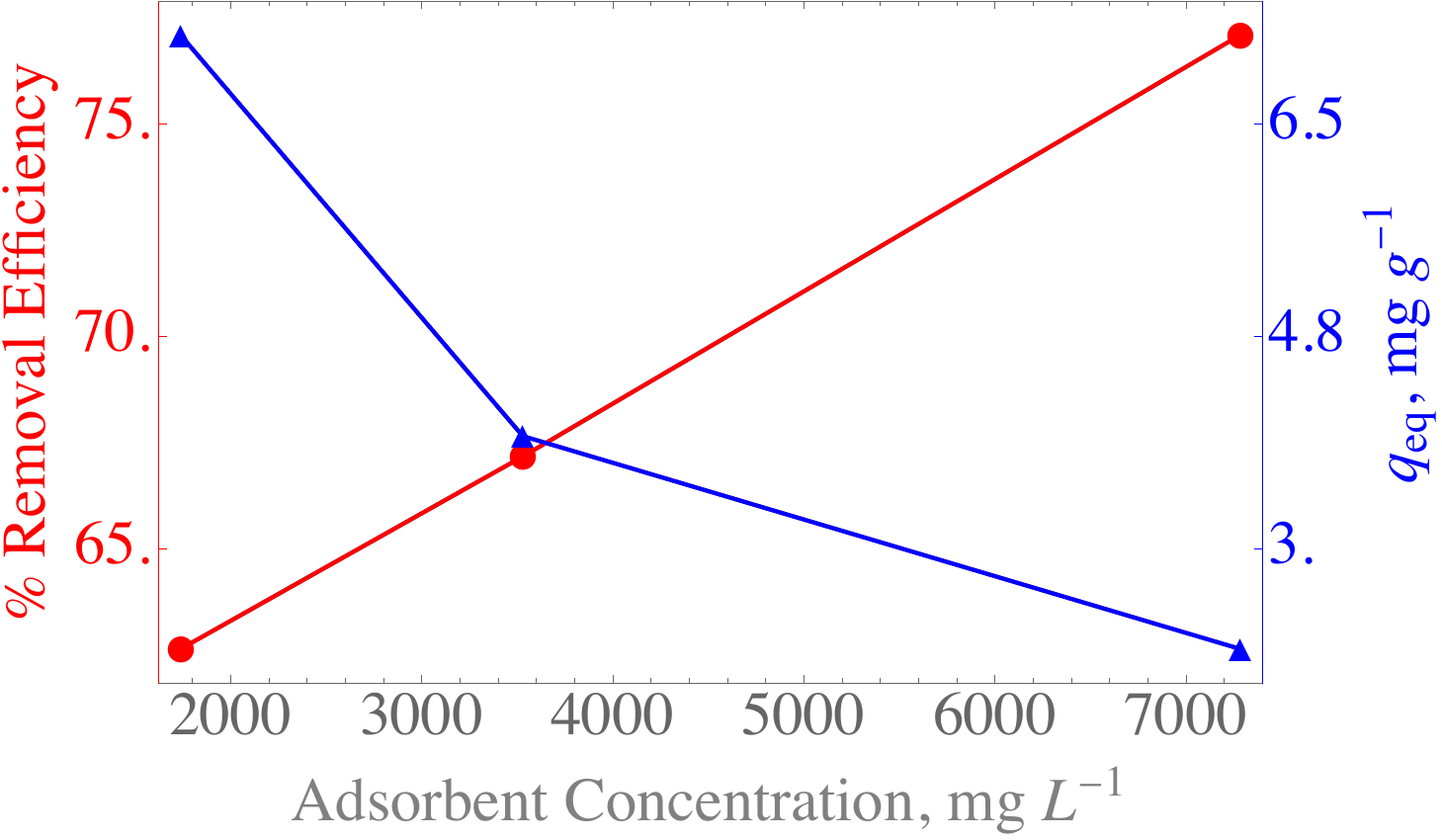}\\
    b) \includegraphics[width=0.6\linewidth]{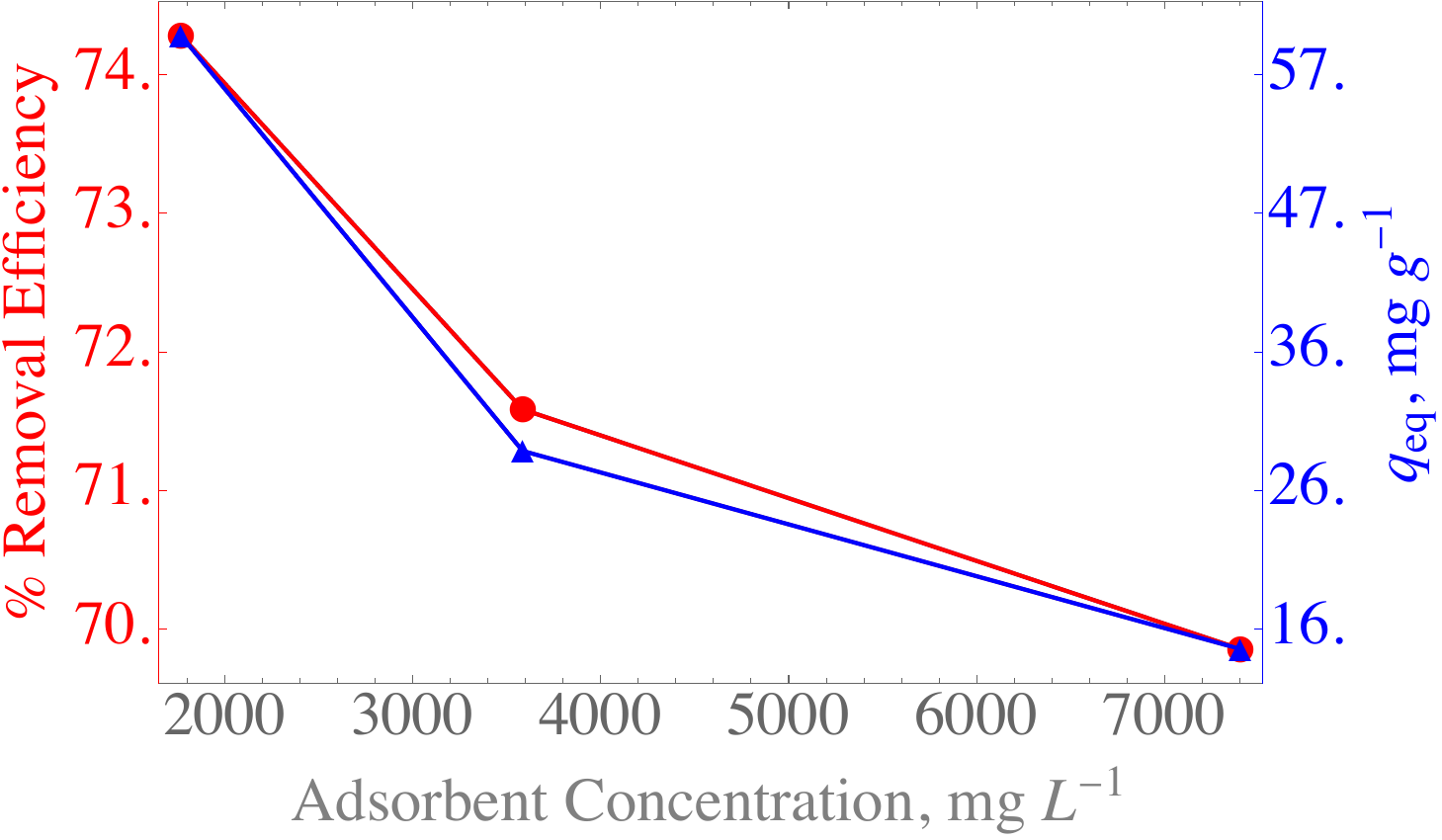}
    \caption{Efficiency of As removal { and adsorption capacity} at stable equilibrium as a function of 
    { adsorbent concentration for a) 1 As(V) molecule and b) 7 As(V) molecules} in the presence of 3, 6, and 12 particles of GO (1800, 3600, and 7250 mg L$^{-1}$, respectively).}
 \label{fig:effect_dose}
\end{figure}

Fig.~\ref{fig:effect_dose} shows how the graphene (i.e., absorbent) concentration affects the As removal efficiency and adsorption capacity ($q_{eq}$) at stable equilibrium for the case of 1 As and 7 arsenic molecules.  For the case of 1 {  As(V) molecule (Fig.~\ref{fig:effect_dose} a))}, as the GO concentration increases, the removal efficiency increases while the adsorption capacity per unit mass of adsorbent decreases. This relationship is typical in adsorption processes~\cite{singhSystematicStudyArsenic2022} in which the capacity of the adsorbent is initially utilized more effectively but decreases with increased adsorbent dose as, in this case, GO particles agglomerate or fewer of the available sites participate in contaminant binding. In contrast, for the case of 7 {  As(V) molecules (Fig. ~\ref{fig:effect_dose} b))}, both the removal efficiency and the adsorption capacity at stable equilibrium decrease. This reduction in removal efficiency is related to the fact that, with an excess of As available in water, not all of it is actually attracted to the surface of GO, hence reducing the removal efficiency. Understanding this balance is crucial for efficiently designing and implementing water purification treatments.

{ 
Thus, the SEAQT solution yields a unique, thermodynamically consistent trajectory, and the resulting adsorption kinetics emerge without any \emph{a priori} assumption about the reaction mechanism, arising instead as a direct consequence of the underlying molecular interactions and the SEA priniple.
 It is also worth mentioning that, although the present work focuses on As (V), the SEAQT formalism can
readily accommodate additional aqueous species such as As (III) and their
pH-dependent protonation states.  Doing so requires only an enlargement of the
density-of-states to include the corresponding energy–particle-number
eigenlevels and the introduction of extra conservation constraints (total
arsenic and hydrogen ions).}
\begin{figure}[htbp]
    \centering
    \includegraphics[width=0.6\linewidth]{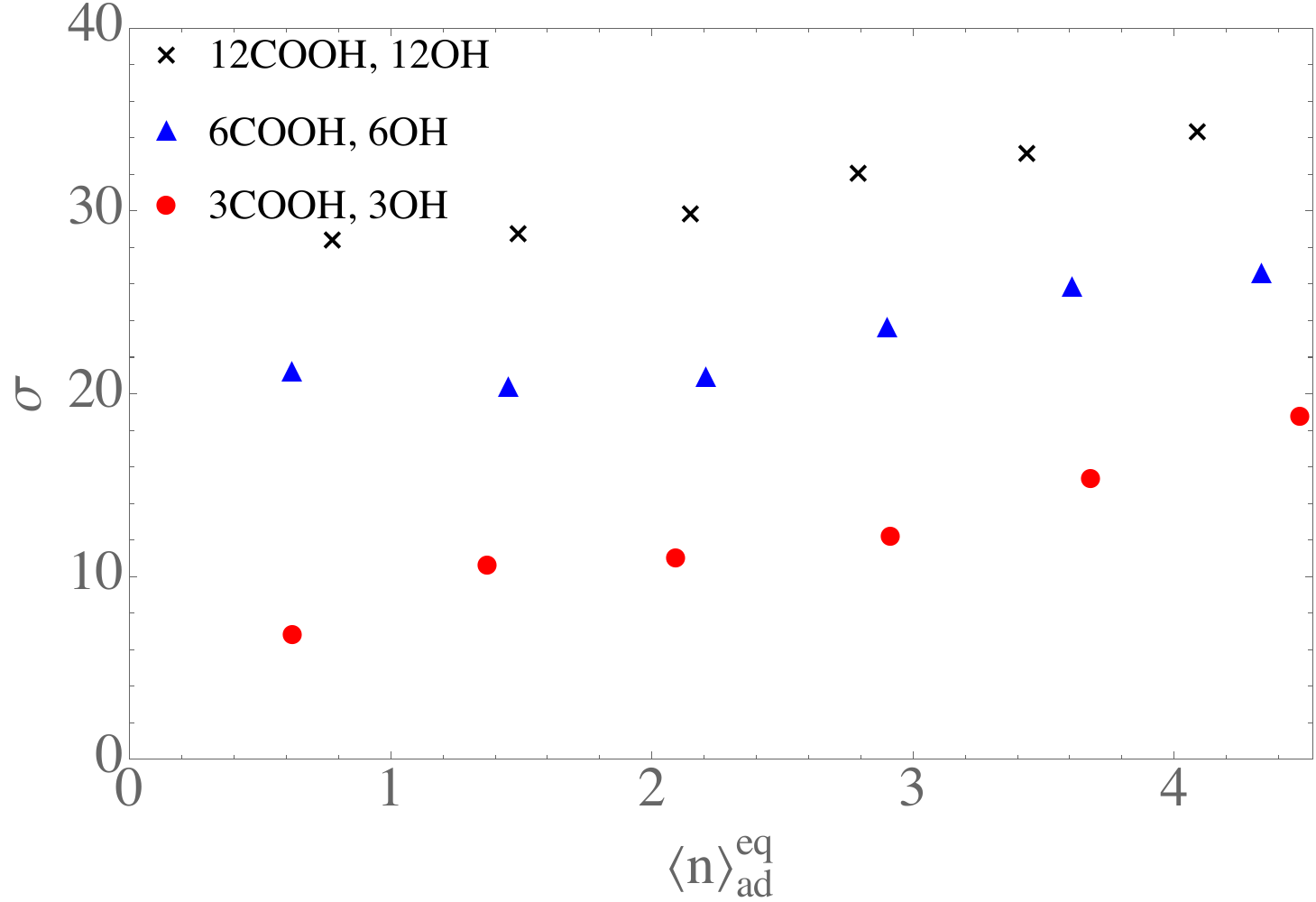}
    \caption{Total entropy production, $\sigma$, versus the number of adsorbed {  As(V) molecules} for 3 (red), 6 (blue) and 12 (black) each of the functional groups, COOH$^-$ and OH$^-$. The individual points starting on the left and and moving to the right correspond to initial As concentrations of 1 to 6 molecules.}.
    \label{fig:Gen_Ent}
\end{figure}

\begin{figure}[htbp]
    \centering
    a)\includegraphics[width=0.6\linewidth]{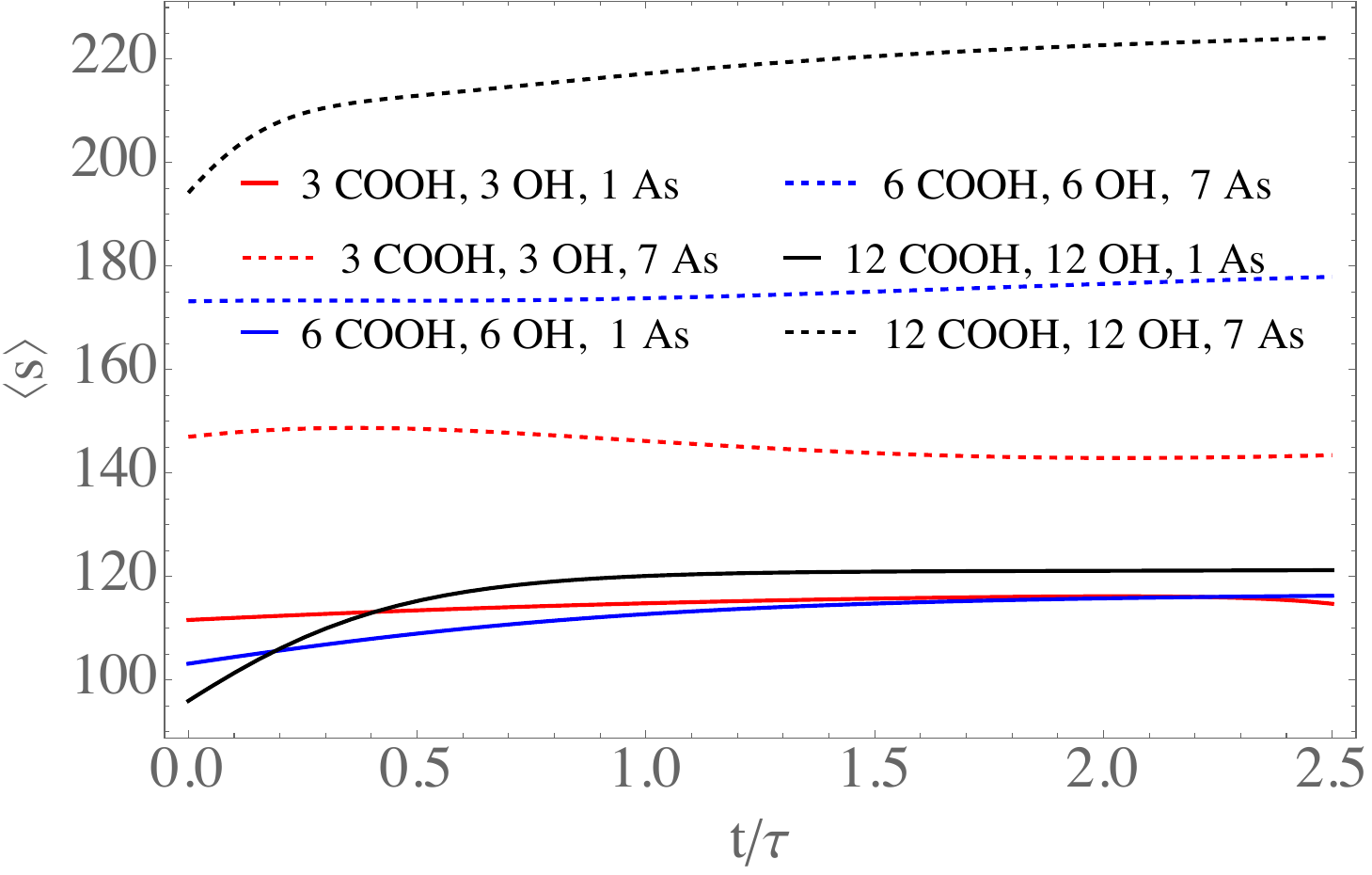}
    b)\includegraphics[width=0.6\linewidth]{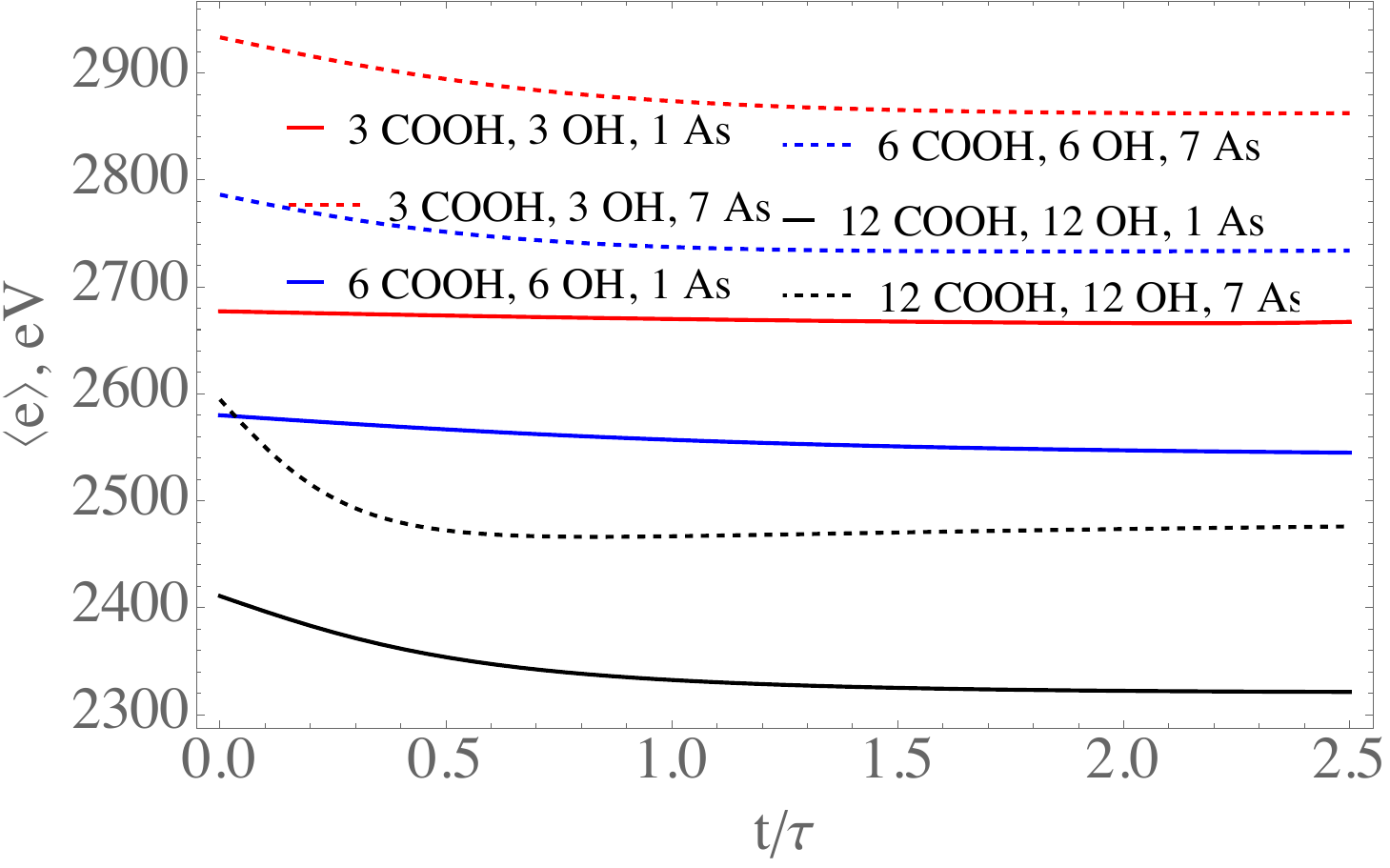}
    \caption{a) The expected entropy, $\langle s \rangle$, as a function of dimensionless time for 3 (black), 6 (red) and 12 (blue) each of the functional groups COOH$^-$ and OH$^-$; b) The expected energy, $\langle e \rangle$, as a function of dimensionless time for 3 (black), 6 (red) and 12 (blue) each of the functional groups COOH$^-$ and OH$^-$. For both a) and b), the solid and dashed curves are for an initial As concentration of 1 and 7 molecules, respectively. }
    \label{fig:Ent_En}
\end{figure}

\subsection{Thermodynamic properties}

In addition to providing kinetic information, the SEAQT framework also gives direct access to state properties like the time evolution of entropy and energy as well as the rate of entropy production, $\dot{\sigma}$, whichis a measure of irreversibilities present in the system.  Fig.~\ref{fig:Gen_Ent} shows how the total entropy production in going from some initial state to stable equilibrium varies with the number of adsorbed {  As(V) molecules}. The adsorption of {  As(V) molecules} onto GO is expected to be sensitive to the number of functional groups on the GO surface, so the As adsorption capacity should increase with the number of functional groups per GO particle. As the GO mass increases, the number of these functional groups available also increases for different numbers of the binding functional groups. Fig.~\ref{fig:Gen_Ent} demonstrates that greater As adsorption and greater numbers of functional groups both tend to increase the entropy produced during (non-equilibrium) adsorption. This trend can be rationalized in terms of the decrease in the system energy that drives adsorption due to the system's interaction with the thermal reservoir. In general, higher numbers of adsorbed {  As(V) molecules} correspond to lower energy levels (Fig.~\ref{fig:MachineLearning} c) and these lower energy levels have higher degeneracies (Fig.~\ref{fig:REWL} b)). These higher degeneracies increase the entropy production rate. Thus, the greater the number of {  As(V) molecules} that are captured during the adsorption process, the greater the deviation from equilibrium. This trend holds when the number of adsorbed {  As(V) molecules} is driven by an increased concentration of As, or more numerous functional groups. 

The time variation of entropy and energy during the (non-equilibrium) process are shown in Figs.~\ref{fig:Ent_En} a) and b), respectively.  The changes in $\langle s \rangle$ and $\langle e \rangle$ suggest the fastest adsorption (for a given As concentration) occurs initially when there is a high concentration of the functional groups (12 COOH$^-$ and 12 OH$^-$). This rapid adsorption corresponds to the highest entropy generation (greatest deviation from equilibrium) in Fig.~\ref{fig:Gen_Ent}.

\subsection{Influence of adsorbent mass}
Increasing the number of functional groups in the system (for a fixed As concentration) increases the binding sites for As, and intuition suggests the total number of adsorbed {  As(V) molecules} will increase.  It is also reasonable to suppose the rate of As adsorption will increase too.  Somewhat surprisingly, the SEAQT model suggests thatthe adsorption rate is much more sensitive to the number of functional groups than the eventual equilibrium number of adsorbed {  As(V) molecules}. 

\begin{figure}[htbp]
    \centering

    a)\includegraphics[width=0.6\linewidth]{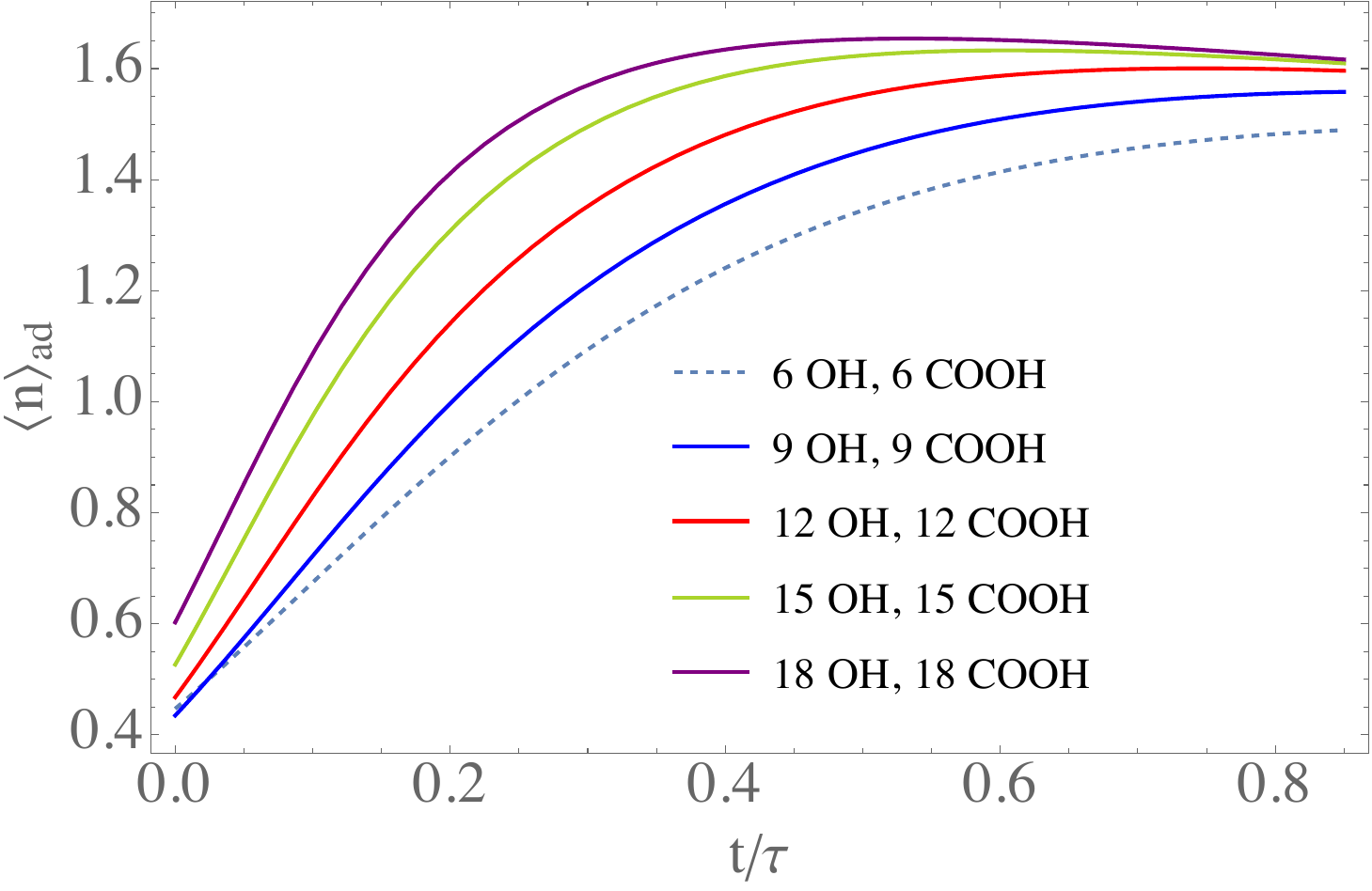}
    b)\includegraphics[width=0.6\linewidth]{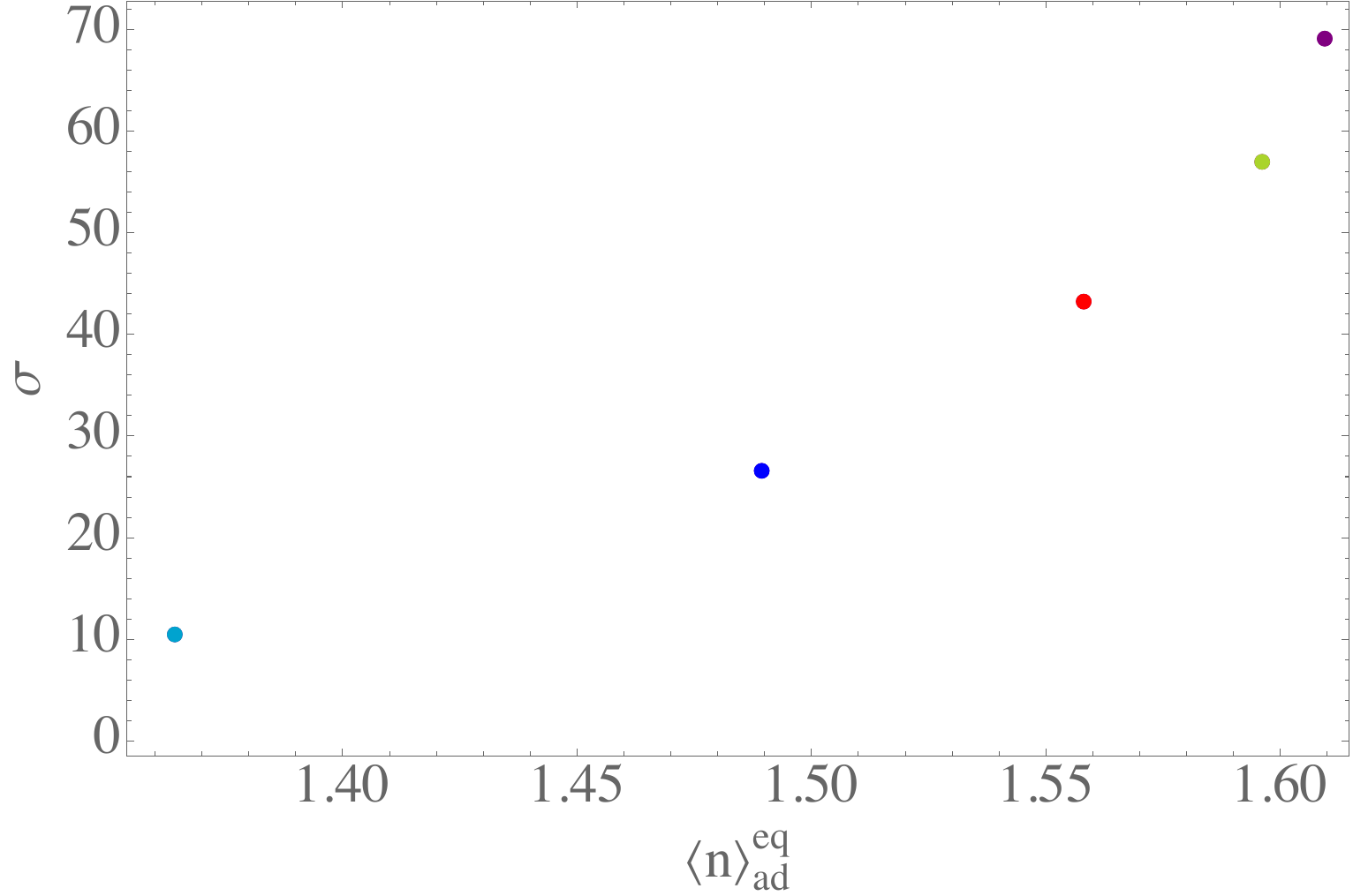}
    \caption{a) Evolution of the number of adsorbed {  As(V) molecules} on a 20 $\times$ 20 lattice with 2 {  As(V) molecules} and 1 molecule of H$^+$; Each curve corresponds to a different concentration of the COOH$^-$ and OH$^-$ functional groups; b) total entropy production that occurs during the evolution to stable equilibrium as a function of the adsorbed As particle number.  }
    \label{fig:GO_Kinetics}
\end{figure}

Fig.~\ref{fig:GO_Kinetics} a) shows how the number of adsorbed {  As(V) molecules} varies with time from an initially non-equilibrium state to stable equilibrium for five different concentrations of the functional groups (from 6 COOH$^-$ and 6 OH$^-$ to 18 COOH$^-$ and 18 OH$^-$) for a lattice size of 20 $\times$ 20, with 2 {  As(V) molecules} and 1 molecule of H$^+$.  The final stable equilibrium number of {  As(V) molecules} (the value reached on the far right of each curve) varies only between 1.4 and about 1.6 molecules --- a modest difference for the concentration range of functional groups considered. However, the {\it rate} at which these stable equilibrium adsorptions are reached varies significantly with the number of functional groups present.  For 6 COOH$^-$ and 6 OH$^-$, a dimensionless time of about 0.7 is needed to approach the equilibrium adsorption number of {  As(V) molecules}. For a solution with 18 COOH$^-$ and 18 OH$^-$ (an approximately 9\% increase in adsorbent mass) the equilibrium As adsorption is reached in about a third of the time (a dimensionless time of 0.25). Although increasing the adsorbent mass has only a modest effect on the amount of As adsorbed, it accelerates the As adsorption substantially.  This observation underscores the importance of optimizing the adsorbent mass to balance the effectiveness and efficiency of the adsorption processes. Additionally, it demonstrates the significance of the non-equilibrium path to rapidly remove  contaminants like As from the water. This insight is crucial for developing more effective and efficient water treatment systems, taking into account both equilibrium and non-equilibrium considerations to achieve optimal adsorption performance.

As to the total entropy produced for each of the concentrations of the functional groups, it is shown in Fig.~\ref{fig:GO_Kinetics} b) and increases with the number of functional groups. This trend confirms the connection noted above between a larger deviation from equilibrium and faster adsorption kinetics.

\begin{figure}[htbp]
    \centering
 
    a)\includegraphics[width=0.6\linewidth]{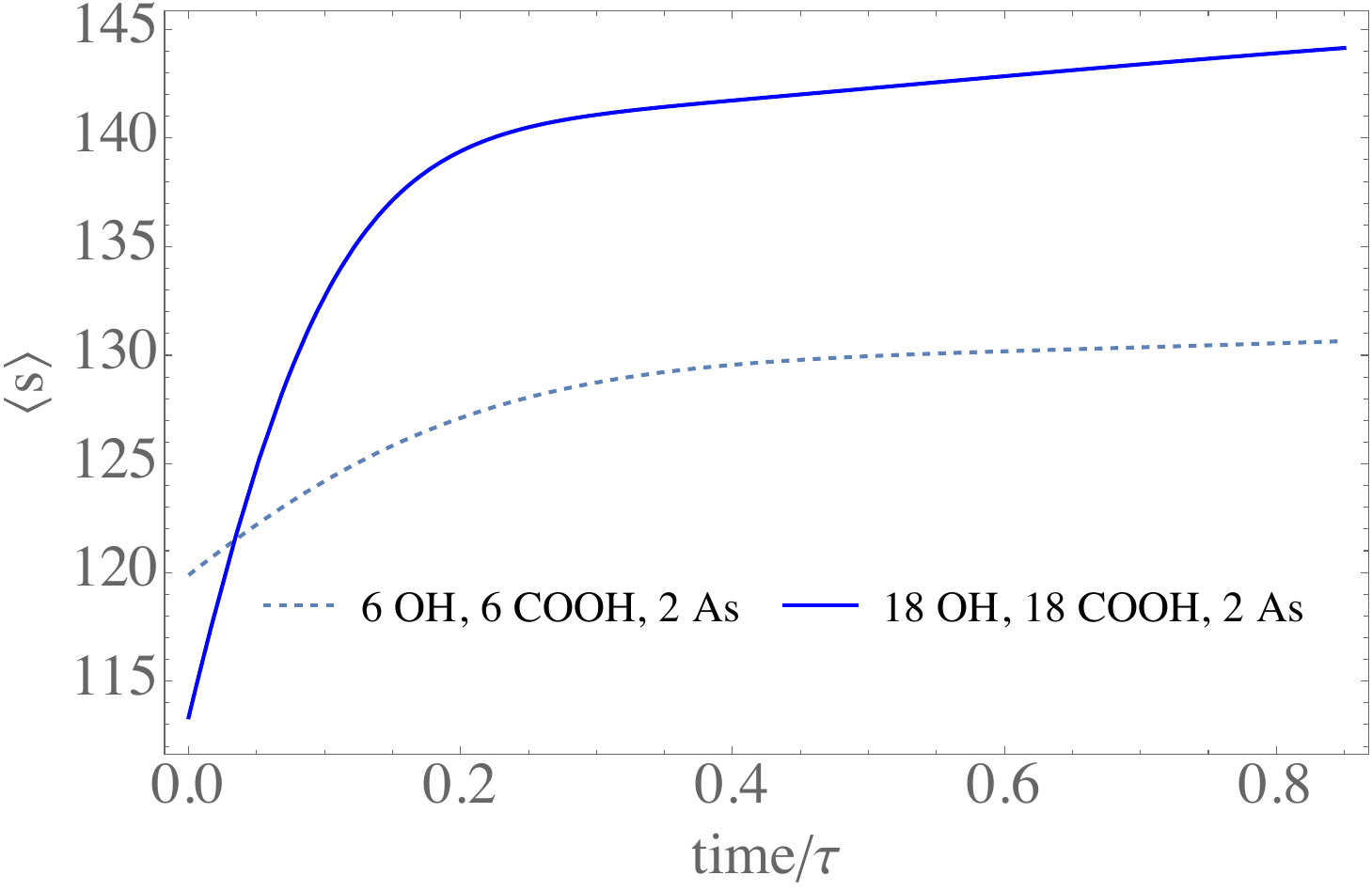}
  
     b)\includegraphics[width=0.6\linewidth]{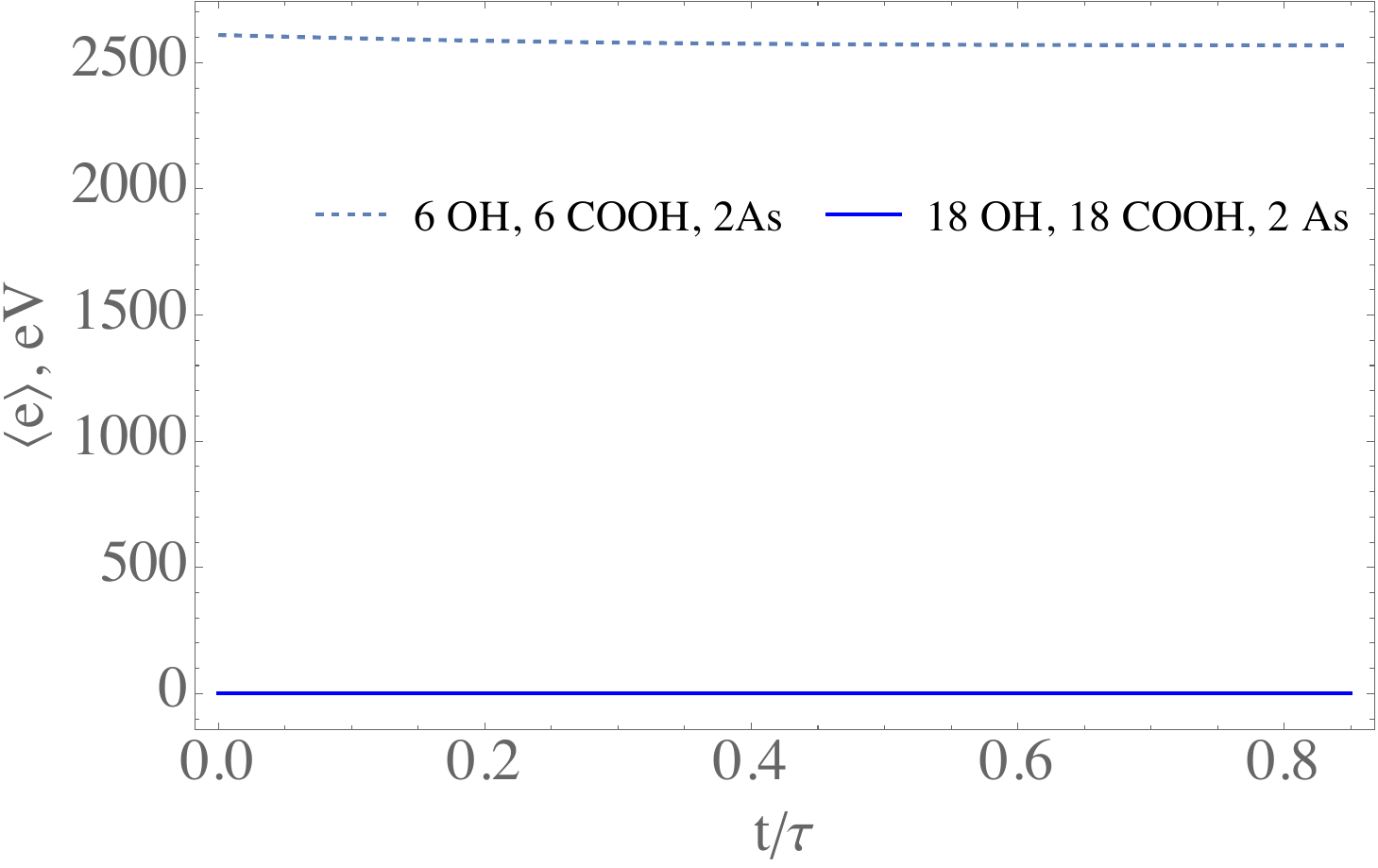}
     
\caption{a) The expected entropy, $\langle s \rangle$, as a function of dimensionless time for 6 and 18 each (dashed and solid curves, respectively) of the functional groups, COOH$^-$ and OH$^-$; b) the expected energy, $\langle e \rangle$, as a function of dimensionless time for 6 and 18 each (dashed and solid curves, respectively) of the functional groups, COOH$^-$ and OH$^-$;  for both a) and b), the system is a 20 $\times$ 20 lattice with 2 {  As(V) molecules} and 1 molecule of H$^+$. }
    \label{fig:Gen_Ent_As}
\end{figure}

The time evolution of the expected entropy and energy for a 20 $\times$ 20 lattice with 2 {  As(V) molecules} and 1 molecule of H$^+$ is shown in Fig.~\ref{fig:Gen_Ent_As} for two cases:
6 COOH$^-$ and 6 OH$^-$ functional groups and 18 COOH$^-$ and 18 OH$^-$ functional groups. 
As seen in Fig.~\ref{fig:Gen_Ent_As} a), the entropy for both cases initially increases rapidly and then much more slowly. These entropy changes result from a heat interaction with the thermal reservoir, which decreases the entropy due to the cooling that occurs, and from the entropy production, which increases the entropy and results from a redistribution of the energy among the energy eigenlevels available to the adsorption process.  Given that the system energy for the two concentrations of functional groups decreases only slightly throughout the adsorption process (Fig.~\ref{fig:Gen_Ent_As} b)), the entropy decrease resulting from the heat interaction is very small compared to theentropy produced by the adsorption process. Again, the greater the concentration of functional groups, the greater the deviation of the adsorption process from equilibrium and the faster the kinetics of the initial adsorption.

\subsection{Comparisons with classical models (stable equilibrium limit)}
\label{sec:equi}

\begin{figure}[htbp]
    \centering
    \includegraphics[width=0.6\linewidth]{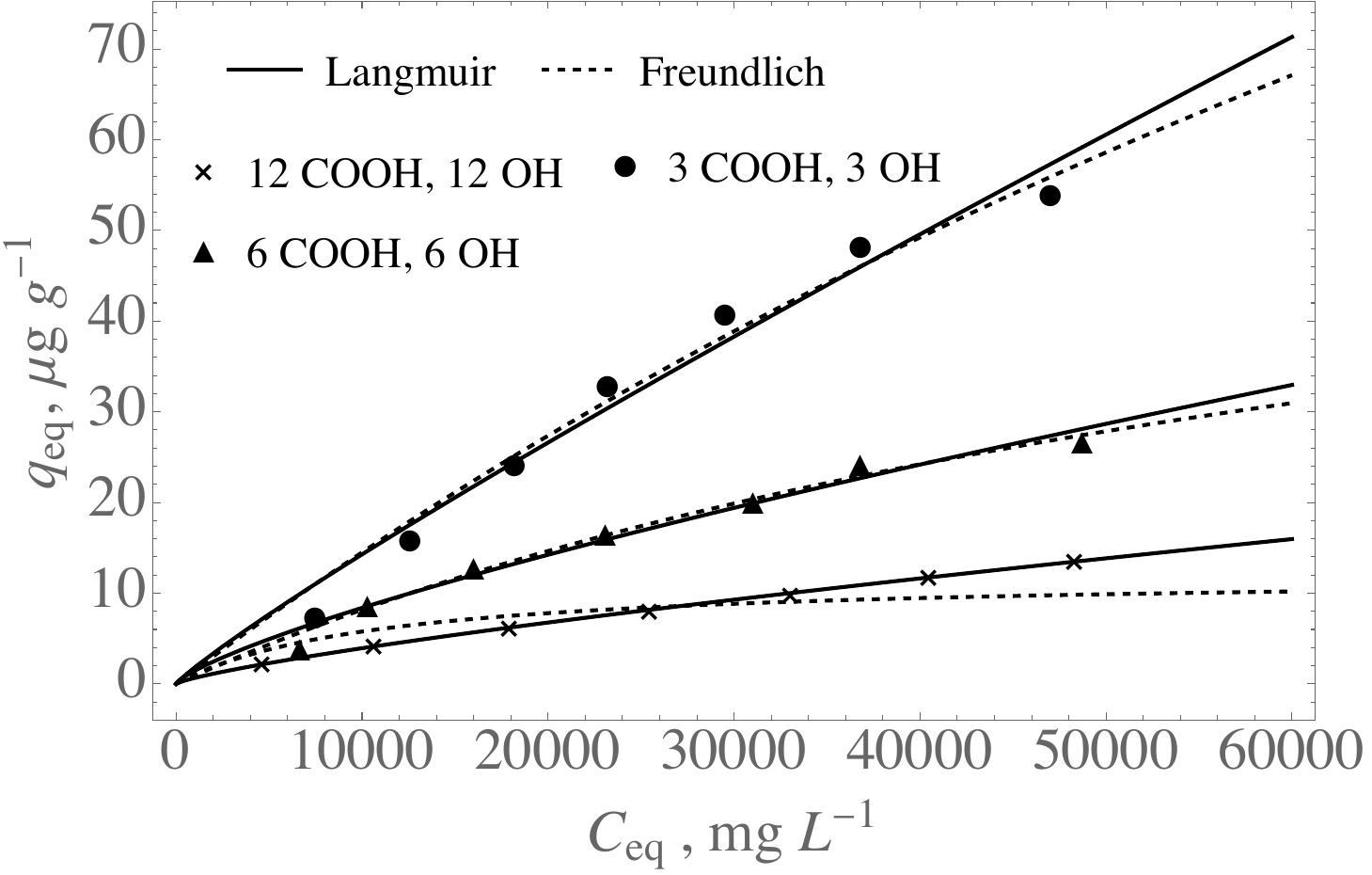}  
    \caption{Stable equilibrium concentrations for the time evolution by 12, 6 and 3 molecules of GO. The individual data are SEAQT predictions, the solid and dotted curves represent fitted curves using the Langmuir adsorption isotherm and the Freundlich model, respectively. 
    }
    \label{fig:equilibrium_1}
\end{figure}

Since the SEAQT framework describes adsorption kinetics that eventually end at stable equilibrium, it is interesting to compare its equilibrium predictions with classical adsorption models.  Fig.~\ref{fig:equilibrium_1} shows the stable equilibrium adsorption capacities predicted by the SEAQT equations of motion as a function of As concentration for the three adsorbent loadings shown in Fig.~\ref{fig:Evol}.  The three sets of data points represent three levels of GO adsorbent loading: 3, 6, and 12 molecules each of COOH$^-$ and 12 OH$^{-}$. The adsorption capacity, $q_{eq}$, for these adsorbent loadings are obtained from the particle numbers, $n_{\text{GO}-\text{OH}^-}$ and $n_{\text{GO}-\text{COOH}^-}$, via Eq.~(\ref{eq:conv}). The individual points are the SEAQT-predicted adsorption capacities at equilibrium for solutions with increasing As concentrations ($C_{\text {eq}}$). The solid curve represents a fit to the Langmuir adsorption isotherm~\cite{elmorsiEquilibriumIsothermsKinetic2011}  and the dotted currve is a fit to the empirical Freundlich model~\cite{ayaweiSynthesisCharacterizationApplication2015}.

For the greatest GO loading (12 COOH$^-$ and 12 OH$^{-}$ in Fig.~\ref{fig:equilibrium_1}, the adsorption capacity predicted by the Langmuir isotherm and the SEAQT at equilibrium (solid blue curve and the blue $\times$'s, respectively) vary almost linearly with the As concentration in water. This kind of trend suggests this level of GO loading has enough available adsorbant sites to accommodate increases in the system As concentration. In other words, the adsorption capacity is proportional to the As concentration in water for the plotted range of As, presumably because the adsorbant is not yet approaching saturation, i.e., there are more than enough functional sites to adsorb additional {  As(V) molecules}. The Freundlich model (red dashed line), on the other hand, systematically overestimates adsorption capacity at low As concentrations and underestimates it at high As concentrations.  

For the intermediate GO loading (6 COOH$^-$ and 6 OH$^{-}$ in Fig.~\ref{fig:equilibrium_1}, the Langmuir, Freundlich and SEAQT models agree well for $C_{\text eq}>10,000$, suggesting an approach to adsorption saturation. In situations where the adsorbent concentration is sufficient to reach saturation, the adsorption capacity is expected to bend toward a plateau associated with the adsorption limit.

At the lowest GO loading (3 COOH$^-$ and 3 OH$^{-}$ in Fig.~\ref{fig:equilibrium_1}, the Langmuir isotherm and the Freundlich model both predict significantly greater adsorption than the SEAQT framework at stable equilibrium.  Interestingly, the greatest deviation of the SEAQT approach from the conventional equilibrium models occurs at low adsorbent loading.  One possible explanation for this divergence is the onset of multilayer adsorption. For example, the adsorption capacity data shown in Fig.~\ref{fig:equilibrium_1} was fit to the Langmuir and Freundlich models with correlation coefficients, $R^2$, of 0.999, 0.995 and 0.994 for 12, 6, and 3 GO functional groups, respectively, which is suggestive of the appearance of the multilayer adsorption phenomenon at lower As concentrations. 

Note that multilayer adsorption is a common explanation for deviations of segregation data from equilibrium models, but the deviation highlighted here is between equilibrium models and the SEAQT framework.  The SEAQT framework is a mechanism-free description that is equally applicable for single and multi-layer segregation, so the deviation of SEAQT from standard equilibrium models at low adsorbent loadings in Fig.~\ref{fig:equilibrium_1} and low As concentrations is at least consistent with the onset of multi-layer adsorption. The wide applicability of the SEAQT framework demonstrates its value as a flexible tool for modeling adsorption processes of pollutants on GO.  It has the potential to predict pollutant adsorption properties before the synthesis of the adsorbents, and it may be useful for optimizing the design of adsorbent materials.

\label{sec:val}

\section{$\text{p}$H Variation}
\label{sec:pH}

\begin{figure}[htbp]
 \centering
    a)\includegraphics[width=0.48\linewidth]{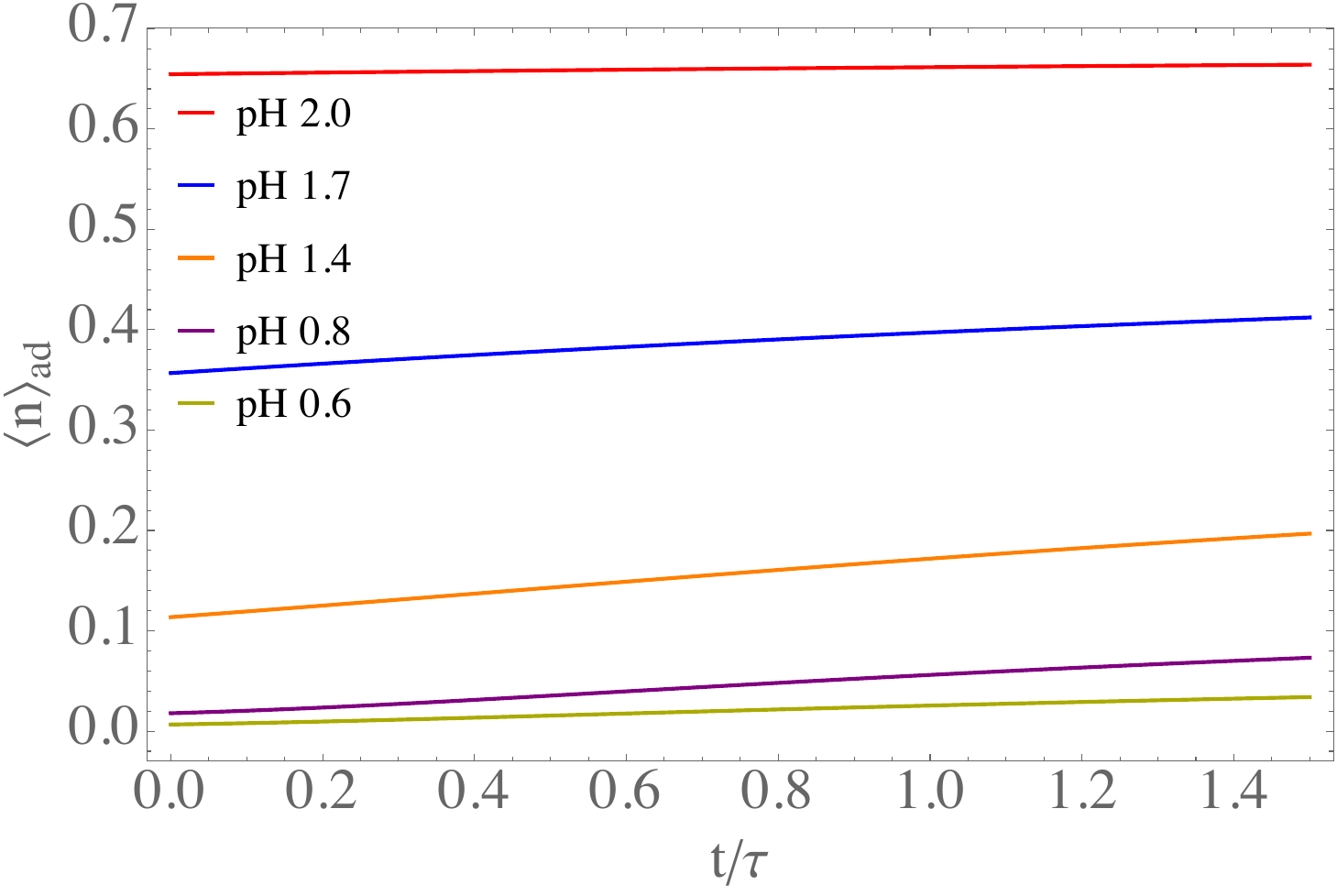}
    b)\includegraphics[width=0.48\linewidth]{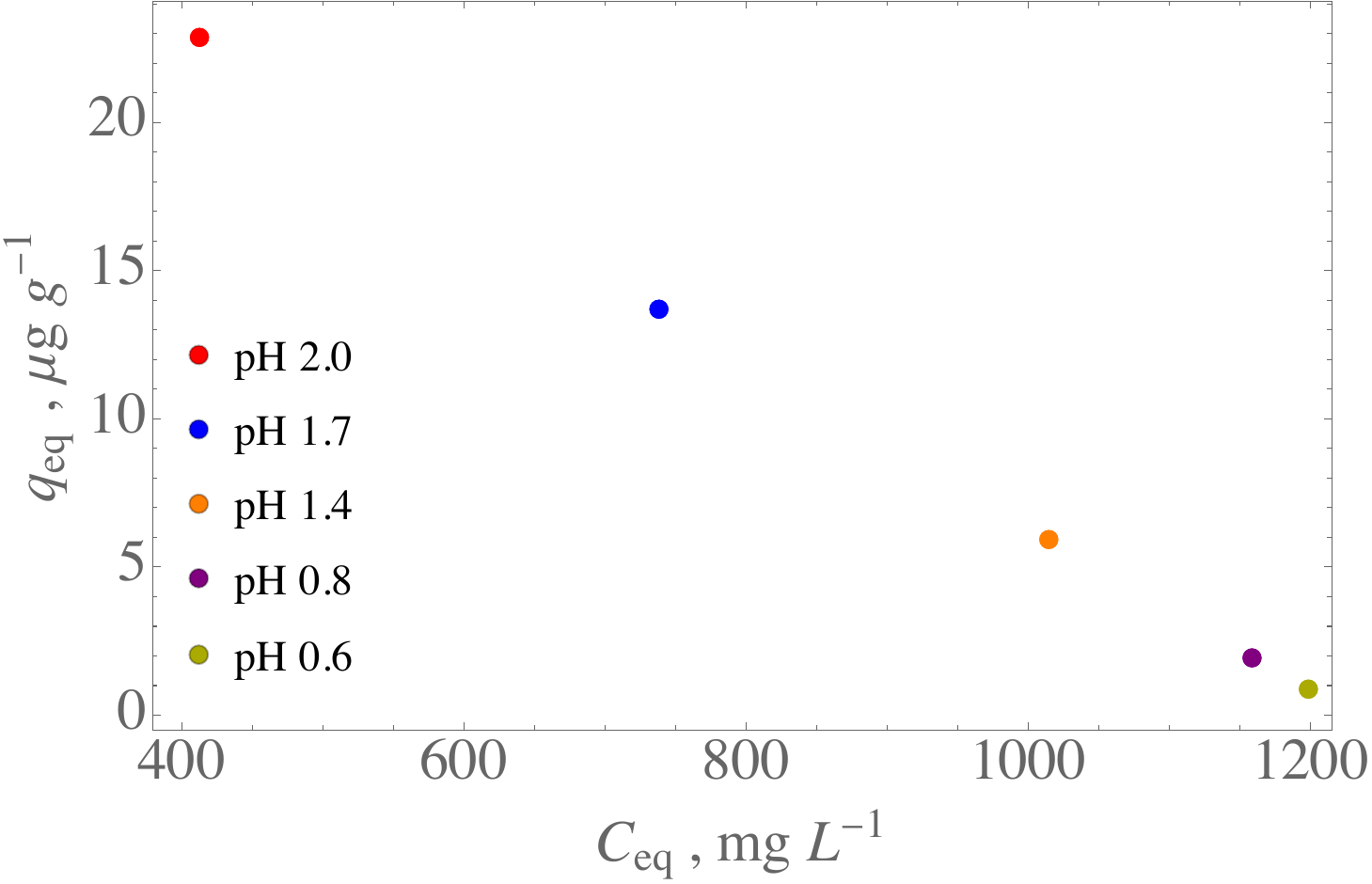}
    c)\includegraphics[width=0.48\linewidth]{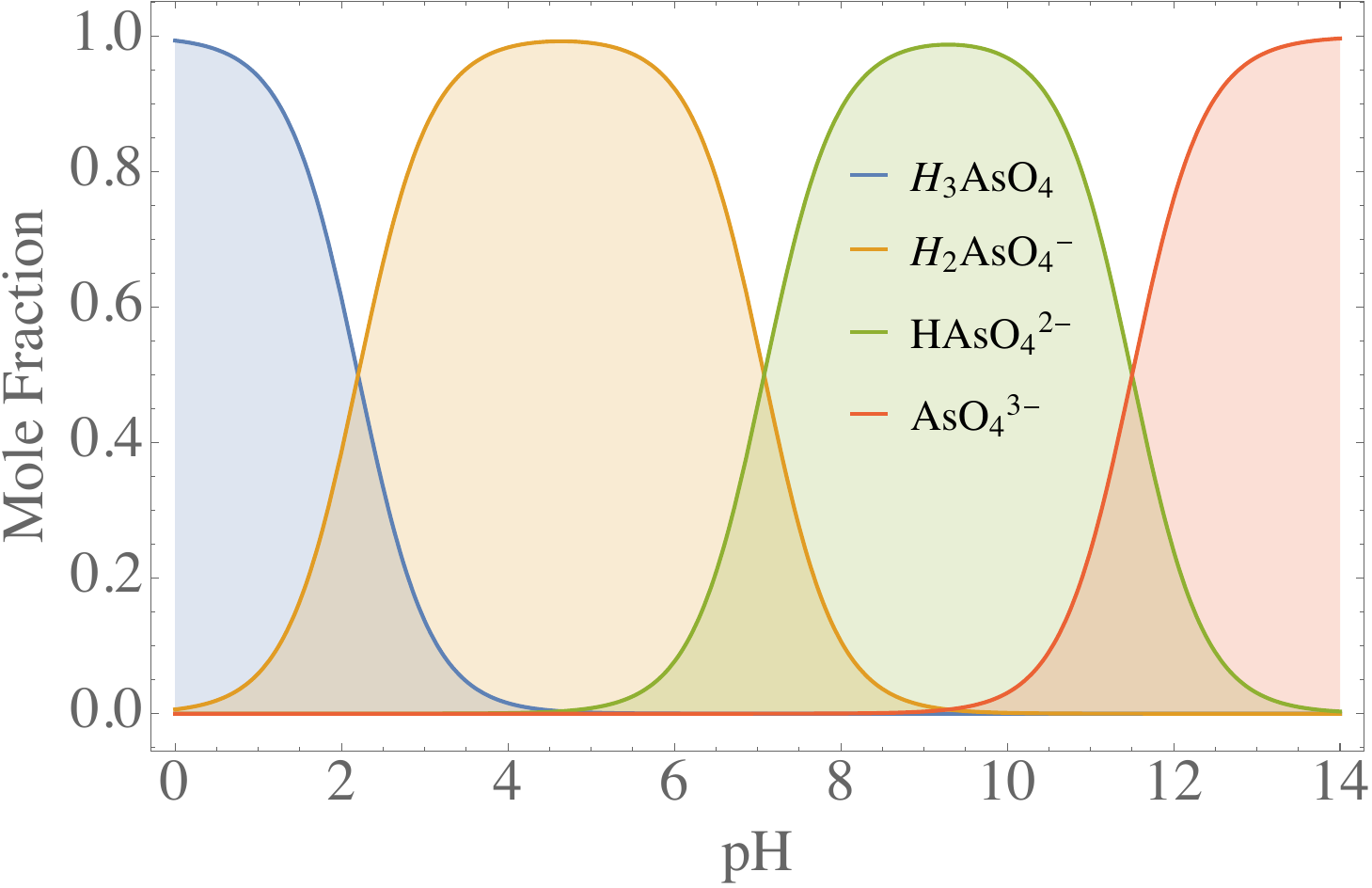}

  \caption{a) SEAQT evolution of the expected number of adsorbed {  As(V) molecules} by 6 GO functional groups at a pH of 0.6 (olive),  0.8 (purple),  1.4 (orange),  1.7 (blue), and 2.0 (red); b) stable equilibrium adsorption capacity versus the As concentration at equilibrium for the different pH (0.6 (olive),  0.8 (purple),  1.4 (orange),  1.7 (blue), and 2.0 (red)) evolutions; c) stable equilibrium mole fractions for the different As species as a function of pH.}
  \label{fig:pH}
\end{figure}

Fig.~\ref{fig:pH}a), shows the SEAQT predicted kinetics of As adsorption on 2 GO functional groups for a range of pH conditions: 0.6, 0.8, 1.4, 1.7, and 2.0. These conditions are represented on a 80 $\times$ 80 lattice, with 1 {  As(V) molecule} and variable H$^+$ content of 30, 17, 5, 2, and 1 molecules. It is evident in Figs.~\ref{fig:pH}a) and b) that the adsorption of As becomes less effective as the pH of the solution decreases. A significant increase in the predicted As adsorption at a pH of 2.0 over that at a pH of 0.6 is seen at both non-equilibrium (Figs.~\ref{fig:pH}a)) and stable equilibrium (Figs.~\ref{fig:pH}b)). 

This enhanced adsorption at higher pH levels (2.0 versus 0.6) is closely associated with the predominant form of As in solution, which directly depends upon the medium. The reduced adsorption efficiency at the lowest range of pH arises from a lower effective charge on {  As(V) molecules} in solution as the pH decreases. Fig.~(\ref{fig:pH}) c) shows how the charge state of the stable As specie changes with pH.  This diagram illustrates that the predominant As species at equilibrium shifts toward less negatively-charged forms as the pH decreases. In solutions with pH below 2, the expected dominant As species is H$_3$AsO$_4$, whose neutrality effectively eliminates the electrostatic interaction with the available adsorption sites on GO. 

Although the pair-potential model used to build the energy eigenstructure in the SEAQT framework does not explicitly consider the form of the {  As(V) molecule}, it does, nevertheless, accurately reflect the reduced electrostatic interaction between As and the adsorption sites at very low pH. This trend is shown by the declining SEAQT predicted adsorption capacity with decreasing pH (below 2) in Fig.~(\ref{fig:pH}) b). In the energy eigenstructure model, the more numerous H$^+$ ions in the lowest pH solutions presumably arrange themselves spatially to shield the charge on unadsorbed {  As(V) molecules} so that their effective charge is reduced and their attraction to the GO adsorption sites is greatly diminished.  As pH increases, this shielding of the As charge by H$^+$ decreases, the electrostatic interaction with the adsorbent increases, and As removal becomes more effective (adsorption capacity rises). Thus, the predictions of the SEAQT framework with a simple pair-potential model is consistent with equilibrium solution chemistry (Fig.~(\ref{fig:pH}) c)) and the intuitive logic that more positively charged As species are more easily removed from solution by electrostatic interactions with GO.

%At a pH below 2, the expected dominant arsenic species is H$_3$AsO$_4$, whose neutrality effectively eliminates the electrostatic interaction with the available adsorption sites on GO.  This trend is confirmed by the points with  a pH lower than 2 that are predicted using the SEAQT equation of motion. By adjusting the pH to 2 or above, there is a shift in the predominant chemical form of arsenic to H$_2$AsO$_4^-$, an ion with a charge (see Fig.~\ref{fig:pH} c)). This larger electrostatic interaction resulting from the greater charge increases the reactivity of As with the adsorption sites on the GO, significantly improving the efficiency of As adsorption.

Understanding how pH affects the chemical form of As in solution is crucial for optimizing the adsorption processes and designing more effective removal strategies through the leveraging of the electrostatic and chemical properties of both the adsorbate and the adsorbent. Additionally, it emphasizes the importance of considering the intrinsic properties and behavior of the adsorbent material at the design stage, enabling the development of more selective and efficient materials for specific contaminant capture.

{  
Although regeneration of the GO adsorbent lies beyond the scope of the present study, the SEAQT formalism could, in principle, be applied to simulate desorption by adjusting the boundary conditions (e.g., increasing solution pH or introducing competitive anions) to shift the system away from the adsorbed equilibrium. Such an approach could be used in future work to model regeneration kinetics and help identify operating conditions for multi-cycle reuse.
}

\section{Conclusions}
\label{sec:con}

{  The SEAQT framework is applied to modeling the non-equilibrium adsorption of arsenic from water onto graphene oxide, while explicitly including As(V), H$^+$, H$_2$O, and the two dominant surface groups of graphene oxide ($\mathrm{GO\!-\!OH}$ and $\mathrm{GO\!-\!COOH}$). Given only the bulk As(V) concentration, solution pH, and GO as inputs, the framework yields the complete time-dependent uptake. At low arsenate levels, the stable equilibrium predictions coincide with the Langmuir isotherm, whereas at higher concentrations, a larger capacity is obtained, indicating a transition to multilayer adsorption beyond the scope of the Langmuir model. Adsorption kinetics are strongly influenced by initial conditions. For example, low $C_0$ and low GO loading produce a monotonic uptake, whereas high $C_0$ and high loading produce an initial overshoot followed by a relaxation to equilibrium. A decline in capacity is also captured as the pH falls below $\approx3$, consistent with reduced electrostatic attraction attributable to protonation.

Because SEAQT imposes no {\it a priori} constraint on mono- or multi-layer  coverage, the framework provides a unified route to optimize adsorbent dose versus target removal. Results indicate diminishing returns once the adsorbate-to-surface ratio drops below a critical value and confirm that maintaining a pH above 2 preserves the electrostatic driving forces. An ANN surrogate trained on the REWL eigenstructure demonstrates how the method can be generalized to experimental domains not explicitly
sampled here.

Future work will refine interaction energies, incorporate additional edge chemistries in GO, and couple the SEAQT solver to continuum mass-transfer models, thereby extending applicability to full-scale water-treatment scenarios.}

\section*{Acknowledgements}

WTR acknowledges helpful discussions on graphene oxide with the TIET (Thapar Institute of Engineering and Technology)-Virginia Tech, Center of Excellence in Emerging Materials. ASR and CDA thank to VT for their kind hospitality during the realization of this project. ASR and CDA also thank CONAHCYT for its financial support through scholarships 252825 and 274533.

%\bibliography{references_z1}

\begin{thebibliography}{79}%
\makeatletter
\providecommand \@ifxundefined [1]{%
 \@ifx{#1\undefined}
}%
\providecommand \@ifnum [1]{%
 \ifnum #1\expandafter \@firstoftwo
 \else \expandafter \@secondoftwo
 \fi
}%
\providecommand \@ifx [1]{%
 \ifx #1\expandafter \@firstoftwo
 \else \expandafter \@secondoftwo
 \fi
}%
\providecommand \natexlab [1]{#1}%
\providecommand \enquote  [1]{``#1''}%
\providecommand \bibnamefont  [1]{#1}%
\providecommand \bibfnamefont [1]{#1}%
\providecommand \citenamefont [1]{#1}%
\providecommand \href@noop [0]{\@secondoftwo}%
\providecommand \href [0]{\begingroup \@sanitize@url \@href}%
\providecommand \@href[1]{\@@startlink{#1}\@@href}%
\providecommand \@@href[1]{\endgroup#1\@@endlink}%
\providecommand \@sanitize@url [0]{\catcode `\\12\catcode `\$12\catcode
  `\&12\catcode `\#12\catcode `\^12\catcode `\_12\catcode `\%12\relax}%
\providecommand \@@startlink[1]{}%
\providecommand \@@endlink[0]{}%
\providecommand \url  [0]{\begingroup\@sanitize@url \@url }%
\providecommand \@url [1]{\endgroup\@href {#1}{\urlprefix }}%
\providecommand \urlprefix  [0]{URL }%
\providecommand \Eprint [0]{\href }%
\providecommand \doibase [0]{https://doi.org/}%
\providecommand \selectlanguage [0]{\@gobble}%
\providecommand \bibinfo  [0]{\@secondoftwo}%
\providecommand \bibfield  [0]{\@secondoftwo}%
\providecommand \translation [1]{[#1]}%
\providecommand \BibitemOpen [0]{}%
\providecommand \bibitemStop [0]{}%
\providecommand \bibitemNoStop [0]{.\EOS\space}%
\providecommand \EOS [0]{\spacefactor3000\relax}%
\providecommand \BibitemShut  [1]{\csname bibitem#1\endcsname}%
\let\auto@bib@innerbib\@empty
%</preamble>
\bibitem [{\citenamefont {Kobya}\ \emph {et~al.}(2020)\citenamefont {Kobya},
  \citenamefont {Soltani}, \citenamefont {Omwene},\ and\ \citenamefont
  {Khataee}}]{kobyareviewdecontaminationarseniccontained2020}%
  \BibitemOpen
  \bibfield  {author} {\bibinfo {author} {\bibfnamefont {M.}~\bibnamefont
  {Kobya}}, \bibinfo {author} {\bibfnamefont {R.~D.~C.}\ \bibnamefont
  {Soltani}}, \bibinfo {author} {\bibfnamefont {P.~I.}\ \bibnamefont
  {Omwene}},\ and\ \bibinfo {author} {\bibfnamefont {A.}~\bibnamefont
  {Khataee}},\ }\bibfield  {title} {\bibinfo {title} {A review on
  decontamination of arsenic-contained water by electrocoagulation: {{Reactor}}
  configurations and operating cost along with removal mechanisms},\ }\href
  {https://doi.org/10.1016/j.eti.2019.100519} {\bibfield  {journal} {\bibinfo
  {journal} {Environmental Technology \& Innovation}\ }\textbf {\bibinfo
  {volume} {17}},\ \bibinfo {pages} {100519} (\bibinfo {year}
  {2020})}\BibitemShut {NoStop}%
\bibitem [{\citenamefont {Ullah}\ \emph {et~al.}(2023)\citenamefont {Ullah},
  \citenamefont {Rashid}, \citenamefont {Ghani}, \citenamefont {Talib},
  \citenamefont {Shahab},\ and\ \citenamefont
  {Lun}}]{ullahArsenicContaminationWater2023}%
  \BibitemOpen
  \bibfield  {author} {\bibinfo {author} {\bibfnamefont {Z.}~\bibnamefont
  {Ullah}}, \bibinfo {author} {\bibfnamefont {A.}~\bibnamefont {Rashid}},
  \bibinfo {author} {\bibfnamefont {J.}~\bibnamefont {Ghani}}, \bibinfo
  {author} {\bibfnamefont {M.~A.}\ \bibnamefont {Talib}}, \bibinfo {author}
  {\bibfnamefont {A.}~\bibnamefont {Shahab}},\ and\ \bibinfo {author}
  {\bibfnamefont {L.}~\bibnamefont {Lun}},\ }\bibfield  {title} {\bibinfo
  {title} {Arsenic {{Contamination}}, {{Water Toxicity}}, {{Source
  Apportionment}}, and {{Potential Health Risk}} in {{Groundwater}} of {{Jhelum
  Basin}}, {{Punjab}}, {{Pakistan}}},\ }\href
  {https://doi.org/10.1007/s12011-022-03139-0} {\bibfield  {journal} {\bibinfo
  {journal} {Biological Trace Element Research}\ }\textbf {\bibinfo {volume}
  {201}},\ \bibinfo {pages} {514} (\bibinfo {year} {2023})}\BibitemShut
  {NoStop}%
\bibitem [{\citenamefont {Adeloju}\ \emph {et~al.}(2021)\citenamefont
  {Adeloju}, \citenamefont {Khan},\ and\ \citenamefont
  {Patti}}]{adelojuArsenicContaminationGroundwater2021}%
  \BibitemOpen
  \bibfield  {author} {\bibinfo {author} {\bibfnamefont {S.~B.}\ \bibnamefont
  {Adeloju}}, \bibinfo {author} {\bibfnamefont {S.}~\bibnamefont {Khan}},\ and\
  \bibinfo {author} {\bibfnamefont {A.~F.}\ \bibnamefont {Patti}},\ }\bibfield
  {title} {\bibinfo {title} {Arsenic {{Contamination}} of {{Groundwater}} and
  {{Its Implications}} for {{Drinking Water Quality}} and {{Human Health}} in
  {{Under-Developed Countries}} and {{Remote Communities}}---{{A Review}}},\
  }\href {https://doi.org/10.3390/app11041926} {\bibfield  {journal} {\bibinfo
  {journal} {Applied Sciences}\ }\textbf {\bibinfo {volume} {11}},\ \bibinfo
  {pages} {1926} (\bibinfo {year} {2021})}\BibitemShut {NoStop}%
\bibitem [{\citenamefont {Rahman}\ \emph {et~al.}(2021)\citenamefont {Rahman},
  \citenamefont {Mondal},\ and\ \citenamefont
  {Fauzia}}]{rahmanArsenicenrichmentits2021}%
  \BibitemOpen
  \bibfield  {author} {\bibinfo {author} {\bibfnamefont {A.}~\bibnamefont
  {Rahman}}, \bibinfo {author} {\bibfnamefont {N.~C.}\ \bibnamefont {Mondal}},\
  and\ \bibinfo {author} {\bibfnamefont {F.}~\bibnamefont {Fauzia}},\
  }\bibfield  {title} {\bibinfo {title} {Arsenic enrichment and its natural
  background in groundwater at the proximity of active floodplains of {{Ganga
  River}}, northern {{India}}},\ }\href
  {https://doi.org/10.1016/j.chemosphere.2020.129096} {\bibfield  {journal}
  {\bibinfo  {journal} {Chemosphere}\ }\textbf {\bibinfo {volume} {265}},\
  \bibinfo {pages} {129096} (\bibinfo {year} {2021})}\BibitemShut {NoStop}%
\bibitem [{\citenamefont {Nguyen}\ \emph {et~al.}(2023)\citenamefont {Nguyen},
  \citenamefont {Nguyen}, \citenamefont {Vigneswaran}, \citenamefont {Ha},\
  and\ \citenamefont {Ratnaweera}}]{nguyenReviewTheoreticalKnowledge2023}%
  \BibitemOpen
  \bibfield  {author} {\bibinfo {author} {\bibfnamefont {T.~H.}\ \bibnamefont
  {Nguyen}}, \bibinfo {author} {\bibfnamefont {T.~V.}\ \bibnamefont {Nguyen}},
  \bibinfo {author} {\bibfnamefont {S.}~\bibnamefont {Vigneswaran}}, \bibinfo
  {author} {\bibfnamefont {N.~T.~H.}\ \bibnamefont {Ha}},\ and\ \bibinfo
  {author} {\bibfnamefont {H.}~\bibnamefont {Ratnaweera}},\ }\bibfield  {title}
  {\bibinfo {title} {A {{Review}} of {{Theoretical Knowledge}} and {{Practical
  Applications}} of {{Iron-Based Adsorbents}} for {{Removing Arsenic}} from
  {{Water}}},\ }\href {https://doi.org/10.3390/min13060741} {\bibfield
  {journal} {\bibinfo  {journal} {Minerals}\ }\textbf {\bibinfo {volume}
  {13}},\ \bibinfo {pages} {741} (\bibinfo {year} {2023})}\BibitemShut
  {NoStop}%
\bibitem [{\citenamefont {Rathi}\ and\ \citenamefont
  {Kumar}(2021)}]{rathireviewsourcesidentification2021}%
  \BibitemOpen
  \bibfield  {author} {\bibinfo {author} {\bibfnamefont {B.~S.}\ \bibnamefont
  {Rathi}}\ and\ \bibinfo {author} {\bibfnamefont {P.~S.}\ \bibnamefont
  {Kumar}},\ }\bibfield  {title} {\bibinfo {title} {A review on sources,
  identification and treatment strategies for the removal of toxic {{Arsenic}}
  from water system},\ }\href {https://doi.org/10.1016/j.jhazmat.2021.126299}
  {\bibfield  {journal} {\bibinfo  {journal} {Journal of Hazardous Materials}\
  }\textbf {\bibinfo {volume} {418}},\ \bibinfo {pages} {126299} (\bibinfo
  {year} {2021})}\BibitemShut {NoStop}%
\bibitem [{\citenamefont {Kumar}\ \emph {et~al.}(2020)\citenamefont {Kumar},
  \citenamefont {Goswami}, \citenamefont {Patel}, \citenamefont {Srivastava},\
  and\ \citenamefont {Das}}]{kumarScenarioperspectivesmechanism2020}%
  \BibitemOpen
  \bibfield  {author} {\bibinfo {author} {\bibfnamefont {M.}~\bibnamefont
  {Kumar}}, \bibinfo {author} {\bibfnamefont {R.}~\bibnamefont {Goswami}},
  \bibinfo {author} {\bibfnamefont {A.~K.}\ \bibnamefont {Patel}}, \bibinfo
  {author} {\bibfnamefont {M.}~\bibnamefont {Srivastava}},\ and\ \bibinfo
  {author} {\bibfnamefont {N.}~\bibnamefont {Das}},\ }\bibfield  {title}
  {\bibinfo {title} {Scenario, perspectives and mechanism of arsenic and
  fluoride {{Co-occurrence}} in the groundwater: {{A}} review},\ }\href
  {https://doi.org/10.1016/j.chemosphere.2020.126126} {\bibfield  {journal}
  {\bibinfo  {journal} {Chemosphere}\ }\textbf {\bibinfo {volume} {249}},\
  \bibinfo {pages} {126126} (\bibinfo {year} {2020})}\BibitemShut {NoStop}%
\bibitem [{\citenamefont {Tabassum}\ \emph {et~al.}(2019)\citenamefont
  {Tabassum}, \citenamefont {Shahid}, \citenamefont {Dumat}, \citenamefont
  {Niazi}, \citenamefont {Khalid}, \citenamefont {Shah}, \citenamefont
  {Imran},\ and\ \citenamefont {Khalid}}]{tabassumHealthriskassessment2019}%
  \BibitemOpen
  \bibfield  {author} {\bibinfo {author} {\bibfnamefont {R.~A.}\ \bibnamefont
  {Tabassum}}, \bibinfo {author} {\bibfnamefont {M.}~\bibnamefont {Shahid}},
  \bibinfo {author} {\bibfnamefont {C.}~\bibnamefont {Dumat}}, \bibinfo
  {author} {\bibfnamefont {N.~K.}\ \bibnamefont {Niazi}}, \bibinfo {author}
  {\bibfnamefont {S.}~\bibnamefont {Khalid}}, \bibinfo {author} {\bibfnamefont
  {N.~S.}\ \bibnamefont {Shah}}, \bibinfo {author} {\bibfnamefont
  {M.}~\bibnamefont {Imran}},\ and\ \bibinfo {author} {\bibfnamefont
  {S.}~\bibnamefont {Khalid}},\ }\bibfield  {title} {\bibinfo {title} {Health
  risk assessment of drinking arsenic-containing groundwater in {{Hasilpur}},
  {{Pakistan}}: Effect of sampling area, depth, and source},\ }\href
  {https://doi.org/10.1007/s11356-018-1276-z} {\bibfield  {journal} {\bibinfo
  {journal} {Environmental Science and Pollution Research}\ }\textbf {\bibinfo
  {volume} {26}},\ \bibinfo {pages} {20018} (\bibinfo {year}
  {2019})}\BibitemShut {NoStop}%
\bibitem [{\citenamefont {Adebayo}\ \emph {et~al.}(2021)\citenamefont
  {Adebayo}, \citenamefont {Abegunrin}, \citenamefont {Awe}, \citenamefont
  {Are}, \citenamefont {Guo}, \citenamefont {Onofua}, \citenamefont
  {Adegbola},\ and\ \citenamefont
  {Ojediran}}]{adebayoGeospatialmappingsuitability2021}%
  \BibitemOpen
  \bibfield  {author} {\bibinfo {author} {\bibfnamefont {T.~B.}\ \bibnamefont
  {Adebayo}}, \bibinfo {author} {\bibfnamefont {T.~P.}\ \bibnamefont
  {Abegunrin}}, \bibinfo {author} {\bibfnamefont {G.~O.}\ \bibnamefont {Awe}},
  \bibinfo {author} {\bibfnamefont {K.~S.}\ \bibnamefont {Are}}, \bibinfo
  {author} {\bibfnamefont {H.}~\bibnamefont {Guo}}, \bibinfo {author}
  {\bibfnamefont {O.~E.}\ \bibnamefont {Onofua}}, \bibinfo {author}
  {\bibfnamefont {G.~A.}\ \bibnamefont {Adegbola}},\ and\ \bibinfo {author}
  {\bibfnamefont {J.~O.}\ \bibnamefont {Ojediran}},\ }\bibfield  {title}
  {\bibinfo {title} {Geospatial mapping and suitability classification of
  groundwater quality for agriculture and domestic uses in a {{Precambrian}}
  basement complex},\ }\href {https://doi.org/10.1016/j.gsd.2020.100497}
  {\bibfield  {journal} {\bibinfo  {journal} {Groundwater for Sustainable
  Development}\ }\textbf {\bibinfo {volume} {12}},\ \bibinfo {pages} {100497}
  (\bibinfo {year} {2021})}\BibitemShut {NoStop}%
\bibitem [{\citenamefont {Malsawmdawngzela}\ \emph {et~al.}(2023)\citenamefont
  {Malsawmdawngzela}, \citenamefont {{Lalhmunsiama}}, \citenamefont {Tiwari},\
  and\ \citenamefont {Lee}}]{malsawmdawngzelaSynthesisnovelclaybased2023}%
  \BibitemOpen
  \bibfield  {author} {\bibinfo {author} {\bibfnamefont {R.}~\bibnamefont
  {Malsawmdawngzela}}, \bibinfo {author} {\bibnamefont {{Lalhmunsiama}}},
  \bibinfo {author} {\bibfnamefont {D.}~\bibnamefont {Tiwari}},\ and\ \bibinfo
  {author} {\bibfnamefont {S.}~\bibnamefont {Lee}},\ }\bibfield  {title}
  {\bibinfo {title} {Synthesis of novel clay-based nanocomposite materials and
  its application in the remediation of arsenic contaminated water},\ }\href
  {https://doi.org/10.1007/s13762-022-04506-z} {\bibfield  {journal} {\bibinfo
  {journal} {International Journal of Environmental Science and Technology}\
  }\textbf {\bibinfo {volume} {20}},\ \bibinfo {pages} {10285} (\bibinfo {year}
  {2023})}\BibitemShut {NoStop}%
\bibitem [{\citenamefont {Tropea}\ \emph {et~al.}(2021)\citenamefont {Tropea},
  \citenamefont {Hynds}, \citenamefont {McDermott}, \citenamefont {Brown},\
  and\ \citenamefont {Majury}}]{tropeaEnvironmentaladaptationcoli2021}%
  \BibitemOpen
  \bibfield  {author} {\bibinfo {author} {\bibfnamefont {E.}~\bibnamefont
  {Tropea}}, \bibinfo {author} {\bibfnamefont {P.}~\bibnamefont {Hynds}},
  \bibinfo {author} {\bibfnamefont {K.}~\bibnamefont {McDermott}}, \bibinfo
  {author} {\bibfnamefont {R.~S.}\ \bibnamefont {Brown}},\ and\ \bibinfo
  {author} {\bibfnamefont {A.}~\bibnamefont {Majury}},\ }\bibfield  {title}
  {\bibinfo {title} {Environmental adaptation of {{{\emph{E}}}}{\emph{. coli}}
  within private groundwater sources in southeastern {{Ontario}}:
  {{Implications}} for groundwater quality monitoring and human health},\
  }\href {https://doi.org/10.1016/j.envpol.2021.117263} {\bibfield  {journal}
  {\bibinfo  {journal} {Environmental Pollution}\ }\textbf {\bibinfo {volume}
  {285}},\ \bibinfo {pages} {117263} (\bibinfo {year} {2021})}\BibitemShut
  {NoStop}%
\bibitem [{\citenamefont {{Natasha}}\ \emph {et~al.}(2021)\citenamefont
  {{Natasha}}, \citenamefont {Shahid}, \citenamefont {Khalid}, \citenamefont
  {Niazi}, \citenamefont {Murtaza}, \citenamefont {Ahmad}, \citenamefont
  {Farooq}, \citenamefont {Zakir}, \citenamefont {Imran},\ and\ \citenamefont
  {Abbas}}]{natashaHealthrisksarsenic2021}%
  \BibitemOpen
  \bibfield  {author} {\bibinfo {author} {\bibnamefont {{Natasha}}}, \bibinfo
  {author} {\bibfnamefont {M.}~\bibnamefont {Shahid}}, \bibinfo {author}
  {\bibfnamefont {S.}~\bibnamefont {Khalid}}, \bibinfo {author} {\bibfnamefont
  {N.~K.}\ \bibnamefont {Niazi}}, \bibinfo {author} {\bibfnamefont
  {B.}~\bibnamefont {Murtaza}}, \bibinfo {author} {\bibfnamefont
  {N.}~\bibnamefont {Ahmad}}, \bibinfo {author} {\bibfnamefont
  {A.}~\bibnamefont {Farooq}}, \bibinfo {author} {\bibfnamefont
  {A.}~\bibnamefont {Zakir}}, \bibinfo {author} {\bibfnamefont
  {M.}~\bibnamefont {Imran}},\ and\ \bibinfo {author} {\bibfnamefont
  {G.}~\bibnamefont {Abbas}},\ }\bibfield  {title} {\bibinfo {title} {Health
  risks of arsenic buildup in soil and food crops after wastewater
  irrigation},\ }\href {https://doi.org/10.1016/j.scitotenv.2021.145266}
  {\bibfield  {journal} {\bibinfo  {journal} {Science of The Total
  Environment}\ }\textbf {\bibinfo {volume} {772}},\ \bibinfo {pages} {145266}
  (\bibinfo {year} {2021})}\BibitemShut {NoStop}%
\bibitem [{\citenamefont {Zhou}\ \emph {et~al.}(2021)\citenamefont {Zhou},
  \citenamefont {Li}, \citenamefont {Chen}, \citenamefont {Dong},\ and\
  \citenamefont {Lu}}]{zhouGroundwaterqualitypotable2021}%
  \BibitemOpen
  \bibfield  {author} {\bibinfo {author} {\bibfnamefont {Y.}~\bibnamefont
  {Zhou}}, \bibinfo {author} {\bibfnamefont {P.}~\bibnamefont {Li}}, \bibinfo
  {author} {\bibfnamefont {M.}~\bibnamefont {Chen}}, \bibinfo {author}
  {\bibfnamefont {Z.}~\bibnamefont {Dong}},\ and\ \bibinfo {author}
  {\bibfnamefont {C.}~\bibnamefont {Lu}},\ }\bibfield  {title} {\bibinfo
  {title} {Groundwater quality for potable and irrigation uses and associated
  health risk in southern part of {{Gu}}'an {{County}}, {{North China
  Plain}}},\ }\href {https://doi.org/10.1007/s10653-020-00553-y} {\bibfield
  {journal} {\bibinfo  {journal} {Environmental Geochemistry and Health}\
  }\textbf {\bibinfo {volume} {43}},\ \bibinfo {pages} {813} (\bibinfo {year}
  {2021})}\BibitemShut {NoStop}%
\bibitem [{\citenamefont
  {Rahidul~Hassan}(2023)}]{rahidulhassanReviewDifferentArsenic2023}%
  \BibitemOpen
  \bibfield  {author} {\bibinfo {author} {\bibfnamefont {H.}~\bibnamefont
  {Rahidul~Hassan}},\ }\bibfield  {title} {\bibinfo {title} {A review on
  different arsenic removal techniques used for decontamination of drinking
  water},\ }\href {https://doi.org/10.1080/26395940.2023.2165964} {\bibfield
  {journal} {\bibinfo  {journal} {Environmental Pollutants and
  Bioavailability}\ }\textbf {\bibinfo {volume} {35}},\ \bibinfo {pages}
  {2165964} (\bibinfo {year} {2023})}\BibitemShut {NoStop}%
\bibitem [{\citenamefont {Litter}\ \emph {et~al.}(2019)\citenamefont {Litter},
  \citenamefont {Ingallinella}, \citenamefont {Olmos}, \citenamefont {Savio},
  \citenamefont {Difeo}, \citenamefont {Botto}, \citenamefont {Torres},
  \citenamefont {Taylor}, \citenamefont {Frangie}, \citenamefont {Herkovits},
  \citenamefont {Schalamuk}, \citenamefont {Gonz{\'a}lez}, \citenamefont
  {Berardozzi}, \citenamefont {Garc{\'i}a~Einschlag}, \citenamefont
  {Bhattacharya},\ and\ \citenamefont
  {Ahmad}}]{litterArsenicArgentinaTechnologies2019}%
  \BibitemOpen
  \bibfield  {author} {\bibinfo {author} {\bibfnamefont {M.~I.}\ \bibnamefont
  {Litter}}, \bibinfo {author} {\bibfnamefont {A.~M.}\ \bibnamefont
  {Ingallinella}}, \bibinfo {author} {\bibfnamefont {V.}~\bibnamefont {Olmos}},
  \bibinfo {author} {\bibfnamefont {M.}~\bibnamefont {Savio}}, \bibinfo
  {author} {\bibfnamefont {G.}~\bibnamefont {Difeo}}, \bibinfo {author}
  {\bibfnamefont {L.}~\bibnamefont {Botto}}, \bibinfo {author} {\bibfnamefont
  {E.~M.~F.}\ \bibnamefont {Torres}}, \bibinfo {author} {\bibfnamefont
  {S.}~\bibnamefont {Taylor}}, \bibinfo {author} {\bibfnamefont
  {S.}~\bibnamefont {Frangie}}, \bibinfo {author} {\bibfnamefont
  {J.}~\bibnamefont {Herkovits}}, \bibinfo {author} {\bibfnamefont
  {I.}~\bibnamefont {Schalamuk}}, \bibinfo {author} {\bibfnamefont {M.~J.}\
  \bibnamefont {Gonz{\'a}lez}}, \bibinfo {author} {\bibfnamefont
  {E.}~\bibnamefont {Berardozzi}}, \bibinfo {author} {\bibfnamefont {F.~S.}\
  \bibnamefont {Garc{\'i}a~Einschlag}}, \bibinfo {author} {\bibfnamefont
  {P.}~\bibnamefont {Bhattacharya}},\ and\ \bibinfo {author} {\bibfnamefont
  {A.}~\bibnamefont {Ahmad}},\ }\bibfield  {title} {\bibinfo {title} {Arsenic
  in {{Argentina}}: {{Technologies}} for arsenic removal from groundwater
  sources, investment costs and waste management practices},\ }\href
  {https://doi.org/10.1016/j.scitotenv.2019.06.358} {\bibfield  {journal}
  {\bibinfo  {journal} {Science of The Total Environment}\ }\textbf {\bibinfo
  {volume} {690}},\ \bibinfo {pages} {778} (\bibinfo {year}
  {2019})}\BibitemShut {NoStop}%
\bibitem [{\citenamefont {{Joya-C{\'a}rdenas}}\ \emph
  {et~al.}(2022)\citenamefont {{Joya-C{\'a}rdenas}}, \citenamefont
  {{Rodr{\'i}guez-Caicedo}}, \citenamefont {{Gallegos-Mu{\~n}oz}},
  \citenamefont {Zanor}, \citenamefont {{Caycedo-Garc{\'i}a}}, \citenamefont
  {{Damian-Ascencio}},\ and\ \citenamefont
  {{Salda{\~n}a-Robles}}}]{joya-cardenasGrapheneBasedAdsorbentsArsenic2022}%
  \BibitemOpen
  \bibfield  {author} {\bibinfo {author} {\bibfnamefont {D.~R.}\ \bibnamefont
  {{Joya-C{\'a}rdenas}}}, \bibinfo {author} {\bibfnamefont {J.~P.}\
  \bibnamefont {{Rodr{\'i}guez-Caicedo}}}, \bibinfo {author} {\bibfnamefont
  {A.}~\bibnamefont {{Gallegos-Mu{\~n}oz}}}, \bibinfo {author} {\bibfnamefont
  {G.~A.}\ \bibnamefont {Zanor}}, \bibinfo {author} {\bibfnamefont {M.~S.}\
  \bibnamefont {{Caycedo-Garc{\'i}a}}}, \bibinfo {author} {\bibfnamefont
  {C.~E.}\ \bibnamefont {{Damian-Ascencio}}},\ and\ \bibinfo {author}
  {\bibfnamefont {A.}~\bibnamefont {{Salda{\~n}a-Robles}}},\ }\bibfield
  {title} {\bibinfo {title} {Graphene-{{Based Adsorbents}} for {{Arsenic}},
  {{Fluoride}}, and {{Chromium Adsorption}}: {{Synthesis Methods Review}}},\
  }\href {https://doi.org/10.3390/nano12223942} {\bibfield  {journal} {\bibinfo
   {journal} {Nanomaterials}\ }\textbf {\bibinfo {volume} {12}},\ \bibinfo
  {pages} {3942} (\bibinfo {year} {2022})}\BibitemShut {NoStop}%
\bibitem [{\citenamefont {Ansari}\ \emph {et~al.}(2019)\citenamefont {Ansari},
  \citenamefont {Azamat},\ and\ \citenamefont
  {Khataee}}]{ansariSeparationPerchloratesAqueous2019}%
  \BibitemOpen
  \bibfield  {author} {\bibinfo {author} {\bibfnamefont {P.}~\bibnamefont
  {Ansari}}, \bibinfo {author} {\bibfnamefont {J.}~\bibnamefont {Azamat}},\
  and\ \bibinfo {author} {\bibfnamefont {A.}~\bibnamefont {Khataee}},\
  }\bibfield  {title} {\bibinfo {title} {Separation of perchlorates from
  aqueous solution using functionalized graphene oxide nanosheets: A
  computational study},\ }\href {https://doi.org/10.1007/s10853-018-3045-2}
  {\bibfield  {journal} {\bibinfo  {journal} {Journal of Materials Science}\
  }\textbf {\bibinfo {volume} {54}},\ \bibinfo {pages} {2289} (\bibinfo {year}
  {2019})}\BibitemShut {NoStop}%
\bibitem [{\citenamefont {{Reynosa-Mart{\'i}nez}}\ \emph
  {et~al.}(2020)\citenamefont {{Reynosa-Mart{\'i}nez}}, \citenamefont {Tovar},
  \citenamefont {Gallegos}, \citenamefont {{Rodr{\'i}guez-Mel{\'e}ndez}},
  \citenamefont {{Torres-Cadena}}, \citenamefont
  {{Mondrag{\'o}n-Sol{\'o}rzano}}, \citenamefont {{Barroso-Flores}},
  \citenamefont {{Alvarez-Lemus}}, \citenamefont {Montalvo},\ and\
  \citenamefont
  {{L{\'o}pez-Honorato}}}]{reynosa-martinezEffectDegreeOxidation2020}%
  \BibitemOpen
  \bibfield  {author} {\bibinfo {author} {\bibfnamefont {A.~C.}\ \bibnamefont
  {{Reynosa-Mart{\'i}nez}}}, \bibinfo {author} {\bibfnamefont {G.~N.}\
  \bibnamefont {Tovar}}, \bibinfo {author} {\bibfnamefont {W.~R.}\ \bibnamefont
  {Gallegos}}, \bibinfo {author} {\bibfnamefont {H.}~\bibnamefont
  {{Rodr{\'i}guez-Mel{\'e}ndez}}}, \bibinfo {author} {\bibfnamefont
  {R.}~\bibnamefont {{Torres-Cadena}}}, \bibinfo {author} {\bibfnamefont
  {G.}~\bibnamefont {{Mondrag{\'o}n-Sol{\'o}rzano}}}, \bibinfo {author}
  {\bibfnamefont {J.}~\bibnamefont {{Barroso-Flores}}}, \bibinfo {author}
  {\bibfnamefont {M.~A.}\ \bibnamefont {{Alvarez-Lemus}}}, \bibinfo {author}
  {\bibfnamefont {V.~G.}\ \bibnamefont {Montalvo}},\ and\ \bibinfo {author}
  {\bibfnamefont {E.}~\bibnamefont {{L{\'o}pez-Honorato}}},\ }\bibfield
  {title} {\bibinfo {title} {Effect of the degree of oxidation of graphene
  oxide on {{As}}({{III}}) adsorption},\ }\href
  {https://doi.org/10.1016/j.jhazmat.2019.121440} {\bibfield  {journal}
  {\bibinfo  {journal} {Journal of Hazardous Materials}\ }\textbf {\bibinfo
  {volume} {384}},\ \bibinfo {pages} {121440} (\bibinfo {year}
  {2020})}\BibitemShut {NoStop}%
\bibitem [{\citenamefont {Zhang}\ \emph {et~al.}(2015)\citenamefont {Zhang},
  \citenamefont {Dabbs}, \citenamefont {Liu}, \citenamefont {Aksay},
  \citenamefont {Car},\ and\ \citenamefont
  {Selloni}}]{zhangCombinedEffectsFunctional2015}%
  \BibitemOpen
  \bibfield  {author} {\bibinfo {author} {\bibfnamefont {C.}~\bibnamefont
  {Zhang}}, \bibinfo {author} {\bibfnamefont {D.~M.}\ \bibnamefont {Dabbs}},
  \bibinfo {author} {\bibfnamefont {L.-M.}\ \bibnamefont {Liu}}, \bibinfo
  {author} {\bibfnamefont {I.~A.}\ \bibnamefont {Aksay}}, \bibinfo {author}
  {\bibfnamefont {R.}~\bibnamefont {Car}},\ and\ \bibinfo {author}
  {\bibfnamefont {A.}~\bibnamefont {Selloni}},\ }\bibfield  {title} {\bibinfo
  {title} {Combined {{Effects}} of {{Functional Groups}}, {{Lattice Defects}},
  and {{Edges}} in the {{Infrared Spectra}} of {{Graphene Oxide}}},\ }\href
  {https://doi.org/10.1021/acs.jpcc.5b02727} {\bibfield  {journal} {\bibinfo
  {journal} {The Journal of Physical Chemistry C}\ }\textbf {\bibinfo {volume}
  {119}},\ \bibinfo {pages} {18167} (\bibinfo {year} {2015})}\BibitemShut
  {NoStop}%
\bibitem [{\citenamefont {He}\ \emph {et~al.}(2023)\citenamefont {He},
  \citenamefont {Zheng}, \citenamefont {Ni}, \citenamefont {Shen},
  \citenamefont {Xu}, \citenamefont {Tian}, \citenamefont {Zou}, \citenamefont
  {Lei}, \citenamefont {He},\ and\ \citenamefont
  {Liu}}]{heDesignOptimizationNovel2023}%
  \BibitemOpen
  \bibfield  {author} {\bibinfo {author} {\bibfnamefont {J.}~\bibnamefont
  {He}}, \bibinfo {author} {\bibfnamefont {H.}~\bibnamefont {Zheng}}, \bibinfo
  {author} {\bibfnamefont {F.}~\bibnamefont {Ni}}, \bibinfo {author}
  {\bibfnamefont {F.}~\bibnamefont {Shen}}, \bibinfo {author} {\bibfnamefont
  {M.}~\bibnamefont {Xu}}, \bibinfo {author} {\bibfnamefont {D.}~\bibnamefont
  {Tian}}, \bibinfo {author} {\bibfnamefont {J.}~\bibnamefont {Zou}}, \bibinfo
  {author} {\bibfnamefont {Y.}~\bibnamefont {Lei}}, \bibinfo {author}
  {\bibfnamefont {Y.}~\bibnamefont {He}},\ and\ \bibinfo {author}
  {\bibfnamefont {Y.}~\bibnamefont {Liu}},\ }\bibfield  {title} {\bibinfo
  {title} {Design and optimization of a novel {{Y-Fe-GO}} magnetic adsorbent
  for simultaneous removal of tetracycline and arsenic and adsorption
  mechanisms},\ }\href {https://doi.org/10.1016/j.cej.2022.141195} {\bibfield
  {journal} {\bibinfo  {journal} {Chemical Engineering Journal}\ }\textbf
  {\bibinfo {volume} {457}},\ \bibinfo {pages} {141195} (\bibinfo {year}
  {2023})}\BibitemShut {NoStop}%
\bibitem [{\citenamefont {{Joya-C{\'a}rdenas}}\ \emph
  {et~al.}(2024)\citenamefont {{Joya-C{\'a}rdenas}}, \citenamefont
  {{Rodr{\'i}guez-Caicedo}}, \citenamefont {{Corona-Rivera}}, \citenamefont
  {{Salda{\~n}a-Robles}}, \citenamefont {{Dami{\'a}n-Ascencio}},\ and\
  \citenamefont {{Salda{\~n}a-Robles}}}]{joya-cardenasRemovalPresenceCr2024}%
  \BibitemOpen
  \bibfield  {author} {\bibinfo {author} {\bibfnamefont {D.}~\bibnamefont
  {{Joya-C{\'a}rdenas}}}, \bibinfo {author} {\bibfnamefont {J.}~\bibnamefont
  {{Rodr{\'i}guez-Caicedo}}}, \bibinfo {author} {\bibfnamefont
  {M.}~\bibnamefont {{Corona-Rivera}}}, \bibinfo {author} {\bibfnamefont
  {N.}~\bibnamefont {{Salda{\~n}a-Robles}}}, \bibinfo {author} {\bibfnamefont
  {C.}~\bibnamefont {{Dami{\'a}n-Ascencio}}},\ and\ \bibinfo {author}
  {\bibfnamefont {A.}~\bibnamefont {{Salda{\~n}a-Robles}}},\ }\bibfield
  {title} {\bibinfo {title} {Removal of {{As}}({{V}}) in the presence of
  {{Cr}}({{VI}}) in contaminated water from the {{Bajio}} region of {{Mexico}}
  using ferrihydrite-functionalized graphene oxide ({{GOFH}}): {{A}} case
  study},\ }\href {https://doi.org/10.1016/j.emcon.2024.100312} {\bibfield
  {journal} {\bibinfo  {journal} {Emerging Contaminants}\ }\textbf {\bibinfo
  {volume} {10}},\ \bibinfo {pages} {100312} (\bibinfo {year}
  {2024})}\BibitemShut {NoStop}%
\bibitem [{\citenamefont {Hossain}\ \emph {et~al.}(2024)\citenamefont
  {Hossain}, \citenamefont {Yasmin},\ and\ \citenamefont
  {Kabir}}]{hossainCosteffectiveSynthesisMagnetic2024}%
  \BibitemOpen
  \bibfield  {author} {\bibinfo {author} {\bibfnamefont {M.~S.}\ \bibnamefont
  {Hossain}}, \bibinfo {author} {\bibfnamefont {S.}~\bibnamefont {Yasmin}},\
  and\ \bibinfo {author} {\bibfnamefont {M.~H.}\ \bibnamefont {Kabir}},\
  }\bibfield  {title} {\bibinfo {title} {Cost-effective synthesis of magnetic
  graphene oxide nanocomposite from waste battery for the removal of arsenic
  from aqueous solutions: {{Adsorption}} mechanism with {{DFT}} calculation},\
  }\href {https://doi.org/10.1016/j.jscs.2024.101873} {\bibfield  {journal}
  {\bibinfo  {journal} {Journal of Saudi Chemical Society}\ }\textbf {\bibinfo
  {volume} {28}},\ \bibinfo {pages} {101873} (\bibinfo {year}
  {2024})}\BibitemShut {NoStop}%
\bibitem [{\citenamefont {{Vazquez-Jaime}}\ \emph {et~al.}(2020)\citenamefont
  {{Vazquez-Jaime}}, \citenamefont {{Arcibar-Orozco}}, \citenamefont
  {{Damian-Ascencio}}, \citenamefont {{Salda{\~n}a-Robles}}, \citenamefont
  {{Mart{\'i}nez-Rosales}}, \citenamefont {{Salda{\~n}a-Robles}},\ and\
  \citenamefont {{Cano-Andrade}}}]{vazquez-jaimeEffectiveRemovalArsenic2020}%
  \BibitemOpen
  \bibfield  {author} {\bibinfo {author} {\bibfnamefont {M.}~\bibnamefont
  {{Vazquez-Jaime}}}, \bibinfo {author} {\bibfnamefont {J.~A.}\ \bibnamefont
  {{Arcibar-Orozco}}}, \bibinfo {author} {\bibfnamefont {C.~E.}\ \bibnamefont
  {{Damian-Ascencio}}}, \bibinfo {author} {\bibfnamefont {A.~L.}\ \bibnamefont
  {{Salda{\~n}a-Robles}}}, \bibinfo {author} {\bibfnamefont {M.}~\bibnamefont
  {{Mart{\'i}nez-Rosales}}}, \bibinfo {author} {\bibfnamefont {A.}~\bibnamefont
  {{Salda{\~n}a-Robles}}},\ and\ \bibinfo {author} {\bibfnamefont
  {S.}~\bibnamefont {{Cano-Andrade}}},\ }\bibfield  {title} {\bibinfo {title}
  {Effective removal of arsenic from an aqueous solution by
  ferrihydrite/goethite graphene oxide composites using the modified
  {{Hummers}} method},\ }\href {https://doi.org/10.1016/j.jece.2020.104416}
  {\bibfield  {journal} {\bibinfo  {journal} {Journal of Environmental Chemical
  Engineering}\ }\textbf {\bibinfo {volume} {8}},\ \bibinfo {pages} {104416}
  (\bibinfo {year} {2020})}\BibitemShut {NoStop}%
\bibitem [{\citenamefont {Lingamdinne}\ \emph {et~al.}(2021)\citenamefont
  {Lingamdinne}, \citenamefont {Lee}, \citenamefont {Choi}, \citenamefont
  {Lebaka}, \citenamefont {Durbaka},\ and\ \citenamefont
  {Koduru}}]{lingamdinnePotentialMagneticHollow2021}%
  \BibitemOpen
  \bibfield  {author} {\bibinfo {author} {\bibfnamefont {L.~P.}\ \bibnamefont
  {Lingamdinne}}, \bibinfo {author} {\bibfnamefont {S.}~\bibnamefont {Lee}},
  \bibinfo {author} {\bibfnamefont {J.-S.}\ \bibnamefont {Choi}}, \bibinfo
  {author} {\bibfnamefont {V.~R.}\ \bibnamefont {Lebaka}}, \bibinfo {author}
  {\bibfnamefont {V.~R.~P.}\ \bibnamefont {Durbaka}},\ and\ \bibinfo {author}
  {\bibfnamefont {J.~R.}\ \bibnamefont {Koduru}},\ }\bibfield  {title}
  {\bibinfo {title} {Potential of the magnetic hollow sphere nanocomposite
  (graphene oxide-gadolinium oxide) for arsenic removal from real field water
  and antimicrobial applications},\ }\href
  {https://doi.org/10.1016/j.jhazmat.2020.123882} {\bibfield  {journal}
  {\bibinfo  {journal} {Journal of Hazardous Materials}\ }\textbf {\bibinfo
  {volume} {402}},\ \bibinfo {pages} {123882} (\bibinfo {year}
  {2021})}\BibitemShut {NoStop}%
\bibitem [{\citenamefont {Choi}\ \emph {et~al.}(2020)\citenamefont {Choi},
  \citenamefont {Lingamdinne}, \citenamefont {Yang}, \citenamefont {Chang},\
  and\ \citenamefont {Koduru}}]{choiFabricationChitosanGraphene2020}%
  \BibitemOpen
  \bibfield  {author} {\bibinfo {author} {\bibfnamefont {J.-S.}\ \bibnamefont
  {Choi}}, \bibinfo {author} {\bibfnamefont {L.~P.}\ \bibnamefont
  {Lingamdinne}}, \bibinfo {author} {\bibfnamefont {J.-K.}\ \bibnamefont
  {Yang}}, \bibinfo {author} {\bibfnamefont {Y.-Y.}\ \bibnamefont {Chang}},\
  and\ \bibinfo {author} {\bibfnamefont {J.~R.}\ \bibnamefont {Koduru}},\
  }\bibfield  {title} {\bibinfo {title} {Fabrication of chitosan/graphene
  oxide-gadolinium nanorods as a novel nanocomposite for arsenic removal from
  aqueous solutions},\ }\href {https://doi.org/10.1016/j.molliq.2020.114410}
  {\bibfield  {journal} {\bibinfo  {journal} {Journal of Molecular Liquids}\
  }\textbf {\bibinfo {volume} {320}},\ \bibinfo {pages} {114410} (\bibinfo
  {year} {2020})}\BibitemShut {NoStop}%
\bibitem [{\citenamefont {Tabatabaiee~Bafrooee}\ \emph
  {et~al.}(2021)\citenamefont {Tabatabaiee~Bafrooee}, \citenamefont {Moniri},
  \citenamefont {Ahmad~Panahi}, \citenamefont {Miralinaghi},\ and\
  \citenamefont
  {Hasani}}]{tabatabaieebafrooeeEthylenediamineFunctionalizedMagnetic2021}%
  \BibitemOpen
  \bibfield  {author} {\bibinfo {author} {\bibfnamefont {A.~A.}\ \bibnamefont
  {Tabatabaiee~Bafrooee}}, \bibinfo {author} {\bibfnamefont {E.}~\bibnamefont
  {Moniri}}, \bibinfo {author} {\bibfnamefont {H.}~\bibnamefont
  {Ahmad~Panahi}}, \bibinfo {author} {\bibfnamefont {M.}~\bibnamefont
  {Miralinaghi}},\ and\ \bibinfo {author} {\bibfnamefont {A.~H.}\ \bibnamefont
  {Hasani}},\ }\bibfield  {title} {\bibinfo {title} {Ethylenediamine
  functionalized magnetic graphene oxide ({{Fe3O4}}@{{GO-EDA}}) as an
  efficient~adsorbent in {{Arsenic}}({{III}}) decontamination from aqueous
  solution},\ }\href {https://doi.org/10.1007/s11164-020-04368-5} {\bibfield
  {journal} {\bibinfo  {journal} {Research on Chemical Intermediates}\ }\textbf
  {\bibinfo {volume} {47}},\ \bibinfo {pages} {1397} (\bibinfo {year}
  {2021})}\BibitemShut {NoStop}%
\bibitem [{\citenamefont {Ma}\ \emph {et~al.}(2022)\citenamefont {Ma},
  \citenamefont {Zheng}, \citenamefont {Yang}, \citenamefont {Wu},
  \citenamefont {Dong}, \citenamefont {Gao},\ and\ \citenamefont
  {Zhao}}]{maComputationalStudyAdsorption2022}%
  \BibitemOpen
  \bibfield  {author} {\bibinfo {author} {\bibfnamefont {K.}~\bibnamefont
  {Ma}}, \bibinfo {author} {\bibfnamefont {D.}~\bibnamefont {Zheng}}, \bibinfo
  {author} {\bibfnamefont {W.}~\bibnamefont {Yang}}, \bibinfo {author}
  {\bibfnamefont {C.}~\bibnamefont {Wu}}, \bibinfo {author} {\bibfnamefont
  {S.}~\bibnamefont {Dong}}, \bibinfo {author} {\bibfnamefont {Z.}~\bibnamefont
  {Gao}},\ and\ \bibinfo {author} {\bibfnamefont {X.}~\bibnamefont {Zhao}},\
  }\bibfield  {title} {\bibinfo {title} {A computational study on the
  adsorption of arsenic pollutants on graphene-based single-atom iron
  adsorbents},\ }\href {https://doi.org/10.1039/D1CP02170B} {\bibfield
  {journal} {\bibinfo  {journal} {Physical Chemistry Chemical Physics}\
  }\textbf {\bibinfo {volume} {24}},\ \bibinfo {pages} {13156} (\bibinfo {year}
  {2022})}\BibitemShut {NoStop}%
\bibitem [{\citenamefont {Gazzari}\ and\ \citenamefont
  {{Cort{\'e}s-Arriagada}}(2019)}]{gazzariInteractionTrivalentArsenic2019}%
  \BibitemOpen
  \bibfield  {author} {\bibinfo {author} {\bibfnamefont {S.}~\bibnamefont
  {Gazzari}}\ and\ \bibinfo {author} {\bibfnamefont {D.}~\bibnamefont
  {{Cort{\'e}s-Arriagada}}},\ }\bibfield  {title} {\bibinfo {title}
  {Interaction of trivalent arsenic on different topologies of {{Fe-doped}}
  graphene nanosheets at water environments: {{A}} computational study},\
  }\href {https://doi.org/10.1016/j.molliq.2019.111137} {\bibfield  {journal}
  {\bibinfo  {journal} {Journal of Molecular Liquids}\ }\textbf {\bibinfo
  {volume} {289}},\ \bibinfo {pages} {111137} (\bibinfo {year}
  {2019})}\BibitemShut {NoStop}%
\bibitem [{\citenamefont {Srivastava}\ \emph {et~al.}(2017)\citenamefont
  {Srivastava}, \citenamefont {Kommu}, \citenamefont {Sinha},\ and\
  \citenamefont {Singh}}]{srivastavaRemovalArsenicIons2017}%
  \BibitemOpen
  \bibfield  {author} {\bibinfo {author} {\bibfnamefont {R.}~\bibnamefont
  {Srivastava}}, \bibinfo {author} {\bibfnamefont {A.}~\bibnamefont {Kommu}},
  \bibinfo {author} {\bibfnamefont {N.}~\bibnamefont {Sinha}},\ and\ \bibinfo
  {author} {\bibfnamefont {J.~K.}\ \bibnamefont {Singh}},\ }\bibfield  {title}
  {\bibinfo {title} {Removal of arsenic ions using hexagonal boron nitride and
  graphene nanosheets: A molecular dynamics study},\ }\href
  {https://doi.org/10.1080/08927022.2017.1321754} {\bibfield  {journal}
  {\bibinfo  {journal} {Molecular Simulation}\ }\textbf {\bibinfo {volume}
  {43}},\ \bibinfo {pages} {985} (\bibinfo {year} {2017})}\BibitemShut
  {NoStop}%
\bibitem [{\citenamefont {Kusaba}\ \emph {et~al.}(2019)\citenamefont {Kusaba},
  \citenamefont {Li}, \citenamefont {Kempisty}, \citenamefont {{von
  Spakovsky}},\ and\ \citenamefont
  {Kangawa}}]{kusabaCH4AdsorptionProbability2019}%
  \BibitemOpen
  \bibfield  {author} {\bibinfo {author} {\bibfnamefont {A.}~\bibnamefont
  {Kusaba}}, \bibinfo {author} {\bibfnamefont {G.}~\bibnamefont {Li}}, \bibinfo
  {author} {\bibfnamefont {P.}~\bibnamefont {Kempisty}}, \bibinfo {author}
  {\bibfnamefont {M.}~\bibnamefont {{von Spakovsky}}},\ and\ \bibinfo {author}
  {\bibfnamefont {Y.}~\bibnamefont {Kangawa}},\ }\bibfield  {title} {\bibinfo
  {title} {{{CH4 Adsorption Probability}} on {{GaN}}(0001) and (000-1) during
  {{Metalorganic Vapor Phase Epitaxy}} and {{Its Relationship}} to {{Carbon
  Contamination}} in the {{Films}}},\ }\href
  {https://doi.org/10.3390/ma12060972} {\bibfield  {journal} {\bibinfo
  {journal} {Materials}\ }\textbf {\bibinfo {volume} {12}},\ \bibinfo {pages}
  {972} (\bibinfo {year} {2019})}\BibitemShut {NoStop}%
\bibitem [{\citenamefont {Kusaba}\ \emph {et~al.}(2017)\citenamefont {Kusaba},
  \citenamefont {Li}, \citenamefont {Von~Spakovsky}, \citenamefont {Kangawa},\
  and\ \citenamefont {Kakimoto}}]{kusabaModelingNonEquilibriumProcess2017}%
  \BibitemOpen
  \bibfield  {author} {\bibinfo {author} {\bibfnamefont {A.}~\bibnamefont
  {Kusaba}}, \bibinfo {author} {\bibfnamefont {G.}~\bibnamefont {Li}}, \bibinfo
  {author} {\bibfnamefont {M.~R.}\ \bibnamefont {Von~Spakovsky}}, \bibinfo
  {author} {\bibfnamefont {Y.}~\bibnamefont {Kangawa}},\ and\ \bibinfo {author}
  {\bibfnamefont {K.}~\bibnamefont {Kakimoto}},\ }\bibfield  {title} {\bibinfo
  {title} {Modeling the {{Non-Equilibrium Process}} of the {{Chemical
  Adsorption}} of {{Ammonia}} on {{GaN}}(0001) {{Reconstructed Surfaces Based}}
  on {{Steepest-Entropy-Ascent Quantum Thermodynamics}}},\ }\href
  {https://doi.org/10.3390/ma10080948} {\bibfield  {journal} {\bibinfo
  {journal} {Materials}\ }\textbf {\bibinfo {volume} {10}},\ \bibinfo {pages}
  {948} (\bibinfo {year} {2017})}\BibitemShut {NoStop}%
\bibitem [{\citenamefont {McDonald}\ \emph
  {et~al.}(2023{\natexlab{a}})\citenamefont {McDonald}, \citenamefont
  {Von~Spakovsky},\ and\ \citenamefont
  {Reynolds}}]{mcdonaldPredictingnonequilibriumfolding2023}%
  \BibitemOpen
  \bibfield  {author} {\bibinfo {author} {\bibfnamefont {J.}~\bibnamefont
  {McDonald}}, \bibinfo {author} {\bibfnamefont {M.~R.}\ \bibnamefont
  {Von~Spakovsky}},\ and\ \bibinfo {author} {\bibfnamefont {W.~T.}\
  \bibnamefont {Reynolds}},\ }\bibfield  {title} {\bibinfo {title} {Predicting
  non-equilibrium folding behavior of polymer chains using the
  steepest-entropy-ascent quantum thermodynamic framework},\ }\href
  {https://doi.org/10.1063/5.0137444} {\bibfield  {journal} {\bibinfo
  {journal} {The Journal of Chemical Physics}\ }\textbf {\bibinfo {volume}
  {158}},\ \bibinfo {pages} {104904} (\bibinfo {year}
  {2023}{\natexlab{a}})}\BibitemShut {NoStop}%
\bibitem [{\citenamefont {McDonald}\ \emph {et~al.}(2024)\citenamefont
  {McDonald}, \citenamefont {{von Spakovsky}},\ and\ \citenamefont
  {Reynolds}}]{mcdonaldPredictingIonSequestration2024}%
  \BibitemOpen
  \bibfield  {author} {\bibinfo {author} {\bibfnamefont {J.}~\bibnamefont
  {McDonald}}, \bibinfo {author} {\bibfnamefont {M.~R.}\ \bibnamefont {{von
  Spakovsky}}},\ and\ \bibinfo {author} {\bibfnamefont {W.~T.}\ \bibnamefont
  {Reynolds}},\ }\bibfield  {title} {\bibinfo {title} {Predicting {{Ion
  Sequestration}} in {{Charged Polymers}} with the {{Steepest-Entropy-Ascent
  Quantum Thermodynamic Framework}}},\ }\href
  {https://doi.org/10.3390/nano14050458} {\bibfield  {journal} {\bibinfo
  {journal} {Nanomaterials}\ }\textbf {\bibinfo {volume} {14}},\ \bibinfo
  {pages} {458} (\bibinfo {year} {2024})}\BibitemShut {NoStop}%
\bibitem [{\citenamefont {McDonald}\ \emph
  {et~al.}(2023{\natexlab{b}})\citenamefont {McDonald}, \citenamefont
  {Von~Spakovsky},\ and\ \citenamefont
  {Reynolds}}]{mcdonaldPredictingPolymerBrush2023}%
  \BibitemOpen
  \bibfield  {author} {\bibinfo {author} {\bibfnamefont {J.}~\bibnamefont
  {McDonald}}, \bibinfo {author} {\bibfnamefont {M.~R.}\ \bibnamefont
  {Von~Spakovsky}},\ and\ \bibinfo {author} {\bibfnamefont {W.~T.}\
  \bibnamefont {Reynolds}},\ }\bibfield  {title} {\bibinfo {title} {Predicting
  {{Polymer Brush Behavior}} in {{Solvents Using}} the
  {{Steepest-Entropy-Ascent Quantum Thermodynamic Framework}}},\ }\href
  {https://doi.org/10.1021/acs.jpcb.3c02713} {\bibfield  {journal} {\bibinfo
  {journal} {The Journal of Physical Chemistry B}\ }\textbf {\bibinfo {volume}
  {127}},\ \bibinfo {pages} {10370} (\bibinfo {year}
  {2023}{\natexlab{b}})}\BibitemShut {NoStop}%
\bibitem [{\citenamefont {McDonald}\ \emph {et~al.}(2022)\citenamefont
  {McDonald}, \citenamefont {{von Spakovsky}},\ and\ \citenamefont
  {Reynolds}}]{mcdonaldEntropydrivenmicrostructureevolution2022}%
  \BibitemOpen
  \bibfield  {author} {\bibinfo {author} {\bibfnamefont {J.}~\bibnamefont
  {McDonald}}, \bibinfo {author} {\bibfnamefont {M.~R.}\ \bibnamefont {{von
  Spakovsky}}},\ and\ \bibinfo {author} {\bibfnamefont {W.~T.}\ \bibnamefont
  {Reynolds}},\ }\bibfield  {title} {\bibinfo {title} {Entropy-driven
  microstructure evolution predicted with the steepest-entropy-ascent quantum
  thermodynamic framework},\ }\href
  {https://doi.org/10.1016/j.actamat.2022.118163} {\bibfield  {journal}
  {\bibinfo  {journal} {Acta Materialia}\ }\textbf {\bibinfo {volume} {237}},\
  \bibinfo {pages} {118163} (\bibinfo {year} {2022})}\BibitemShut {NoStop}%
\bibitem [{\citenamefont {Yamada}\ \emph
  {et~al.}(2019{\natexlab{a}})\citenamefont {Yamada}, \citenamefont {{von
  Spakovsky}},\ and\ \citenamefont
  {Reynolds}}]{yamadaLowtemperatureAtomisticSpin2019}%
  \BibitemOpen
  \bibfield  {author} {\bibinfo {author} {\bibfnamefont {R.}~\bibnamefont
  {Yamada}}, \bibinfo {author} {\bibfnamefont {M.~R.}\ \bibnamefont {{von
  Spakovsky}}},\ and\ \bibinfo {author} {\bibfnamefont {W.~T.}\ \bibnamefont
  {Reynolds}},\ }\bibfield  {title} {\bibinfo {title} {Low-temperature
  atomistic spin relaxation and non-equilibrium intensive properties using
  steepest-entropy-ascent quantum-inspired thermodynamics modeling},\ }\href
  {https://doi.org/10.1088/1361-648X/ab4014} {\bibfield  {journal} {\bibinfo
  {journal} {Journal of Physics: Condensed Matter}\ }\textbf {\bibinfo {volume}
  {31}},\ \bibinfo {pages} {505901} (\bibinfo {year}
  {2019}{\natexlab{a}})}\BibitemShut {NoStop}%
\bibitem [{\citenamefont {Yamada}\ \emph
  {et~al.}(2019{\natexlab{b}})\citenamefont {Yamada}, \citenamefont {{von
  Spakovsky}},\ and\ \citenamefont
  {Reynolds}}]{yamadaPredictingcontinuousdiscontinuous2019}%
  \BibitemOpen
  \bibfield  {author} {\bibinfo {author} {\bibfnamefont {R.}~\bibnamefont
  {Yamada}}, \bibinfo {author} {\bibfnamefont {M.~R.}\ \bibnamefont {{von
  Spakovsky}}},\ and\ \bibinfo {author} {\bibfnamefont {W.~T.}\ \bibnamefont
  {Reynolds}},\ }\bibfield  {title} {\bibinfo {title} {Predicting continuous
  and discontinuous phase decompositions using steepest-entropy-ascent quantum
  thermodynamics},\ }\href {https://doi.org/10.1103/PhysRevE.99.052121}
  {\bibfield  {journal} {\bibinfo  {journal} {Physical Review E}\ }\textbf
  {\bibinfo {volume} {99}},\ \bibinfo {pages} {052121} (\bibinfo {year}
  {2019}{\natexlab{b}})}\BibitemShut {NoStop}%
\bibitem [{\citenamefont {Yamada}\ \emph {et~al.}(2020)\citenamefont {Yamada},
  \citenamefont {{von Spakovsky}},\ and\ \citenamefont
  {Reynolds}}]{yamadaKineticpathwaysordering2020}%
  \BibitemOpen
  \bibfield  {author} {\bibinfo {author} {\bibfnamefont {R.}~\bibnamefont
  {Yamada}}, \bibinfo {author} {\bibfnamefont {M.~R.}\ \bibnamefont {{von
  Spakovsky}}},\ and\ \bibinfo {author} {\bibfnamefont {W.~T.}\ \bibnamefont
  {Reynolds}},\ }\bibfield  {title} {\bibinfo {title} {Kinetic pathways of
  ordering and phase separation using classical solid state models within the
  steepest-entropy-ascent quantum thermodynamic framework},\ }\href
  {https://doi.org/10.1016/j.actamat.2019.10.002} {\bibfield  {journal}
  {\bibinfo  {journal} {Acta Materialia}\ }\textbf {\bibinfo {volume} {182}},\
  \bibinfo {pages} {87} (\bibinfo {year} {2020})}\BibitemShut {NoStop}%
\bibitem [{\citenamefont {Yamada}\ \emph {et~al.}(2018)\citenamefont {Yamada},
  \citenamefont {{von Spakovsky}},\ and\ \citenamefont
  {Reynolds}}]{yamadamethodpredictingnonequilibrium2018}%
  \BibitemOpen
  \bibfield  {author} {\bibinfo {author} {\bibfnamefont {R.}~\bibnamefont
  {Yamada}}, \bibinfo {author} {\bibfnamefont {M.~R.}\ \bibnamefont {{von
  Spakovsky}}},\ and\ \bibinfo {author} {\bibfnamefont {W.~T.}\ \bibnamefont
  {Reynolds}},\ }\bibfield  {title} {\bibinfo {title} {A method for predicting
  non-equilibrium thermal expansion using steepest-entropy-ascent quantum
  thermodynamics},\ }\href {https://doi.org/10.1088/1361-648X/aad072}
  {\bibfield  {journal} {\bibinfo  {journal} {Journal of Physics: Condensed
  Matter}\ }\textbf {\bibinfo {volume} {30}},\ \bibinfo {pages} {325901}
  (\bibinfo {year} {2018})}\BibitemShut {NoStop}%
\bibitem [{\citenamefont {Goswami}\ \emph {et~al.}(2021)\citenamefont
  {Goswami}, \citenamefont {Bielitz}, \citenamefont {Verbridge},\ and\
  \citenamefont {{von Spakovsky}}}]{goswamithermodynamicscalinglaw2021}%
  \BibitemOpen
  \bibfield  {author} {\bibinfo {author} {\bibfnamefont {I.}~\bibnamefont
  {Goswami}}, \bibinfo {author} {\bibfnamefont {R.}~\bibnamefont {Bielitz}},
  \bibinfo {author} {\bibfnamefont {S.~S.}\ \bibnamefont {Verbridge}},\ and\
  \bibinfo {author} {\bibfnamefont {M.~R.}\ \bibnamefont {{von Spakovsky}}},\
  }\bibfield  {title} {\bibinfo {title} {A thermodynamic scaling law for
  electrically perturbed lipid membranes: {{Validation}} with steepest entropy
  ascent framework},\ }\href {https://doi.org/10.1016/j.bioelechem.2021.107800}
  {\bibfield  {journal} {\bibinfo  {journal} {Bioelectrochemistry}\ }\textbf
  {\bibinfo {volume} {140}},\ \bibinfo {pages} {107800} (\bibinfo {year}
  {2021})}\BibitemShut {NoStop}%
\bibitem [{\citenamefont {{von Spakovsky}}\ \emph {et~al.}(2020)\citenamefont
  {{von Spakovsky}}, \citenamefont {Schlosser}, \citenamefont {Martin},\ and\
  \citenamefont {Josyula}}]{vonspakovskyPredictingChemicalKinetics2020}%
  \BibitemOpen
  \bibfield  {author} {\bibinfo {author} {\bibfnamefont {M.~R.}\ \bibnamefont
  {{von Spakovsky}}}, \bibinfo {author} {\bibfnamefont {C.}~\bibnamefont
  {Schlosser}}, \bibinfo {author} {\bibfnamefont {J.~B.}\ \bibnamefont
  {Martin}},\ and\ \bibinfo {author} {\bibfnamefont {E.}~\bibnamefont
  {Josyula}},\ }\bibfield  {title} {\bibinfo {title} {Predicting the {{Chemical
  Kinetics}} of {{Air}} at {{High Temperatures Using Steepest-Entropy-Ascent
  Quantum Thermodynamics}}},\ }in\ \href {https://doi.org/10.2514/6.2020-3274}
  {\emph {\bibinfo {booktitle} {{{AIAA AVIATION}} 2020 {{FORUM}}}}}\ (\bibinfo
  {publisher} {{American Institute of Aeronautics and Astronautics}},\ \bibinfo
  {address} {VIRTUAL EVENT},\ \bibinfo {year} {2020})\BibitemShut {NoStop}%
\bibitem [{\citenamefont {{Monta{\~n}ez-Barrera}}\ \emph
  {et~al.}(2022)\citenamefont {{Monta{\~n}ez-Barrera}}, \citenamefont {{von
  Spakovsky}}, \citenamefont {Damian~Ascencio},\ and\ \citenamefont
  {{Cano-Andrade}}}]{montanez-barreraDecoherencepredictionssuperconducting2022}%
  \BibitemOpen
  \bibfield  {author} {\bibinfo {author} {\bibfnamefont {J.~A.}\ \bibnamefont
  {{Monta{\~n}ez-Barrera}}}, \bibinfo {author} {\bibfnamefont {M.~R.}\
  \bibnamefont {{von Spakovsky}}}, \bibinfo {author} {\bibfnamefont {C.~E.}\
  \bibnamefont {Damian~Ascencio}},\ and\ \bibinfo {author} {\bibfnamefont
  {S.}~\bibnamefont {{Cano-Andrade}}},\ }\bibfield  {title} {\bibinfo {title}
  {Decoherence predictions in a superconducting quantum processor using the
  steepest-entropy-ascent quantum thermodynamics framework},\ }\href
  {https://doi.org/10.1103/PhysRevA.106.032426} {\bibfield  {journal} {\bibinfo
   {journal} {Physical Review A}\ }\textbf {\bibinfo {volume} {106}},\ \bibinfo
  {pages} {032426} (\bibinfo {year} {2022})}\BibitemShut {NoStop}%
\bibitem [{\citenamefont {{Monta{\~n}ez-Barrera}}\ \emph
  {et~al.}(2020)\citenamefont {{Monta{\~n}ez-Barrera}}, \citenamefont
  {{Damian-Ascencio}}, \citenamefont {{von Spakovsky}},\ and\ \citenamefont
  {{Cano-Andrade}}}]{montanez-barreraLossofentanglementpredictioncontrolledphase2020}%
  \BibitemOpen
  \bibfield  {author} {\bibinfo {author} {\bibfnamefont {J.~A.}\ \bibnamefont
  {{Monta{\~n}ez-Barrera}}}, \bibinfo {author} {\bibfnamefont {C.~E.}\
  \bibnamefont {{Damian-Ascencio}}}, \bibinfo {author} {\bibfnamefont {M.~R.}\
  \bibnamefont {{von Spakovsky}}},\ and\ \bibinfo {author} {\bibfnamefont
  {S.}~\bibnamefont {{Cano-Andrade}}},\ }\bibfield  {title} {\bibinfo {title}
  {Loss-of-entanglement prediction of a controlled-phase gate in the framework
  of steepest-entropy-ascent quantum thermodynamics},\ }\href
  {https://doi.org/10.1103/PhysRevA.101.052336} {\bibfield  {journal} {\bibinfo
   {journal} {Physical Review A}\ }\textbf {\bibinfo {volume} {101}},\ \bibinfo
  {pages} {052336} (\bibinfo {year} {2020})}\BibitemShut {NoStop}%
\bibitem [{\citenamefont {{Cano-Andrade}}\ \emph {et~al.}(2015)\citenamefont
  {{Cano-Andrade}}, \citenamefont {Beretta},\ and\ \citenamefont {{von
  Spakovsky}}}]{cano-andradeSteepestentropyascentquantumthermodynamic2015}%
  \BibitemOpen
  \bibfield  {author} {\bibinfo {author} {\bibfnamefont {S.}~\bibnamefont
  {{Cano-Andrade}}}, \bibinfo {author} {\bibfnamefont {G.~P.}\ \bibnamefont
  {Beretta}},\ and\ \bibinfo {author} {\bibfnamefont {M.~R.}\ \bibnamefont
  {{von Spakovsky}}},\ }\bibfield  {title} {\bibinfo {title}
  {Steepest-entropy-ascent quantum thermodynamic modeling of decoherence in two
  different microscopic composite systems},\ }\href
  {https://doi.org/10.1103/PhysRevA.91.013848} {\bibfield  {journal} {\bibinfo
  {journal} {Physical Review A}\ }\textbf {\bibinfo {volume} {91}},\ \bibinfo
  {pages} {013848} (\bibinfo {year} {2015})}\BibitemShut {NoStop}%
\bibitem [{\citenamefont {Li}\ \emph {et~al.}(2018)\citenamefont {Li},
  \citenamefont {{von Spakovsky}},\ and\ \citenamefont
  {Hin}}]{liSteepestentropyascent2018}%
  \BibitemOpen
  \bibfield  {author} {\bibinfo {author} {\bibfnamefont {G.}~\bibnamefont
  {Li}}, \bibinfo {author} {\bibfnamefont {M.~R.}\ \bibnamefont {{von
  Spakovsky}}},\ and\ \bibinfo {author} {\bibfnamefont {C.}~\bibnamefont
  {Hin}},\ }\bibfield  {title} {\bibinfo {title} {Steepest entropy ascent
  quantum thermodynamic model of electron and phonon transport},\ }\href
  {https://doi.org/10.1103/PhysRevB.97.024308} {\bibfield  {journal} {\bibinfo
  {journal} {Physical Review B}\ }\textbf {\bibinfo {volume} {97}},\ \bibinfo
  {pages} {024308} (\bibinfo {year} {2018})}\BibitemShut {NoStop}%
\bibitem [{\citenamefont {Beretta}(2014)}]{berettaSteepestEntropyAscent2014}%
  \BibitemOpen
  \bibfield  {author} {\bibinfo {author} {\bibfnamefont {G.~P.}\ \bibnamefont
  {Beretta}},\ }\bibfield  {title} {\bibinfo {title} {Steepest entropy ascent
  model for far-nonequilibrium thermodynamics: {{Unified}} implementation of
  the maximum entropy production principle},\ }\href
  {https://doi.org/10.1103/PhysRevE.90.042113} {\bibfield  {journal} {\bibinfo
  {journal} {Physical Review E}\ }\textbf {\bibinfo {volume} {90}},\ \bibinfo
  {pages} {042113} (\bibinfo {year} {2014})}\BibitemShut {NoStop}%
\bibitem [{\citenamefont {Beretta}(2020)}]{berettaFourthLawThermodynamics2020}%
  \BibitemOpen
  \bibfield  {author} {\bibinfo {author} {\bibfnamefont {G.~P.}\ \bibnamefont
  {Beretta}},\ }\bibfield  {title} {\bibinfo {title} {The fourth law of
  thermodynamics: Steepest entropy ascent},\ }\href
  {https://doi.org/10.1098/rsta.2019.0168} {\bibfield  {journal} {\bibinfo
  {journal} {Philosophical Transactions of the Royal Society A: Mathematical,
  Physical and Engineering Sciences}\ }\textbf {\bibinfo {volume} {378}},\
  \bibinfo {pages} {20190168} (\bibinfo {year} {2020})}\BibitemShut {NoStop}%
\bibitem [{\citenamefont
  {Beretta}(2010)}]{berettaMaximumEntropyProduction2010}%
  \BibitemOpen
  \bibfield  {author} {\bibinfo {author} {\bibfnamefont {G.~P.}\ \bibnamefont
  {Beretta}},\ }\bibfield  {title} {\bibinfo {title} {Maximum entropy
  production rate in quantum thermodynamics},\ }\href
  {https://doi.org/10.1088/1742-6596/237/1/012004} {\bibfield  {journal}
  {\bibinfo  {journal} {Journal of Physics: Conference Series}\ }\textbf
  {\bibinfo {volume} {237}},\ \bibinfo {pages} {012004} (\bibinfo {year}
  {2010})}\BibitemShut {NoStop}%
\bibitem [{\citenamefont {{Saldana-Robles}}\ \emph {et~al.}(2025)\citenamefont
  {{Saldana-Robles}}, \citenamefont {Damian}, \citenamefont {Reynolds},\ and\
  \citenamefont {Spakovsky}}]{saldana-roblesModelpredictingadsorption2025}%
  \BibitemOpen
  \bibfield  {author} {\bibinfo {author} {\bibfnamefont {A.}~\bibnamefont
  {{Saldana-Robles}}}, \bibinfo {author} {\bibfnamefont {C.}~\bibnamefont
  {Damian}}, \bibinfo {author} {\bibfnamefont {W.~T.}\ \bibnamefont
  {Reynolds}},\ and\ \bibinfo {author} {\bibfnamefont {M.~R.~V.}\ \bibnamefont
  {Spakovsky}},\ }\bibfield  {title} {\bibinfo {title} {Model for predicting
  adsorption isotherms and the kinetics of adsorption via
  steepest-entropy-ascent quantum thermodynamics},\ }\href
  {https://doi.org/10.1007/s10450-025-00629-0} {\bibfield  {journal} {\bibinfo
  {journal} {Adsorption}\ }\textbf {\bibinfo {volume} {31}},\ \bibinfo {pages}
  {76} (\bibinfo {year} {2025})}\BibitemShut {NoStop}%
\bibitem [{\citenamefont {Vogel}\ \emph {et~al.}(2013)\citenamefont {Vogel},
  \citenamefont {Li}, \citenamefont {W{\"u}st},\ and\ \citenamefont
  {Landau}}]{vogelGenericHierarchicalFramework2013}%
  \BibitemOpen
  \bibfield  {author} {\bibinfo {author} {\bibfnamefont {T.}~\bibnamefont
  {Vogel}}, \bibinfo {author} {\bibfnamefont {Y.~W.}\ \bibnamefont {Li}},
  \bibinfo {author} {\bibfnamefont {T.}~\bibnamefont {W{\"u}st}},\ and\
  \bibinfo {author} {\bibfnamefont {D.~P.}\ \bibnamefont {Landau}},\ }\bibfield
   {title} {\bibinfo {title} {Generic, {{Hierarchical Framework}} for
  {{Massively Parallel Wang-Landau Sampling}}},\ }\href
  {https://doi.org/10.1103/PhysRevLett.110.210603} {\bibfield  {journal}
  {\bibinfo  {journal} {Physical Review Letters}\ }\textbf {\bibinfo {volume}
  {110}},\ \bibinfo {pages} {210603} (\bibinfo {year} {2013})}\BibitemShut
  {NoStop}%
\bibitem [{\citenamefont {Vogel}\ \emph {et~al.}(2014)\citenamefont {Vogel},
  \citenamefont {Li}, \citenamefont {W{\"u}st},\ and\ \citenamefont
  {Landau}}]{vogelScalableReplicaexchangeFramework2014}%
  \BibitemOpen
  \bibfield  {author} {\bibinfo {author} {\bibfnamefont {T.}~\bibnamefont
  {Vogel}}, \bibinfo {author} {\bibfnamefont {Y.~W.}\ \bibnamefont {Li}},
  \bibinfo {author} {\bibfnamefont {T.}~\bibnamefont {W{\"u}st}},\ and\
  \bibinfo {author} {\bibfnamefont {D.~P.}\ \bibnamefont {Landau}},\ }\bibfield
   {title} {\bibinfo {title} {Scalable replica-exchange framework for
  {{Wang-Landau}} sampling},\ }\href
  {https://doi.org/10.1103/PhysRevE.90.023302} {\bibfield  {journal} {\bibinfo
  {journal} {Physical Review E}\ }\textbf {\bibinfo {volume} {90}},\ \bibinfo
  {pages} {023302} (\bibinfo {year} {2014})}\BibitemShut {NoStop}%
\bibitem [{\citenamefont {Vogel}\ \emph {et~al.}(2018)\citenamefont {Vogel},
  \citenamefont {Wai~Li},\ and\ \citenamefont
  {P~Landau}}]{vogelPracticalGuideReplicaexchange2018}%
  \BibitemOpen
  \bibfield  {author} {\bibinfo {author} {\bibfnamefont {T.}~\bibnamefont
  {Vogel}}, \bibinfo {author} {\bibfnamefont {Y.}~\bibnamefont {Wai~Li}},\ and\
  \bibinfo {author} {\bibfnamefont {D.}~\bibnamefont {P~Landau}},\ }\bibfield
  {title} {\bibinfo {title} {A practical guide to replica-exchange
  {{Wang}}---{{Landau}} simulations},\ }\href
  {https://doi.org/10.1088/1742-6596/1012/1/012003} {\bibfield  {journal}
  {\bibinfo  {journal} {Journal of Physics: Conference Series}\ }\textbf
  {\bibinfo {volume} {1012}},\ \bibinfo {pages} {012003} (\bibinfo {year}
  {2018})}\BibitemShut {NoStop}%
\bibitem [{\citenamefont {Ghaedi}\ and\ \citenamefont
  {Vafaei}(2017)}]{ghaediApplicationsartificialneural2017}%
  \BibitemOpen
  \bibfield  {author} {\bibinfo {author} {\bibfnamefont {A.~M.}\ \bibnamefont
  {Ghaedi}}\ and\ \bibinfo {author} {\bibfnamefont {A.}~\bibnamefont
  {Vafaei}},\ }\bibfield  {title} {\bibinfo {title} {Applications of artificial
  neural networks for adsorption removal of dyes from aqueous solution: {{A}}
  review},\ }\href {https://doi.org/10.1016/j.cis.2017.04.015} {\bibfield
  {journal} {\bibinfo  {journal} {Advances in Colloid and Interface Science}\
  }\textbf {\bibinfo {volume} {245}},\ \bibinfo {pages} {20} (\bibinfo {year}
  {2017})}\BibitemShut {NoStop}%
\bibitem [{\citenamefont {Zhang}\ \emph {et~al.}(2019)\citenamefont {Zhang},
  \citenamefont {Dai}, \citenamefont {Liu}, \citenamefont {Peng}, \citenamefont
  {Du}, \citenamefont {Chang}, \citenamefont {Ali}, \citenamefont {Naz},\ and\
  \citenamefont {Saroj}}]{zhangAppraisalCuii2019}%
  \BibitemOpen
  \bibfield  {author} {\bibinfo {author} {\bibfnamefont {Y.}~\bibnamefont
  {Zhang}}, \bibinfo {author} {\bibfnamefont {M.}~\bibnamefont {Dai}}, \bibinfo
  {author} {\bibfnamefont {K.}~\bibnamefont {Liu}}, \bibinfo {author}
  {\bibfnamefont {C.}~\bibnamefont {Peng}}, \bibinfo {author} {\bibfnamefont
  {Y.}~\bibnamefont {Du}}, \bibinfo {author} {\bibfnamefont {Q.}~\bibnamefont
  {Chang}}, \bibinfo {author} {\bibfnamefont {I.}~\bibnamefont {Ali}}, \bibinfo
  {author} {\bibfnamefont {I.}~\bibnamefont {Naz}},\ and\ \bibinfo {author}
  {\bibfnamefont {D.~P.}\ \bibnamefont {Saroj}},\ }\bibfield  {title} {\bibinfo
  {title} {Appraisal of {{Cu}}( {\textsc{ii}} ) adsorption by graphene oxide
  and its modelling {\emph{via}} artificial neural network},\ }\href
  {https://doi.org/10.1039/C9RA06079K} {\bibfield  {journal} {\bibinfo
  {journal} {RSC Advances}\ }\textbf {\bibinfo {volume} {9}},\ \bibinfo {pages}
  {30240} (\bibinfo {year} {2019})}\BibitemShut {NoStop}%
\bibitem [{\citenamefont {Das}\ and\ \citenamefont
  {Mishra}(2021)}]{dasArtificialneuralnetwork2021}%
  \BibitemOpen
  \bibfield  {author} {\bibinfo {author} {\bibfnamefont {S.}~\bibnamefont
  {Das}}\ and\ \bibinfo {author} {\bibfnamefont {S.}~\bibnamefont {Mishra}},\
  }\bibfield  {title} {\bibinfo {title} {Artificial neural network ({{ANN}})
  approach for prediction and modeling of breakthrough curve analysis of
  fixed-bed adsorption of iron ions from aqueous solution by activated carbon
  from {{{\emph{Limonia}}}}{\emph{ acidissima}} shell},\ }\href
  {https://doi.org/10.1515/ijcre-2021-0053} {\bibfield  {journal} {\bibinfo
  {journal} {International Journal of Chemical Reactor Engineering}\ }\textbf
  {\bibinfo {volume} {19}},\ \bibinfo {pages} {1197} (\bibinfo {year}
  {2021})}\BibitemShut {NoStop}%
\bibitem [{\citenamefont {Yusuf}\ \emph {et~al.}(2020)\citenamefont {Yusuf},
  \citenamefont {Song},\ and\ \citenamefont {Li}}]{yusufFixedbedcolumn2020}%
  \BibitemOpen
  \bibfield  {author} {\bibinfo {author} {\bibfnamefont {M.}~\bibnamefont
  {Yusuf}}, \bibinfo {author} {\bibfnamefont {K.}~\bibnamefont {Song}},\ and\
  \bibinfo {author} {\bibfnamefont {L.}~\bibnamefont {Li}},\ }\bibfield
  {title} {\bibinfo {title} {Fixed bed column and artificial neural network
  model to predict heavy metals adsorption dynamic on surfactant decorated
  graphene},\ }\href {https://doi.org/10.1016/j.colsurfa.2019.124076}
  {\bibfield  {journal} {\bibinfo  {journal} {Colloids and Surfaces A:
  Physicochemical and Engineering Aspects}\ }\textbf {\bibinfo {volume}
  {585}},\ \bibinfo {pages} {124076} (\bibinfo {year} {2020})}\BibitemShut
  {NoStop}%
\bibitem [{\citenamefont {Deylami}\ \emph {et~al.}(2023)\citenamefont
  {Deylami}, \citenamefont {Sabzevari}, \citenamefont {Ghaedi}, \citenamefont
  {Azqhandi},\ and\ \citenamefont
  {Marahel}}]{deylamiEfficientphotodegradationdisulfine2023}%
  \BibitemOpen
  \bibfield  {author} {\bibinfo {author} {\bibfnamefont {{\relax
  Sh}.}~\bibnamefont {Deylami}}, \bibinfo {author} {\bibfnamefont {M.~H.}\
  \bibnamefont {Sabzevari}}, \bibinfo {author} {\bibfnamefont {M.}~\bibnamefont
  {Ghaedi}}, \bibinfo {author} {\bibfnamefont {M.~A.}\ \bibnamefont
  {Azqhandi}},\ and\ \bibinfo {author} {\bibfnamefont {F.}~\bibnamefont
  {Marahel}},\ }\bibfield  {title} {\bibinfo {title} {Efficient
  photodegradation of disulfine blue dye and {{Tetracycline}} over {{Robust}}
  and {{Green}} g-{{CN}}/{{Ag3VO4}}/{{PAN}} nanofibers: {{Experimental}}
  design, {{RSM}}, {{RBF-NN}} and {{ANFIS}} modeling},\ }\href
  {https://doi.org/10.1016/j.psep.2022.10.080} {\bibfield  {journal} {\bibinfo
  {journal} {Process Safety and Environmental Protection}\ }\textbf {\bibinfo
  {volume} {169}},\ \bibinfo {pages} {71} (\bibinfo {year} {2023})}\BibitemShut
  {NoStop}%
\bibitem [{\citenamefont {Pournamdari}\ \emph {et~al.}(2024)\citenamefont
  {Pournamdari}, \citenamefont {Niknam}, \citenamefont {Davoudi},\ and\
  \citenamefont {Khazali}}]{pournamdariResponsesurfacemethodology2024}%
  \BibitemOpen
  \bibfield  {author} {\bibinfo {author} {\bibfnamefont {E.}~\bibnamefont
  {Pournamdari}}, \bibinfo {author} {\bibfnamefont {L.}~\bibnamefont {Niknam}},
  \bibinfo {author} {\bibfnamefont {S.}~\bibnamefont {Davoudi}},\ and\ \bibinfo
  {author} {\bibfnamefont {F.}~\bibnamefont {Khazali}},\ }\bibfield  {title}
  {\bibinfo {title} {Response surface methodology, and artificial neural
  network model for removal of textile dye {{Reactive Yellow}} 105 from
  wastewater using {{Zeolitic Imidazolate-67}} modified by
  {{Fe}}{\textsubscript{3}} {{O}}{\textsubscript{4}} nanoparticles},\ }\href
  {https://doi.org/10.1080/15226514.2023.2226217} {\bibfield  {journal}
  {\bibinfo  {journal} {International Journal of Phytoremediation}\ }\textbf
  {\bibinfo {volume} {26}},\ \bibinfo {pages} {98} (\bibinfo {year}
  {2024})}\BibitemShut {NoStop}%
\bibitem [{\citenamefont {Mombeni~Goodajdar}\ \emph {et~al.}(2023)\citenamefont
  {Mombeni~Goodajdar}, \citenamefont {Marahel}, \citenamefont {Niknam},
  \citenamefont {Pournamdari},\ and\ \citenamefont
  {Mousavi}}]{mombenigoodajdarUltrasonicAssistedNeural2023}%
  \BibitemOpen
  \bibfield  {author} {\bibinfo {author} {\bibfnamefont {B.}~\bibnamefont
  {Mombeni~Goodajdar}}, \bibinfo {author} {\bibfnamefont {F.}~\bibnamefont
  {Marahel}}, \bibinfo {author} {\bibfnamefont {L.}~\bibnamefont {Niknam}},
  \bibinfo {author} {\bibfnamefont {E.}~\bibnamefont {Pournamdari}},\ and\
  \bibinfo {author} {\bibfnamefont {E.}~\bibnamefont {Mousavi}},\ }\bibfield
  {title} {\bibinfo {title} {Ultrasonic {{Assisted}} and {{Neural Network
  Model}} for {{Adsorption}} of of humic acid ({{HAs}}) by {{Synthesised
  CM-$\beta$-CD-Fe3O4NPs}} from {{Aqueous Solutions}}},\ }\href
  {https://doi.org/10.1080/03067319.2021.1940159} {\bibfield  {journal}
  {\bibinfo  {journal} {International Journal of Environmental Analytical
  Chemistry}\ }\textbf {\bibinfo {volume} {103}},\ \bibinfo {pages} {5483}
  (\bibinfo {year} {2023})}\BibitemShut {NoStop}%
\bibitem [{\citenamefont {Geramizadegan}\ \emph {et~al.}(2023)\citenamefont
  {Geramizadegan}, \citenamefont {Niknam},\ and\ \citenamefont
  {Pournamdari}}]{geramizadeganMolecularlyImprintedPolymers2023}%
  \BibitemOpen
  \bibfield  {author} {\bibinfo {author} {\bibfnamefont {A.}~\bibnamefont
  {Geramizadegan}}, \bibinfo {author} {\bibfnamefont {L.}~\bibnamefont
  {Niknam}},\ and\ \bibinfo {author} {\bibfnamefont {E.}~\bibnamefont
  {Pournamdari}},\ }\bibfield  {title} {\bibinfo {title} {Molecularly
  {{Imprinted Polymers}} for {{Selective Extraction}} and {{Determination}} of
  {{Toxic Herbicide Bentazon}} in {{Water Samples Using Liquid Chromatography}}
  and {{Assessment}} of {{Mean Square Error Using Artficial Neural Network
  Model}}},\ }\href {https://doi.org/10.1134/S1061934823050052} {\bibfield
  {journal} {\bibinfo  {journal} {Journal of Analytical Chemistry}\ }\textbf
  {\bibinfo {volume} {78}},\ \bibinfo {pages} {572} (\bibinfo {year}
  {2023})}\BibitemShut {NoStop}%
\bibitem [{\citenamefont {Ansari}\ \emph {et~al.}(2020)\citenamefont {Ansari},
  \citenamefont {Azamat},\ and\ \citenamefont
  {Khataee}}]{ansariComputationalStudyRemoval2020}%
  \BibitemOpen
  \bibfield  {author} {\bibinfo {author} {\bibfnamefont {P.}~\bibnamefont
  {Ansari}}, \bibinfo {author} {\bibfnamefont {J.}~\bibnamefont {Azamat}},\
  and\ \bibinfo {author} {\bibfnamefont {A.}~\bibnamefont {Khataee}},\
  }\bibfield  {title} {\bibinfo {title} {Computational study on the removal of
  trihalomethanes from water using functionalized graphene oxide membranes},\
  }\href {https://doi.org/10.1016/j.chemphys.2019.110589} {\bibfield  {journal}
  {\bibinfo  {journal} {Chemical Physics}\ }\textbf {\bibinfo {volume} {531}},\
  \bibinfo {pages} {110589} (\bibinfo {year} {2020})}\BibitemShut {NoStop}%
\bibitem [{\citenamefont {Tanaka}\ \emph {et~al.}(2013)\citenamefont {Tanaka},
  \citenamefont {Takahashi}, \citenamefont {Yamaguchi}, \citenamefont {Kim},
  \citenamefont {Zheng},\ and\ \citenamefont
  {Sakamitsu}}]{tanakaDifferenceDiffusionCoefficients2013}%
  \BibitemOpen
  \bibfield  {author} {\bibinfo {author} {\bibfnamefont {M.}~\bibnamefont
  {Tanaka}}, \bibinfo {author} {\bibfnamefont {Y.}~\bibnamefont {Takahashi}},
  \bibinfo {author} {\bibfnamefont {N.}~\bibnamefont {Yamaguchi}}, \bibinfo
  {author} {\bibfnamefont {K.-W.}\ \bibnamefont {Kim}}, \bibinfo {author}
  {\bibfnamefont {G.}~\bibnamefont {Zheng}},\ and\ \bibinfo {author}
  {\bibfnamefont {M.}~\bibnamefont {Sakamitsu}},\ }\bibfield  {title} {\bibinfo
  {title} {The difference of diffusion coefficients in water for arsenic
  compounds at various {{pH}} and its dominant factors implied by molecular
  simulations},\ }\href {https://doi.org/10.1016/j.gca.2012.12.004} {\bibfield
  {journal} {\bibinfo  {journal} {Geochimica et Cosmochimica Acta}\ }\textbf
  {\bibinfo {volume} {105}},\ \bibinfo {pages} {360} (\bibinfo {year}
  {2013})}\BibitemShut {NoStop}%
\bibitem [{\citenamefont {Jorgensen}\ \emph {et~al.}(1996)\citenamefont
  {Jorgensen}, \citenamefont {Maxwell},\ and\ \citenamefont
  {{Tirado-Rives}}}]{jorgensenDevelopmentTestingOPLS1996}%
  \BibitemOpen
  \bibfield  {author} {\bibinfo {author} {\bibfnamefont {W.~L.}\ \bibnamefont
  {Jorgensen}}, \bibinfo {author} {\bibfnamefont {D.~S.}\ \bibnamefont
  {Maxwell}},\ and\ \bibinfo {author} {\bibfnamefont {J.}~\bibnamefont
  {{Tirado-Rives}}},\ }\bibfield  {title} {\bibinfo {title} {Development and
  {{Testing}} of the {{OPLS All-Atom Force Field}} on {{Conformational
  Energetics}} and {{Properties}} of {{Organic Liquids}}},\ }\href
  {https://doi.org/10.1021/ja9621760} {\bibfield  {journal} {\bibinfo
  {journal} {Journal of the American Chemical Society}\ }\textbf {\bibinfo
  {volume} {118}},\ \bibinfo {pages} {11225} (\bibinfo {year}
  {1996})}\BibitemShut {NoStop}%
\bibitem [{\citenamefont {Wang}\ and\ \citenamefont
  {Landau}(2001{\natexlab{a}})}]{wangDeterminingDensityStates2001}%
  \BibitemOpen
  \bibfield  {author} {\bibinfo {author} {\bibfnamefont {F.}~\bibnamefont
  {Wang}}\ and\ \bibinfo {author} {\bibfnamefont {D.~P.}\ \bibnamefont
  {Landau}},\ }\bibfield  {title} {\bibinfo {title} {Determining the density of
  states for classical statistical models: {{A}} random walk algorithm to
  produce a flat histogram},\ }\href
  {https://doi.org/10.1103/PhysRevE.64.056101} {\bibfield  {journal} {\bibinfo
  {journal} {Physical Review E}\ }\textbf {\bibinfo {volume} {64}},\ \bibinfo
  {pages} {056101} (\bibinfo {year} {2001}{\natexlab{a}})}\BibitemShut
  {NoStop}%
\bibitem [{\citenamefont {Wang}\ and\ \citenamefont
  {Landau}(2001{\natexlab{b}})}]{wangEfficientMultipleRangeRandom2001}%
  \BibitemOpen
  \bibfield  {author} {\bibinfo {author} {\bibfnamefont {F.}~\bibnamefont
  {Wang}}\ and\ \bibinfo {author} {\bibfnamefont {D.~P.}\ \bibnamefont
  {Landau}},\ }\bibfield  {title} {\bibinfo {title} {Efficient,
  {{Multiple-Range Random Walk Algorithm}} to {{Calculate}} the {{Density}} of
  {{States}}},\ }\href {https://doi.org/10.1103/PhysRevLett.86.2050} {\bibfield
   {journal} {\bibinfo  {journal} {Physical Review Letters}\ }\textbf {\bibinfo
  {volume} {86}},\ \bibinfo {pages} {2050} (\bibinfo {year}
  {2001}{\natexlab{b}})}\BibitemShut {NoStop}%
\bibitem [{\citenamefont {Li}\ \emph {et~al.}(2014)\citenamefont {Li},
  \citenamefont {Vogel}, \citenamefont {W{\"u}st},\ and\ \citenamefont
  {Landau}}]{liNewParadigmPetascale2014}%
  \BibitemOpen
  \bibfield  {author} {\bibinfo {author} {\bibfnamefont {Y.~W.}\ \bibnamefont
  {Li}}, \bibinfo {author} {\bibfnamefont {T.}~\bibnamefont {Vogel}}, \bibinfo
  {author} {\bibfnamefont {T.}~\bibnamefont {W{\"u}st}},\ and\ \bibinfo
  {author} {\bibfnamefont {D.~P.}\ \bibnamefont {Landau}},\ }\bibfield  {title}
  {\bibinfo {title} {A new paradigm for petascale {{Monte Carlo}} simulation:
  {{Replica}} exchange {{Wang-Landau}} sampling},\ }\href
  {https://doi.org/10.1088/1742-6596/510/1/012012} {\bibfield  {journal}
  {\bibinfo  {journal} {Journal of Physics: Conference Series}\ }\textbf
  {\bibinfo {volume} {510}},\ \bibinfo {pages} {012012} (\bibinfo {year}
  {2014})}\BibitemShut {NoStop}%
\bibitem [{\citenamefont {Li}\ and\ \citenamefont
  {Von~Spakovsky}(2016)}]{liSteepestentropyascentQuantumThermodynamic2016}%
  \BibitemOpen
  \bibfield  {author} {\bibinfo {author} {\bibfnamefont {G.}~\bibnamefont
  {Li}}\ and\ \bibinfo {author} {\bibfnamefont {M.~R.}\ \bibnamefont
  {Von~Spakovsky}},\ }\bibfield  {title} {\bibinfo {title}
  {Steepest-entropy-ascent quantum thermodynamic modeling of the relaxation
  process of isolated chemically reactive systems using density of states and
  the concept of hypoequilibrium state},\ }\href
  {https://doi.org/10.1103/PhysRevE.93.012137} {\bibfield  {journal} {\bibinfo
  {journal} {Physical Review E}\ }\textbf {\bibinfo {volume} {93}},\ \bibinfo
  {pages} {012137} (\bibinfo {year} {2016})}\BibitemShut {NoStop}%
\bibitem [{\citenamefont {Li}\ and\ \citenamefont {{von
  Spakovsky}}(2016)}]{liGeneralizedThermodynamicRelations2016}%
  \BibitemOpen
  \bibfield  {author} {\bibinfo {author} {\bibfnamefont {G.}~\bibnamefont
  {Li}}\ and\ \bibinfo {author} {\bibfnamefont {M.~R.}\ \bibnamefont {{von
  Spakovsky}}},\ }\bibfield  {title} {\bibinfo {title} {Generalized
  thermodynamic relations for a system experiencing heat and mass diffusion in
  the far-from-equilibrium realm based on steepest entropy ascent},\ }\href
  {https://doi.org/10.1103/PhysRevE.94.032117} {\bibfield  {journal} {\bibinfo
  {journal} {Physical Review E}\ }\textbf {\bibinfo {volume} {94}},\ \bibinfo
  {pages} {032117} (\bibinfo {year} {2016})}\BibitemShut {NoStop}%
\bibitem [{\citenamefont {Li}\ and\ \citenamefont {{von
  Spakovsky}}(2018)}]{liSteepestentropyascentmodelmesoscopic2018}%
  \BibitemOpen
  \bibfield  {author} {\bibinfo {author} {\bibfnamefont {G.}~\bibnamefont
  {Li}}\ and\ \bibinfo {author} {\bibfnamefont {M.~R.}\ \bibnamefont {{von
  Spakovsky}}},\ }\bibfield  {title} {\bibinfo {title} {Steepest-entropy-ascent
  model of mesoscopic quantum systems far from equilibrium along with
  generalized thermodynamic definitions of measurement and reservoir},\ }\href
  {https://doi.org/10.1103/PhysRevE.98.042113} {\bibfield  {journal} {\bibinfo
  {journal} {Physical Review E}\ }\textbf {\bibinfo {volume} {98}},\ \bibinfo
  {pages} {042113} (\bibinfo {year} {2018})}\BibitemShut {NoStop}%
\bibitem [{\citenamefont {Su}\ \emph {et~al.}(2017)\citenamefont {Su},
  \citenamefont {Ye},\ and\ \citenamefont
  {Hmidi}}]{suHighperformanceIronOxide2017}%
  \BibitemOpen
  \bibfield  {author} {\bibinfo {author} {\bibfnamefont {H.}~\bibnamefont
  {Su}}, \bibinfo {author} {\bibfnamefont {Z.}~\bibnamefont {Ye}},\ and\
  \bibinfo {author} {\bibfnamefont {N.}~\bibnamefont {Hmidi}},\ }\bibfield
  {title} {\bibinfo {title} {High-performance iron oxide--graphene oxide
  nanocomposite adsorbents for arsenic removal},\ }\href
  {https://doi.org/10.1016/j.colsurfa.2017.02.065} {\bibfield  {journal}
  {\bibinfo  {journal} {Colloids and Surfaces A: Physicochemical and
  Engineering Aspects}\ }\textbf {\bibinfo {volume} {522}},\ \bibinfo {pages}
  {161} (\bibinfo {year} {2017})}\BibitemShut {NoStop}%
\bibitem [{\citenamefont {Pattanayak}\ \emph {et~al.}(2000)\citenamefont
  {Pattanayak}, \citenamefont {Mondal}, \citenamefont {Mathew},\ and\
  \citenamefont {Lalvani}}]{pattanayakparametricevaluationremoval2000}%
  \BibitemOpen
  \bibfield  {author} {\bibinfo {author} {\bibfnamefont {J.}~\bibnamefont
  {Pattanayak}}, \bibinfo {author} {\bibfnamefont {K.}~\bibnamefont {Mondal}},
  \bibinfo {author} {\bibfnamefont {S.}~\bibnamefont {Mathew}},\ and\ \bibinfo
  {author} {\bibfnamefont {S.}~\bibnamefont {Lalvani}},\ }\bibfield  {title}
  {\bibinfo {title} {A parametric evaluation of the removal of {{As}}({{V}})
  and {{As}}({{III}}) by carbon-based adsorbents},\ }\href
  {https://doi.org/10.1016/S0008-6223(99)00144-X} {\bibfield  {journal}
  {\bibinfo  {journal} {Carbon}\ }\textbf {\bibinfo {volume} {38}},\ \bibinfo
  {pages} {589} (\bibinfo {year} {2000})}\BibitemShut {NoStop}%
\bibitem [{\citenamefont {Zhang}\ and\ \citenamefont
  {Itoh}(2005)}]{zhangIronoxideloadedslag2005}%
  \BibitemOpen
  \bibfield  {author} {\bibinfo {author} {\bibfnamefont {F.-S.}\ \bibnamefont
  {Zhang}}\ and\ \bibinfo {author} {\bibfnamefont {H.}~\bibnamefont {Itoh}},\
  }\bibfield  {title} {\bibinfo {title} {Iron oxide-loaded slag for arsenic
  removal from aqueous system},\ }\href
  {https://doi.org/10.1016/j.chemosphere.2004.12.019} {\bibfield  {journal}
  {\bibinfo  {journal} {Chemosphere}\ }\textbf {\bibinfo {volume} {60}},\
  \bibinfo {pages} {319} (\bibinfo {year} {2005})}\BibitemShut {NoStop}%
\bibitem [{\citenamefont
  {Ghimire}(2003)}]{ghimireAdsorptiveseparationarsenate2003}%
  \BibitemOpen
  \bibfield  {author} {\bibinfo {author} {\bibfnamefont {K.}~\bibnamefont
  {Ghimire}},\ }\bibfield  {title} {\bibinfo {title} {Adsorptive separation of
  arsenate and arsenite anions from aqueous medium by using orange waste},\
  }\href {https://doi.org/10.1016/j.watres.2003.08.029} {\bibfield  {journal}
  {\bibinfo  {journal} {Water Research}\ }\textbf {\bibinfo {volume} {37}},\
  \bibinfo {pages} {4945} (\bibinfo {year} {2003})}\BibitemShut {NoStop}%
\bibitem [{\citenamefont {Xia}\ \emph {et~al.}(2022)\citenamefont {Xia},
  \citenamefont {Zhou}, \citenamefont {Xu}, \citenamefont {Wang}, \citenamefont
  {Lan}, \citenamefont {Fan}, \citenamefont {Wang}, \citenamefont {Liu},
  \citenamefont {Chen}, \citenamefont {Feng}, \citenamefont {Tu}, \citenamefont
  {Yang}, \citenamefont {Chen},\ and\ \citenamefont
  {Fang}}]{xiaUnexpectedlyefficiention2022}%
  \BibitemOpen
  \bibfield  {author} {\bibinfo {author} {\bibfnamefont {X.}~\bibnamefont
  {Xia}}, \bibinfo {author} {\bibfnamefont {F.}~\bibnamefont {Zhou}}, \bibinfo
  {author} {\bibfnamefont {J.}~\bibnamefont {Xu}}, \bibinfo {author}
  {\bibfnamefont {Z.}~\bibnamefont {Wang}}, \bibinfo {author} {\bibfnamefont
  {J.}~\bibnamefont {Lan}}, \bibinfo {author} {\bibfnamefont {Y.}~\bibnamefont
  {Fan}}, \bibinfo {author} {\bibfnamefont {Z.}~\bibnamefont {Wang}}, \bibinfo
  {author} {\bibfnamefont {W.}~\bibnamefont {Liu}}, \bibinfo {author}
  {\bibfnamefont {J.}~\bibnamefont {Chen}}, \bibinfo {author} {\bibfnamefont
  {S.}~\bibnamefont {Feng}}, \bibinfo {author} {\bibfnamefont {Y.}~\bibnamefont
  {Tu}}, \bibinfo {author} {\bibfnamefont {Y.}~\bibnamefont {Yang}}, \bibinfo
  {author} {\bibfnamefont {L.}~\bibnamefont {Chen}},\ and\ \bibinfo {author}
  {\bibfnamefont {H.}~\bibnamefont {Fang}},\ }\bibfield  {title} {\bibinfo
  {title} {Unexpectedly efficient ion desorption of graphene-based materials},\
  }\href {https://doi.org/10.1038/s41467-022-35077-9} {\bibfield  {journal}
  {\bibinfo  {journal} {Nature Communications}\ }\textbf {\bibinfo {volume}
  {13}},\ \bibinfo {pages} {7247} (\bibinfo {year} {2022})}\BibitemShut
  {NoStop}%
\bibitem [{\citenamefont {Javarani}\ \emph {et~al.}(2022)\citenamefont
  {Javarani}, \citenamefont {Malakootian}, \citenamefont {Hassani},\ and\
  \citenamefont {Javid}}]{javaraniRemovalHeavyMetal2022}%
  \BibitemOpen
  \bibfield  {author} {\bibinfo {author} {\bibfnamefont {L.}~\bibnamefont
  {Javarani}}, \bibinfo {author} {\bibfnamefont {M.}~\bibnamefont
  {Malakootian}}, \bibinfo {author} {\bibfnamefont {A.~H.}\ \bibnamefont
  {Hassani}},\ and\ \bibinfo {author} {\bibfnamefont {A.~H.}\ \bibnamefont
  {Javid}},\ }\bibfield  {title} {\bibinfo {title} {Removal of {{Heavy Metal
  Ions}} from {{Water Using Functionalized Carbon Nanosheet}}: {{A
  Density-Functional Theory Study}}},\ }\bibfield  {journal} {\bibinfo
  {journal} {Jundishapur Journal of Health Sciences}\ }\textbf {\bibinfo
  {volume} {14}},\ \href {https://doi.org/10.5812/jjhs.122840}
  {10.5812/jjhs.122840} (\bibinfo {year} {2022})\BibitemShut {NoStop}%
\bibitem [{\citenamefont {Dimakis}\ \emph {et~al.}(2019)\citenamefont
  {Dimakis}, \citenamefont {Salas}, \citenamefont {Gonzalez}, \citenamefont
  {Vadodaria}, \citenamefont {Ruiz},\ and\ \citenamefont
  {Bhatti}}]{dimakisLiNaAdsorption2019}%
  \BibitemOpen
  \bibfield  {author} {\bibinfo {author} {\bibfnamefont {N.}~\bibnamefont
  {Dimakis}}, \bibinfo {author} {\bibfnamefont {I.}~\bibnamefont {Salas}},
  \bibinfo {author} {\bibfnamefont {L.}~\bibnamefont {Gonzalez}}, \bibinfo
  {author} {\bibfnamefont {O.}~\bibnamefont {Vadodaria}}, \bibinfo {author}
  {\bibfnamefont {K.}~\bibnamefont {Ruiz}},\ and\ \bibinfo {author}
  {\bibfnamefont {M.~I.}\ \bibnamefont {Bhatti}},\ }\bibfield  {title}
  {\bibinfo {title} {Li and {{Na Adsorption}} on {{Graphene}} and {{Graphene
  Oxide Examined}} by {{Density Functional Theory}}, {{Quantum Theory}} of
  {{Atoms}} in {{Molecules}}, and {{Electron Localization Function}}},\ }\href
  {https://doi.org/10.3390/molecules24040754} {\bibfield  {journal} {\bibinfo
  {journal} {Molecules}\ }\textbf {\bibinfo {volume} {24}},\ \bibinfo {pages}
  {754} (\bibinfo {year} {2019})}\BibitemShut {NoStop}%
\bibitem [{\citenamefont {Singh}\ \emph {et~al.}(2022)\citenamefont {Singh},
  \citenamefont {Naik}, \citenamefont {U}, \citenamefont {Khan}, \citenamefont
  {Wani}, \citenamefont {Behera}, \citenamefont {Nath}, \citenamefont {Bhati},
  \citenamefont {Singh},\ and\ \citenamefont
  {Ramamurthy}}]{singhSystematicStudyArsenic2022}%
  \BibitemOpen
  \bibfield  {author} {\bibinfo {author} {\bibfnamefont {S.}~\bibnamefont
  {Singh}}, \bibinfo {author} {\bibfnamefont {T.~S. S.~K.}\ \bibnamefont
  {Naik}}, \bibinfo {author} {\bibfnamefont {B.}~\bibnamefont {U}}, \bibinfo
  {author} {\bibfnamefont {N.~A.}\ \bibnamefont {Khan}}, \bibinfo {author}
  {\bibfnamefont {A.~B.}\ \bibnamefont {Wani}}, \bibinfo {author}
  {\bibfnamefont {S.~K.}\ \bibnamefont {Behera}}, \bibinfo {author}
  {\bibfnamefont {B.}~\bibnamefont {Nath}}, \bibinfo {author} {\bibfnamefont
  {S.}~\bibnamefont {Bhati}}, \bibinfo {author} {\bibfnamefont
  {J.}~\bibnamefont {Singh}},\ and\ \bibinfo {author} {\bibfnamefont {P.~C.}\
  \bibnamefont {Ramamurthy}},\ }\bibfield  {title} {\bibinfo {title} {A
  systematic study of arsenic adsorption and removal from aqueous environments
  using novel graphene oxide functionalized {{UiO-66-NDC}} nanocomposites},\
  }\href {https://doi.org/10.1038/s41598-022-18959-2} {\bibfield  {journal}
  {\bibinfo  {journal} {Scientific Reports}\ }\textbf {\bibinfo {volume}
  {12}},\ \bibinfo {pages} {15802} (\bibinfo {year} {2022})}\BibitemShut
  {NoStop}%
\bibitem [{\citenamefont
  {Elmorsi}(2011)}]{elmorsiEquilibriumIsothermsKinetic2011}%
  \BibitemOpen
  \bibfield  {author} {\bibinfo {author} {\bibfnamefont {T.~M.}\ \bibnamefont
  {Elmorsi}},\ }\bibfield  {title} {\bibinfo {title} {Equilibrium {{Isotherms}}
  and {{Kinetic Studies}} of {{Removal}} of {{Methylene Blue Dye}} by
  {{Adsorption}} onto {{Miswak Leaves}} as a {{Natural Adsorbent}}},\ }\href
  {https://doi.org/10.4236/jep.2011.26093} {\bibfield  {journal} {\bibinfo
  {journal} {Journal of Environmental Protection}\ }\textbf {\bibinfo {volume}
  {02}},\ \bibinfo {pages} {817} (\bibinfo {year} {2011})}\BibitemShut
  {NoStop}%
\bibitem [{\citenamefont {Ayawei}\ \emph {et~al.}(2015)\citenamefont {Ayawei},
  \citenamefont {Angaye}, \citenamefont {Wankasi},\ and\ \citenamefont
  {Dikio}}]{ayaweiSynthesisCharacterizationApplication2015}%
  \BibitemOpen
  \bibfield  {author} {\bibinfo {author} {\bibfnamefont {N.}~\bibnamefont
  {Ayawei}}, \bibinfo {author} {\bibfnamefont {S.~S.}\ \bibnamefont {Angaye}},
  \bibinfo {author} {\bibfnamefont {D.}~\bibnamefont {Wankasi}},\ and\ \bibinfo
  {author} {\bibfnamefont {E.~D.}\ \bibnamefont {Dikio}},\ }\bibfield  {title}
  {\bibinfo {title} {Synthesis, {{Characterization}} and {{Application}} of
  {{Mg}}/{{Al Layered Double Hydroxide}} for the {{Degradation}} of {{Congo
  Red}} in {{Aqueous Solution}}},\ }\href
  {https://doi.org/10.4236/ojpc.2015.53007} {\bibfield  {journal} {\bibinfo
  {journal} {Open Journal of Physical Chemistry}\ }\textbf {\bibinfo {volume}
  {5}},\ \bibinfo {pages} {56} (\bibinfo {year} {2015})}\BibitemShut {NoStop}%
\end{thebibliography}

%%%%%%%%%%%%%%%%%%%%

\end{document}